\documentclass[journal=jacsat,manuscript=article]{achemso}

\usepackage{chemformula} 
\usepackage[T1]{fontenc} 
\usepackage{subfigure}

\usepackage[utf8]{inputenc}
\usepackage[T1]{fontenc}
\usepackage{mathptmx}
\usepackage{etoolbox}

\usepackage{derivative}
\usepackage{mathtools}
\usepackage{amsmath}
\usepackage{amsfonts}
\usepackage{bm}
\usepackage{physics}
\usepackage{hyperref}
\usepackage{subfigure}
\usepackage{xcolor}
\usepackage{multirow}

\usepackage{codes}

\newcommand{\rev}[1]{\textcolor{black}{#1}}

\newcommand{\e}{\mathrm{e}}
\providecommand{\abs}[1]{\lvert#1\rvert}

\newcommand{\psith}[1]{\psi_{\theta}^{(#1)}}
\newcommand{\phith}[1]{\phi_{\theta}^{(#1)}}

\author{Mart\'in A. Achondo}
\affiliation{Department of Mechanical Engineering, Universidad T\'ecnica Federico Santa Mar\'ia, Valpara\'iso, Chile.}
\author{Jehanzeb H. Chaudhry}%
 \email{jehanzeb@unm.edu.}
\affiliation{Department of Mathematics and Statistics, University of New Mexico, Albuquerque, NM, United States
}%

\author{Christopher D. Cooper}
\altaffiliation{Centro Cient\'ifico Tecnol\'ogico de Valpara\'iso (CCTVal), Universidad T\'ecnica Federico Santa Mar\'ia, Valpara\'iso, Chile.}
\email{christopher.cooper@usm.cl.}
\affiliation{Department of Mechanical Engineering, Universidad T\'ecnica Federico Santa Mar\'ia, Valpara\'iso, Chile.}

\title[An Investigation of PINN to solve the PBE]{An Investigation of Physics Informed Neural Networks to solve the Poisson-Boltzmann Equation in Molecular Electrostatics}

\abbreviations{IR,NMR,UV}

\begin{document}

\begin{tocentry}

\includegraphics[width=\textwidth]{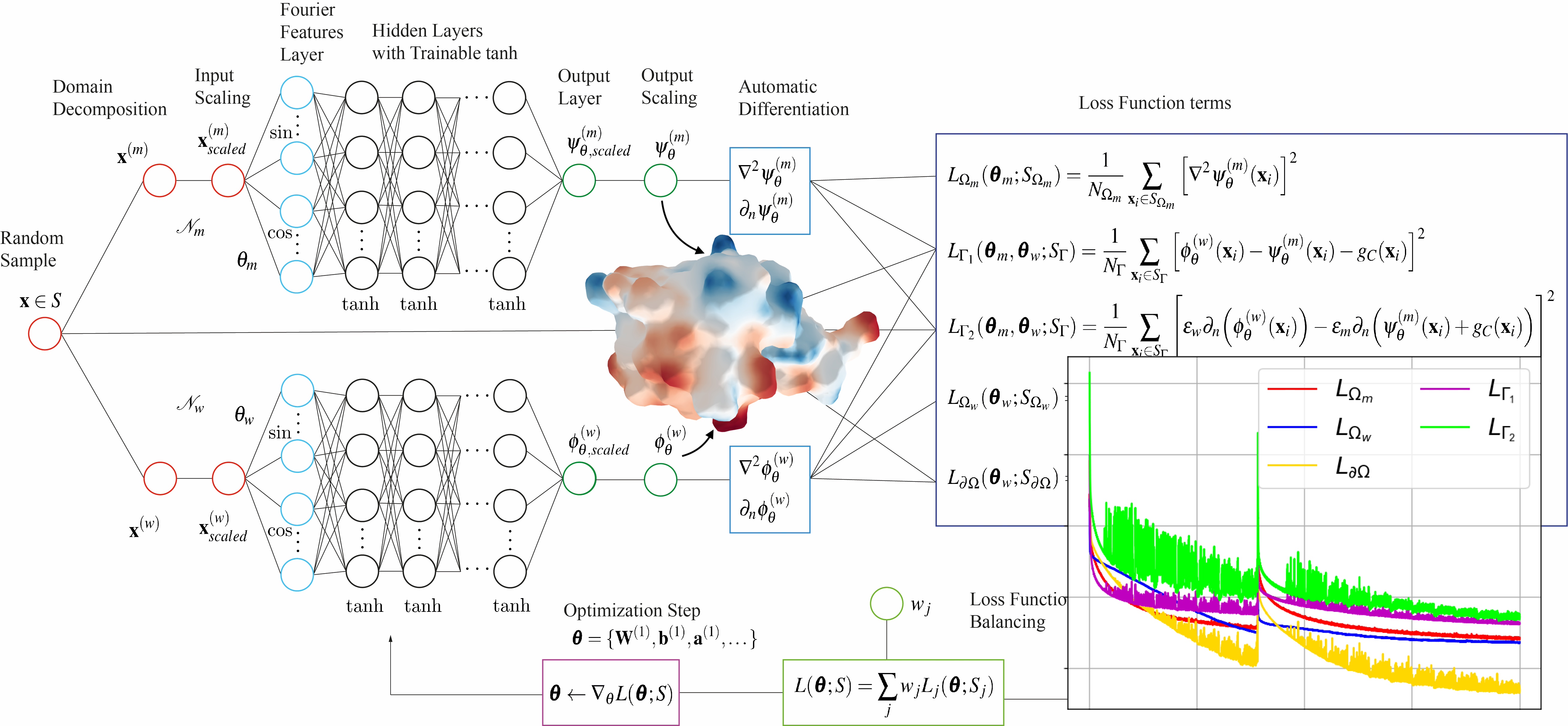}

\end{tocentry}

\begin{abstract}
Physics-informed neural networks (PINN) is a machine learning (ML)-based method to solve partial differential equations that has gained great popularity due to the fast development of ML libraries in the last few years. The Poisson-Boltzmann equation (PBE) is widely used to model mean-field electrostatics in molecular systems, and in this work we present a detailed investigation of the use of PINN to solve the \rev{linear} PBE. Starting from a multidomain  PINN for the \rev{linear} PBE with an interface, we assess the importance of incorporating different features into the neural network architecture. Our  findings indicate that the most accurate architecture  utilizes input and output scaling layers, a random Fourier features layer, trainable activation functions, and a loss balancing algorithm. The accuracy of our implementation is of the order of 10$^{-2}$---$10^{-3}$, which is similar to previous work using PINN to solve other differential equations. We also explore the possibility of incorporating experimental information into the model, and discuss challenges and future work, especially regarding the nonlinear PBE. We are providing an open-source implementation to easily perform computations from a PDB file. We hope this work will motivate application scientists into using PINN to study molecular electrostatics, as ML technology continues to evolve at a high pace. 
\end{abstract}

\section{\label{sec:intro}Introduction}
Implicit solvation is a widely used approach in molecular modeling and considers the solvent as a continuum. This massively decreases the number of degrees of freedom of the system, and hence, the computational cost of mean-field calculations. In electrostatics, a widely used  implicit method is the Poisson-Boltzmann equation (PBE) \cite{roux1999implicit,DecherchiETal2015} solved on a multi-region domain, where the dielectric constant and salt concentration have a sharp variation along interfaces. The PBE has been solved numerically for decades using finite differences  (FDM) \cite{GilsonETal1985,BakerETal2001,rocchia2001extending,boschitsch2011fast,jurrus2018improvements}, finite elements (FEM) \cite{cortis1997numerical,chen2007finite,xie2007new,bond2010first}, boundary elements (BEM) \cite{Shaw1985,YoonLenhoff1990,juffer1991electric,boschitsch2002fast,lu2006order,geng2013treecode,CooperBardhanBarba2014,search2022towards}, (semi) analytical \cite{felberg2017pb,siryk2022arbitrary,jha2023linear,jha2023domain}, and hybrid approaches \cite{boschitsch2004hybrid,ying2018hybrid,bsbbbc2023coupling}, proving useful for a large community. \rev{Even though the PBE is nonlinear, its linearized form is a good approximation in a large range of problems, from low-charge organic molecules to proteins,\cite{fogolari1999biomolecular} and is widely used in the computational chemistry community.\cite{AltmanBardhanWhiteTidor09,CooperBardhanBarba2014,cortis1997numerical,bond2010first,jha2023linear}} 

Scientific machine learning (SciML) has seen tremendous recent interest in utilizing tools from computer science, mathematics and statistics to solve complex scientific and engineering problems. SciML encompasses a wide array of methodologies such as digital twins~\cite{KKHTW22}, data assimilation~\cite{RCSJDCP19}, Bayesian inverse analysis~\cite{CaSo07}, model reduction~\cite{BGW15},  physics-informed machine learning~\cite{karniadakis2021physics} etc.
In particular, a prominent and widely used physics-informed machine learning approach to solve Partial Differential Equations (PDEs) is the  physics-informed neural networks (PINN).\cite{dissanayake1994neural,raissi2019physics,cuomo2022scientific} PINN represents the approximate solution of the differential equation with a neural network, where the network's parameters are optimized using a loss function that contains the PDE residual. PINN has attracted a lot of interest from the computational mathematics community, and has been used in various applications, such as fluid mechanics,\cite{CMWYK21} heat transfer,\cite{CWWPK21} electromagnetism,\cite{BBL23} acoustics,\cite{JPCMS24} Lie-Poisson systems,~\cite{EFSV24} among others.


The definition of the residual gives rise to different variations of PINN, for example, by writing it in variational \cite{yu2018deep,kharazmi2019variational} or boundary integral \cite{lin2021binet,sun2023binn} form.
There are also special forms of PINN that are designed for domain decomposition, for example, extended PINN (XPINN), \cite{jagtap2020extended} conservative PINN (cPINN),\cite{jagtap2020conservative} and distributed PINN (DPINN),\cite{dwivedi2019distributed} among others. These usually use one neural network per region, and they differ in the definition of the loss functions and the treatment of the interface conditions. 
 These methods have been adapted to solve linear elliptic partial differential equations with interfaces \cite{li2020deep,he2022mesh,wu2022inn,ying2023multi,tseng2023cusp,jiang2024generalization,sarma2024interface}, where domain decomposition enhances precision by accounting for particularities of the solution in each subdomain.  
 Additionally, there are theoretical studies that analyze their convergence\cite{WZTL23} and error bounds\cite{jiang2024generalization} for elliptic interface problems.
 Although the aforementioned works vary in their choice of optimization algorithms, parameters, interface conditions, and neural network architectures, they all use a domain decomposition PINN strategy, and we employ a similar strategy in devising a PINN for the \rev{linear} PBE.



The Poisson-Boltzmann model considered in this work is an example of an elliptic interface problem. 
Domain decomposition PINN methods (and others) have been recently applied to the PBE to compute electrostatic potentials and energies in molecular settings,~\cite{LiCaZh20,wu2022inn,wu2023solvingdeeponet,ying2023multi} similar to the present work.
For example, Li {\it et al.},\cite{LiCaZh20} proposed a PINN approach with variational principles to solve the PBE, through a multi-scale deep neural network. In that case, there is only a single neural network, whereas our work explicitly captures the interface by using a multi-domain approach. 
Also, the methodology in the work by Wu and Lu\cite{wu2022inn} is based on using an extended multiple-gradient descent (MGD)~\cite{mgd} algorithm in a multi-domain setting. This may be a difficuly, as optimizers like MGD are not readily available in common software packages like Tensor Flow\footnote{\url{https://www.tensorflow.org/}} or Torch~\footnote{\url{https://pytorch.org/}}. In this work, we developed a PINN method for the PBE utilizing the common Adam optimizer~\cite{KingBa15}. Moreover, Wu and Lu define the solute-solvent interface with the van der Waals surface,~\cite{whitley1998van} as opposed to the solvent-excluded surface (SES),\cite{connolly1985molecular} which prevails in PBE calculations, and is used in the present work. 
An alternative approach to  solve the PBE based  is based on IONet~\cite{wu2023solvingdeeponet}, which is a neural network that learns the differential operator with an interface.  However, this technique requires training with previously computed numerical solutions. 
On the other hand, Ying and co-workers \cite{ying2023multi} presented the multi-scale fusion network (MSFN) approach, that relies on sub-networks that approximate the solution at different frequencies (or scales), and use a least-square type loss function.
\rev{Also, Park and Jo\cite{park2024physics} solved the nonlinear PBE in two dimensions with neural networks, whereas here we focus on three-dimensional molecular structures.}

There are other physics informed machine learning techniques that have been applied to molecular electrostatics and involve the PBE, but are not designed to solve it. For example, PBML\cite{chen2024poisson} is a neural network model that was trained with a large dataset of PB solutions, and provides the solvation free energy from the molecular structure only. Similarly, pyPKa\cite{reis2024pypka} uses PB calculations to feed a machine learning model that estimates pKas.  

Regardless of the important progress of PINN for the PBE, further efforts are needed to understand the impact of different variants offered by PINN towards its applicability in practical cases. The present work intends to fill this gap by extensively testing different sampling strategies, architecture improvements, and loss function treatments to suggest good practices of PINN for the PB equation, and identify pressing challenges moving forward. Along with this analysis, we provide a TensorFlow-based Python code called XPPBE\cite{githubrepo} that computes electrostatic potentials and energies from a PDB structure using an easy interface, for interested application scientists.    

In this work we focus on solving the standard PBE with PINN, however, we believe this approach can pave the way for future advancements of molecular electrostatics simulation, distinguishing itself from traditional methods. 
For instance, PINN can be used to solve inverse problems~\cite{JAGTAP2022Inverse}, where physical properties, like permittivity, are learneable parameters.
Also, PINN gives the possibility to easily incorporate experimental measurements or information from molecular dynamics simulations through a loss function, opening new doors in model development.

\section{\label{sec:methods}Methods}
\subsection{\label{sec:pb}The Poisson-Boltzmann equation}

\begin{figure}
    \centering
    \includegraphics[width=0.4\linewidth]{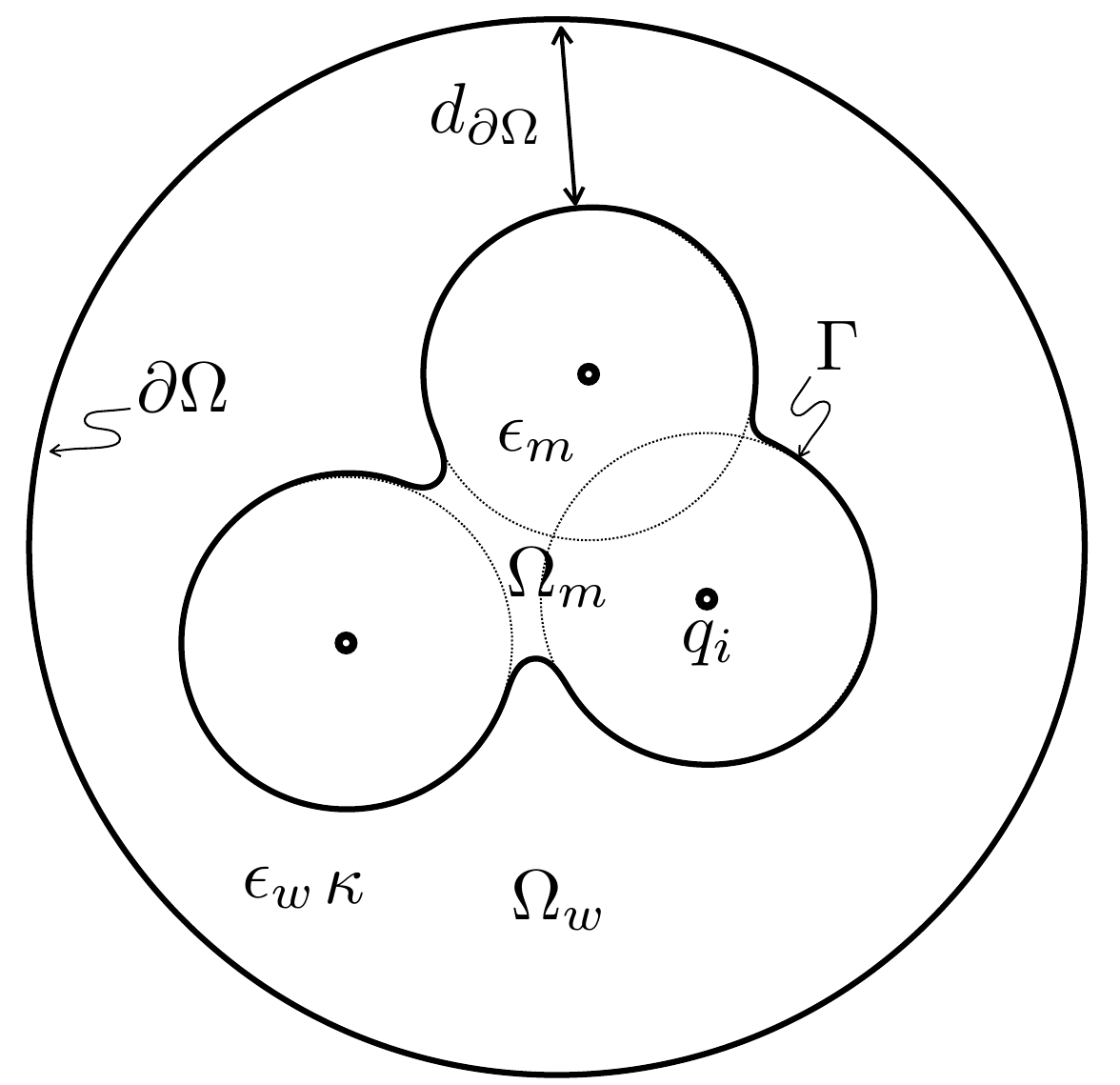}
    \caption{Sketch of molecule for the PB model. $\Omega_m$ is the solute domain (with partial charges $q_i$), $\Omega_w$ the solvent domain, $\Gamma$ the interface, and $\partial\Omega$ the edge of our domain, where we enforce boundary conditions, a minimum distance of $d_{\partial\Omega}$ away from $\Gamma$.}
    \label{fig:molecule}
\end{figure}

When a solute is immersed in a continuum solvent, it can be considered as a low-dielectric cavity ($\Omega_m$), with relative permittivity $\epsilon_m$ and delta-like partial charges ($q_i$), inside an infinite solvent domain ($\Omega_w$), as sketched in Fig. \ref{fig:molecule}. The domain is truncated at $\partial\Omega$ for practical purposes of the numerical method. In biological settings, the solvent is usually water (relative permitivity $\epsilon_w\approx$ 80) with salt ions that move in response to an external electric field. In equilibrium, the free ions arrange according to Boltzmann statistics, giving rise to the Poisson-Boltzmann equation \rev{for symmetric electrolytes}. The electrostatic potential,
\begin{equation}
\phi(\vb{x}) = \begin{cases}
     \phi^{(m)}(\vb{x}), \quad \vb{x} \in \Omega_m,\\
     \phi^{(w)}(\vb{x}), \quad \vb{x} \in \Omega_w,\\
\end{cases}    
\end{equation}
is modeled from the following coupled system of PDEs:
%
\begin{align} 
\label{eq:pbe}
     -\epsilon_{m}\nabla^{2} \phi^{(m)}(\mathbf{x}) = \frac{e}{k_BT\epsilon_0}\sum_{i=1}^{n_{c}}q_{i}\delta\left (\mathbf{x}-\mathbf{x}_{q,i} \right ), &\quad \mathbf{x} \ \in \ \Omega_{m} \nonumber\\
     -\nabla^{2}\phi^{(w)}(\mathbf{x})+\kappa_{w}^{2} \sinh(\phi^{(w)})(\mathbf{x}) = 0, &\quad \mathbf{x} \ \in \ \Omega_{w}  \nonumber \\
     \phi^{(m)}(\mathbf{x}) = \phi^{(w)}(\mathbf{x}),& \quad \mathbf{x} \ \in \ \Gamma  \nonumber \\
     \epsilon_{m}\partial_n \phi^{(m)}(\mathbf{x}) = \epsilon_{w}\partial_n \phi^{(w)}(\mathbf{x}),& \quad \mathbf{x} \ \in \ \Gamma  \nonumber \\
     \phi^{(m)}(\mathbf{x}\to\infty) = 0,&
\end{align}
where $q_i$ is one of the $n_c$ partial charges (represented as a Dirac $\delta$ point charges) in the solute molecule at locations $\vb{x}_{q,i}$, $\kappa_w$ is the inverse of the Debye length, and the electrostatic potential $\phi$ is nondimensionalized by $\frac{k_BT}{e}$ (Boltzmann constant times temperature, divided by the elementary charge).  The interface $\Gamma$ is usually defined either as the solvent accessible, \cite{LeRi71} solvent excluded, \cite{connolly1985molecular} van der Waals \cite{whitley1998van}, or Gaussian \cite{yu2006role} surfaces. In this work, we use the solvent-excluded surface (SES). The symbol $\partial_n$ corresponds to the derivative in the direction of an outward-facing normal to $\Gamma$.

Eq. \eqref{eq:pbe} is challenging to solve numerically because of the singularities at the charge's locations and interface conditions that are enforced in a complex geometry. There are several regularization techniques of the PB equation \cite{Zhou1996,chen2007finite, holst2012adaptive,LeeGengZhao2021} that alleviate the issue of the singularities, for example, by solving for a so-called {\it regular} or {\it reaction} potential, 
which is a difference between the electrostatic potential and the Coulomb potential. The Coulomb potential corresponds to the electrostatic potential due to a collection of point charges in free space, which (nondimensionalized by $\frac{k_BT}{e}$) is
\begin{equation}
g_C(\mathbf{x}) = \frac{e}{k_BT}\sum_i^{n_c}\frac{q_i}{4\pi\epsilon_m\epsilon_0|\mathbf{x}-\mathbf{x}_{q,i}|}.
\end{equation}
%
The regular potential $\psi$ defined on the entire domain $\Omega$ is,
\begin{equation}
    \psi = \phi - g_C, \quad \text{ in } \quad \Omega.
\end{equation}
In our case, we use a regularized version of the PB equation that only decomposes the electrostatic potential into singular and non-singular components in $\Omega_m$. That is, we set 
\begin{equation}
\psi^{(m)}  = \phi^{(m)} - g_C, \quad  \text{ in } \quad \Omega_m. 
\end{equation}
Moreover, in many practical applications, such as proteins, linearizing the hyperbolic sine in the PB equation yields acceptable results. \rev{This leads to the linearized Regularized Poisson-Boltzmann (RPB) equation,}
%
%
%
\begin{align}
    \label{eq:lrpbe}
         \nabla^{2} \psi^{(m)}(\mathbf{x}) = 0, &\quad \mathbf{x} \ \in \ \Omega_{m} \nonumber\\
     -\nabla^{2}\phi^{(w)}(\mathbf{x})+\kappa_{w}^{2}\phi^{(w)}(\mathbf{x}) = 0, &\quad \mathbf{x} \ \in \ \Omega_{w}  \nonumber \\
      \psi^{(m)}(\mathbf{x}) + g_C(\mathbf{x})= \phi^{(w)}(\mathbf{x}),& \quad \mathbf{x} \ \in \ \Gamma  \nonumber \\
     \epsilon_{m}\left(\partial_n \psi^{(m)}(\mathbf{x}) + \partial_n g_C(\mathbf{x})\right)= \epsilon_{w}\partial_n \phi^{(w)}(\mathbf{x}),& \quad \mathbf{x} \ \in \ \Gamma  \nonumber \\
     \phi^{(m)}(\mathbf{x}\to\infty) = 0,&
\end{align}
%
%
%
\rev{which we solve in this work.} The linear PB formulation is widely used, however, it may lead to large errors in highly charged systems, such as nucleic acids. 

The solvation energy is a useful quantity in molecular physics,\cite{che2008freeenergy} which corresponds to the the free energy required to dissolve a molecule (i.e. transfer it from a vacuum into the solvent). The solvation energy has two components: a non-polar one, to generate an uncharged solute-shaped cavity in the solvent, and a polar one, that places the charges in said cavity. The polar component of the solvation  energy is commonly computed from PBE calculations, using the following equation:
\begin{equation}
    \Delta G_{solv} = \frac{1}{2}\sum_i^{n_c}q_i\psi(\vb{x}_{q,i})
\end{equation}
which is valid for the linearized PBE.



\subsection{Neural Networks and PINN}
\label{sec:nn_pinn}
A Neural Network in the context of a PINN  may be considered a function $\mathcal{N}$, parameterized by parameters $\theta$ from inputs $\vb{x} \in \mathbb{R}^{m}$ to outputs $\mathbb{R}^{n}$, where $m$ and $n$ denote the input and output dimensions. There are several neural network architectures, e.g. Multi-layer Perception (MLP), Convolutional  Networks, Recurrent Networks, etc.~\cite{goodfellow2016deep,caterini2018deep} Additionally, there are numerous options for both defining the solution of the PDE using a neural network and also for defining the loss function. One simple setting of using PINN to solve a PDE is to employ a MLP with a mean-squared-error (MSE) loss function and representing the PDE solution by the output of the neural network. This simple setting often forms the  basic building block for solving a PDE using PINN and we describe it next.

Given a so called activation function $\sigma: \mathbb{R} \to \mathbb{R}$, we define $\sigma: \mathbb{R}^d \to \mathbb{R}^d$ by defining its $i$th component as $[\sigma(\vb{x})]_i = \sigma(\vb{x}_i)$ for $\vb{x} \in \mathbb{R}^d$. That is, the activation function is defined component-wise on the input vector. Common choices of the activation function are ReLU, $\tanh$, sigmoid, etc. Given a set of $H+1$ integers $d_i, \, i = 0, \ldots, H$, an $H$-layer  MLP or Feed-Forward Neural network is then defined as $\mathcal{N} (\vb{x} ; \theta) = \vb{x}^{(H)}$, where the output of the $i$th layer $\vb{x}^{(i)}$ is defined recursively as
\begin{equation}\label{eq:nn_output}
\vb{x}^{(i)} = \begin{cases}
\vb{x}, \qquad &i = 0\\
\sigma(W^{(i)}\vb{x}^{(i-1)} + \vb{b}^{(i)}), \qquad &i = 1, \ldots, H-1\\
W^{(H)}\vb{x}^{(H-1)} + \vb{b}^{(H)}, \qquad &i = H
\end{cases}
\end{equation}
Here, the weights $W^{(i)} \in \mathbb{R}^{d_i \times d_{i-1}}$ and the biases $\vb{b}^{(i)} \in \mathbb{R}^{d_i}$ form the set of trainable parameters $\theta$ for the MLP. 

Let $\mathcal{D}(u) = f$ denote a generic PDE operator on a domain $\Omega$, with boundary conditions $\mathcal{B}(u) = g$, for appropriate functions $u$, $f$ and $g$.
Then to solve a  PDE using a neural network like the MLP is to form a loss function  of the form, 
\begin{align}
 &\frac{1}{N_{\Omega}}\sum_{\vb{x}_i \in S_\Omega}|\mathcal{D} (\mathcal{N}(\vb{x}_i;\theta)) - f(\vb{x}_i)|^2 + \nonumber \\
&\frac{1}{N_{\partial \Omega}}\sum_{\vb{x}_i \in S_{\partial \Omega}}|\mathcal{B}(\mathcal{N}(\vb{x}_i;\theta)) - g(\vb{x}_i)|^2.\nonumber
\end{align}
Here the $S_\Omega \subset \Omega$ and $S_{\partial \Omega} \subset \partial \Omega$ are sets containing $N_{\Omega}$ domain and $N_{\partial \Omega}$ boundary collocation points respectively, whereas  $| \cdot |$ refers to the Euclidean norm. Such a loss function is often referred to as the MSE loss. All differential operators acting on $\mathcal{N}(\vb{x};\theta)$ (arising from the PDE), and the derivatives of the loss function itself (needed for its minimization) are  computed via automatic differentiation.\cite{baydin2018automatic} The loss function is then minimized with respect to the parameters $\theta$ to obtain the PINN solution $\mathcal{N}(\vb{x}_i;\theta)$,  approximating the true solution of the PDE. The minimization is usually  carried out by employing a variant of a gradient descent method e.g. SGD or Adams~\cite{goodfellow2016deep}. The convergence properties of PINN in this framework is studied in the work by Shin and co-authors.\cite{shin2020convergence}





\subsection{\label{sec:rpb}A Multidomain PINN for RPB}
\label{sec:basic_pinn_pbe}

In this section we develop a multidomain PINN for the RPB which is significantly more complex than the basic PINN method outlined in \rev{in the previous section}. In particular, it involves a multidomain neural network architecture,  a loss function that accounts for the interface conditions of the RPB,  and  construction of collocation points within the solute, solvent, interface and boundary regions. Later we improve on this PINN architecture with a further series of enhancements.

\subsubsection{A multidomain neural network architecture}
Several recent works have applied Physics-Informed Neural Networks (PINN) to solve elliptic partial differential equations (PDEs) with interfaces.\cite{li2020deep,jiang2024generalization,he2022mesh,wu2022inn,ying2023multi,tseng2023cusp,sarma2024interface} These studies commonly suggest using two independent neural networks, each approximating the solution in a specific subdomain. The interface between the subdomains is handled by adding an additional term in the loss function, ensuring consistency across the boundary. Moreover, convergence and error bounds for such approximations have been established.\cite{jiang2024generalization,WZTL23}

In this work, we follow a similar approach but with certain modifications tailored to the regularized Poisson-Boltzmann Equation given in Eq. \eqref{eq:lrpbe}. Specifically, we design a single neural network architecture with two independent branches (throughout this paper, the term branches will also be referred to as independent neural networks or simply networks). Each branch is responsible for approximating the electrostatic potential within its respective subdomain.
One branch computes the reaction potential within the solute domain ($\mathcal{N}_m$), while the other estimates the total potential in the solvent domain ($\mathcal{N}_w$). At the simplest level, these branches are a simple MLP, however, we outline many improvements to this basic architecture in section \rev{An enhanced PINN architecture for RPB''}.
 It is important to note that these branches output different types of potentials due to the specific form of the regularized PBE used in this work:
\begin{equation} 
    \begin{split} 
    \psi^{(m)}(\vb{x}) \approx \psith{m}(\vb{x}):= \mathcal{N}_m(\vb{x};\bm{\theta}_m) & \\ \phi^{(m)}(\vb{x}) \approx \phith{w}(\vb{x}) :=\mathcal{N}_w(\vb{x};\bm{\theta}_w) \\
\end{split} 
\end{equation}
Here, $\mathcal{N}_m(\vb{x};\bm{\theta}_m)$ and $\mathcal{N}_w(\vb{x};\bm{\theta}_w)$ represent the outputs of the solute and solvent branches, respectively, and $\bm{\theta}_m$ and $\bm{\theta}_w$ are their corresponding trainable parameters. The trainable parameters for the neural network is thus $\bm{\theta} = \bm{\theta}_m \cup \bm{\theta}_w$.
This setup allows for the approximation of the reaction potential at any point within the domain as a combination of these two solutions,
\begin{equation}
\label{eq:def_psi}
    \psi_{\theta}(\vb{x}) = \begin{dcases}
         \psith{m}=\mathcal{N}_m(\vb{x};\bm{\theta}_m) & \vb{x} \in \Omega_m \\
         \psith{w} =\mathcal{N}_w(\vb{x};\bm{\theta}_w)-g_C(\vb{x}) & \vb{x} \in \Omega_w \\
        \psith{\Gamma} = \frac{\psith{m}+\psith{w}}{2} & \vb{x} \in \Gamma\\
     \end{dcases} 
\end{equation}
The total potential is then obtained by simply adding the Coulomb potential to the reaction potential detailed earlier:
\begin{equation}\label{eq:def_phi}
    \phi_\theta(\vb{x}) = \psi_{\theta}(\vb{x}) + g_C(\vb{x})
\end{equation}
%
%
%
The rationale behind using a single neural network with a unified set of parameters $\bm{\theta} = \bm{\theta}_m \cup \bm{\theta}_w$ (instead of two independent neural networks) is that this allows us to employ only one optimizer to minimize the loss function. We have observed that this approach leads to better convergence when solving problems of this type.



\subsubsection{Loss function}
The loss function $L(\bm{\theta};S)$ depends on the parameters $\bm{\theta} = \bm{\theta}_m \cup \bm{\theta}_w$ and the set of collocation points $S$. The set $S$ is divided into subsets corresponding to specific regions: $S = S_{\Omega_m} \cup S_{\Omega_w} \cup S_{\Gamma} \cup S_{\partial \Omega}$. Here, $S_{\Omega_m}$ and $S_{\Omega_w}$ represent the collocation points in the solute and solvent domains, respectively, while $S_{\Gamma}$ refers to the points at the interface, and $S_{\partial \Omega}$ corresponds to the points on the boundary of the solvent domain (see Fig. \ref{fig:molecule}). In $S_{\partial\Omega}$, we need to approximate that the potential decays to zero at infinity ($\phi^{(m)}(\mathbf{x}\to\infty) = 0$ in Eq. \eqref{eq:lrpbe}), which we do by enforcing the following Yukawa potential at those points:  
\begin{equation}
    g_Y(\vb{x}) = \frac{e}{k_BT}\sum_i \frac{q_i\e^{-\kappa \abs{\vb{x} - \vb{x}_{q,i}}}}{4 \pi \epsilon_w\epsilon_0\abs{\vb{x} - \vb{x}_{q,i}}}. 
\end{equation}

Let $N_j$ denote the count of each subset $S_j$.
The loss function $L(\bm{\theta};S)$ is decomposed into,
\begin{equation}\label{ec:loss_elipt_pb}
    \begin{split}
        L(\bm{\theta};S) =w_{\Omega_m}L_{\Omega_m} (\bm{\theta}_m;S_{\Omega_m})+ w_{\Omega_w}  L_{\Omega_w} (\bm{\theta}_w;S_{\Omega_w}) \\ 
        +  w_{\partial  \Omega}L_{\partial \Omega} (\bm{\theta}_w;S_{\partial \Omega}) + w_{\Gamma_1}L_{\Gamma_1} (\bm{\theta}_m,\bm{\theta}_w;S_{\Gamma}) \\ 
        + w_{\Gamma_2}L_{\Gamma_2} (\bm{\theta}_m,\bm{\theta}_w;S_{\Gamma}) 
    \end{split}
    \end{equation}
where 
\begin{align*}
    L_{\Omega_m}(\bm{\theta}_m;S_{\Omega_m}) = \frac{1}{N_{\Omega_m}}\sum_{\vb{x}_i \in S_{\Omega_m}} \Big[&\nabla^2 \psith{m}(\vb{x}_i)   \Big]^2,\\
    L_{\Omega_w}(\bm{\theta}_w;S_{\Omega_w}) = \frac{1}{N_{\Omega_w}}\sum_{\vb{x}_i \in S_{\Omega_w}} \Big[&\nabla^2 \phith{w}(\vb{x}_i) - \kappa^2\phith{w}(\vb{x}_i)  \Big]^2,\\
    L_{\partial \Omega}(\bm{\theta}_w;S_{\partial \Omega}) = \frac{1}{N_{\partial \Omega}}\sum_{\vb{x}_i \in S_{\partial \Omega}} \Bigg[&\phith{w}(\vb{x}_i) - g_Y(\vb{x}_i) \Bigg]^2, \\ 
    L_{\Gamma_1}(\bm{\theta}_m,\bm{\theta}_w;S_{\Gamma}) = \frac{1}{N_{\Gamma}}\sum_{\vb{x}_i \in S_{\Gamma}} \Big[&\phith{w}(\vb{x}_i)  -\psith{m}(\vb{x}_i) - g_C(\vb{x}_i) \Big] ^2,\\
    L_{\Gamma_2}(\bm{\theta}_m,\bm{\theta}_w;S_{\Gamma}) =  \frac{1}{N_\Gamma}\sum_{\vb{x}_i \in S_{\Gamma}} \Bigg[&  \epsilon_w\pdv{}{n}\Big( \phith{w}(\vb{x}_i)\Big) \nonumber\\
    - \epsilon_m \pdv{}{n}&\Big( \psith{m}(\vb{x}_i)+ g_C(\vb{x}_i)\Big)\Bigg]^2.
\end{align*}
Each term in the loss function corresponds to a specific region of the domain: $\Omega_m$ (solute domain), $\Omega_w$ (solvent domain), $\partial \Omega$ (boundary of the solvent domain), and $\Gamma$ (interface between the domains). \rev{Each term in Eq. \eqref{ec:loss_elipt_pb} is weighted by a factor $w_j$, where  the index $j$ varies over $\{\Omega_m, \Omega_w, \partial \Omega, \Gamma_1, \Gamma_2\}$ (referring to each loss term), to balance its contribution according to a loss balancing algorithm detailed later}, ensuring that all components influence the optimization of the parameter set $\bm{\theta}$. Notably, the term $L_{\Omega_m}$ depends only on the parameters of the branch associated with the solute domain, while $L_{\Omega_w}$ and $L_{\partial \Omega}$ depend on the solvent domain. The two interface terms $L_{\Gamma_1}$ and $L_{\Gamma_2}$, which enforce continuity of the potential and the electric displacement respectively, incorporate contributions from both neural networks. 


In addition to the primary loss terms, additional terms can be introduced to incorporate known approximations, experimental results, or other relevant information. For example, a loss term based on known approximation can be defined as:
\begin{equation}\label{eq:loss_alternative}
    L_{data}(\bm{\theta};S_{data}) = \frac{1}{N_{data}}\sum_{\vb{x}_i \in S_{data}} \Big[\phi_{\theta}(\vb{x}_i) -\phi^{\dag}(\vb{x}_i) \Big] ^2  
\end{equation}
This term compares the predicted potential $\phi_{\theta}(\vb{x}_i)$ with known approximations $\phi^{\dag}(\vb{x}_i)$ at the collocation points $S_{data}$.

\subsubsection{Construction of collocation points}
\label{sec:cons_colloc}
\begin{figure}[ht]
    \centering
    \subfigure[Surface mesh]{\includegraphics[width=0.3\textwidth]{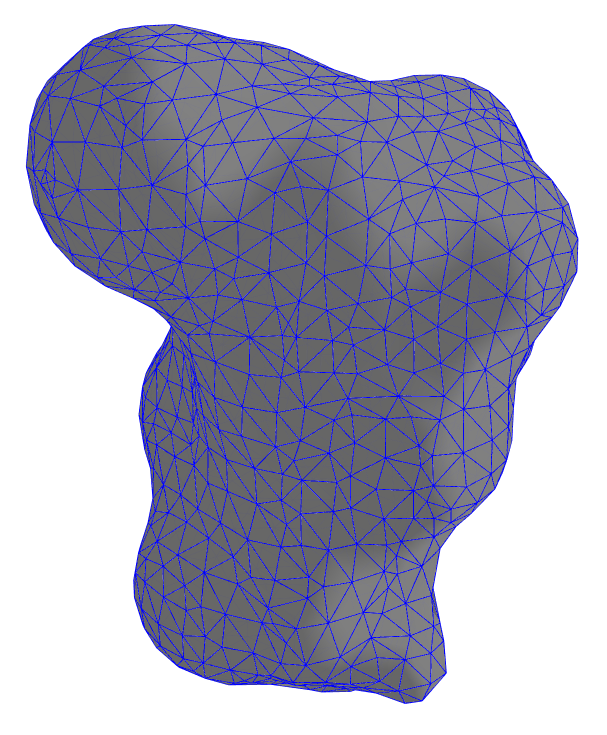}}
     \subfigure[Volumetric meshes]{\includegraphics[width=0.3\textwidth]{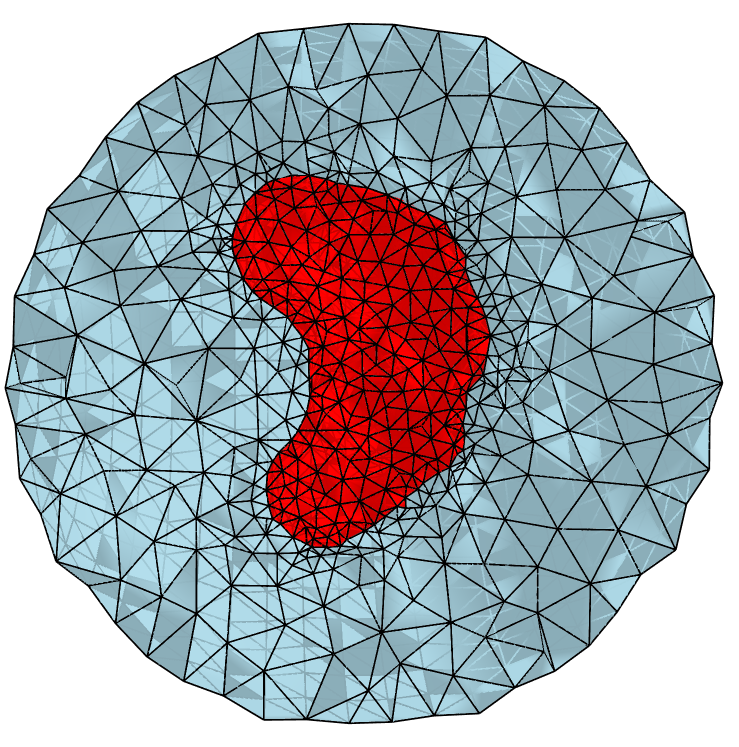}}
    \caption{Examples of (a) surface mesh (b) volumetric meshes \rev{used to generate collocation nodes} for arginine.}
    \label{fig:mesh_img}
\end{figure}

\begin{figure}[ht]
    \centering
    \subfigure[Collocation points in the molecule]{\includegraphics[width=0.4\textwidth]{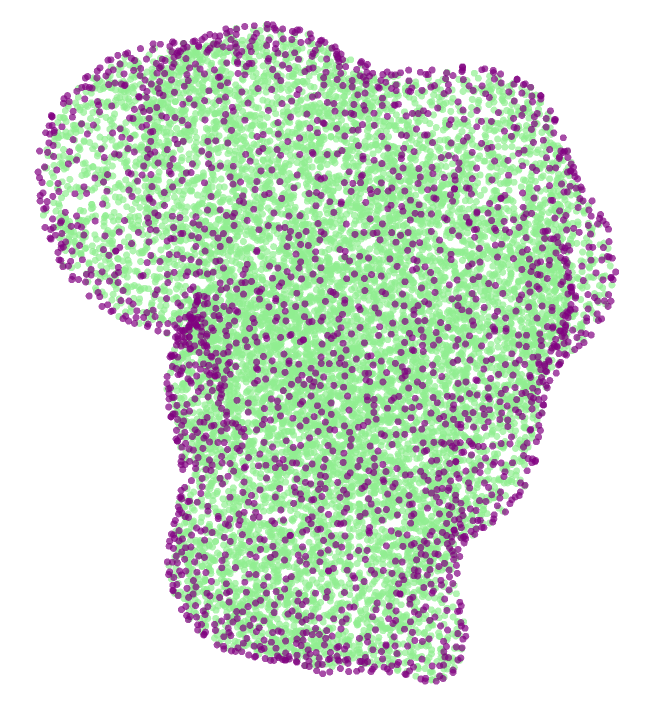}}
    \subfigure[Collocation points in the entire domain]{ \includegraphics[width=0.4\textwidth]{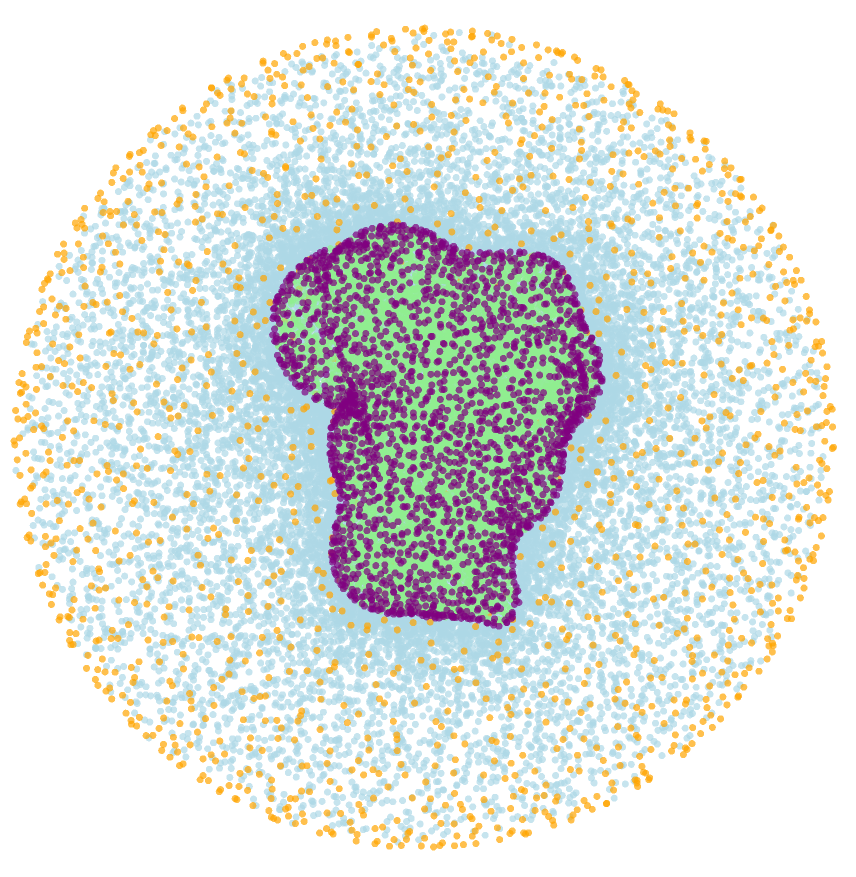}}
    \caption{Examples of the collocation points obtained using the meshes showed in Fig \ref{fig:mesh_img}. Green nodes correspond to $S_{\Omega_m}$, purple nodes are $S_{\Gamma}$, light blue nodes are $S_{\Omega_w}$, and orange nodes are $S_{\partial\Omega}$.}
    \label{fig:mesh_colpoints_img}
\end{figure}

\begin{figure}[ht]
    \centering
     \subfigure[Triangular element]{\includegraphics[width=0.3\textwidth]{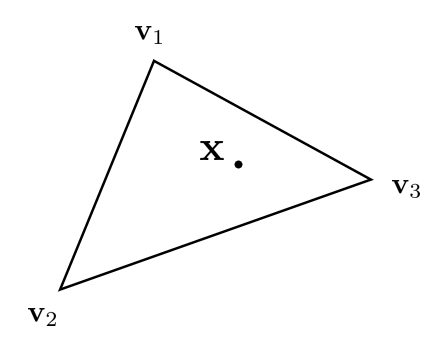}}
    \subfigure[Tetrahedral element]{ \includegraphics[width=0.3\textwidth]{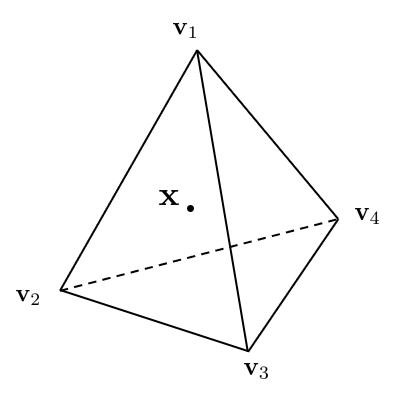}}
    \caption{Schematic of a random point $\mathbf{x}$ inside a (a) triangular element (b) tetrahedral element.}
    \label{fig:colpoints_img}
\end{figure}

 The molecular surface or  interface $\Gamma$, modeled by a solvent-excluded surface (SES), often has a complex geometry and hence it is not straightforward to construct the set of collocation points $S  = S_{\Omega_m} \cup S_{\Omega_w} \cup S_{\Gamma} \cup S_{\partial \Omega}$. 
 \rev{Even though PINN is a mesh-free method, we use surface and volume grids, usually seen in BEM and FEM calculations, to assist in the generation of the collocation nodes.}
We first create a surface triangular mesh of the SES, and then
create a volumetric tetrahedral mesh of the domain $\Omega$ which conforms to the SES (that is, the only intersection between a tetrahedron and the SES is a triangular face of the tetrahedron). Then we consider four sub-meshes:
two tetrahedral sub-meshes corresponding to $\Omega_m$ and $\Omega_w$, and two triangular sub-meshes corresponding to $\Gamma$ and $\partial \Omega$ (see Fig. \ref{fig:mesh_img}). 
Then we select  elements from each sub-mesh, and sample a point per each selected element randomly to generate $S_{\Omega_m}, S_{\Omega_w},  S_{\Gamma}$ and  $S_{\partial \Omega}$, as sketched by Fig. \ref{fig:colpoints_img}. An example of the resulting set of collocation nodes for arginine is presented by Fig. \ref{fig:mesh_colpoints_img}.

This approach allows us to construct random samples without parameterizing the domains, while only requiring the discretization of the molecular surface (SES) which is a well-established process. The distribution of collocation points throughout the domains depends on the distribution of the mesh elements, giving us full control over point density in different regions. This allows us to concentrate more points in areas where residuals are typically higher, such as at the interface. Additionally, the process may  be repeated 
for selected elements to increase the density of collocation points in specific regions. 
Finally, the density of elements in the mesh is chosen such that subdomains with larger size have more elements. This results in an increase of collocation points as the size of the subdomain (solute, solvent, molecular surface and boundary) is increased, similar to the method described by Jiang and co-workers.~\cite{jiang2024generalization}

The approach to construct collocation points is consistent with current research in PINN and can be adapted to include importance sampling techniques \cite{nabian2021efficient} and residual-based adaptive sampling methods \cite{wu2023sampling} by adjusting the probability distribution for sampling within each mesh element.

\subsection{An enhanced PINN architecture for  RPB}
\label{sec:improvements}
\begin{figure*}[ht]
    \centering
    \includegraphics[width=\textwidth]{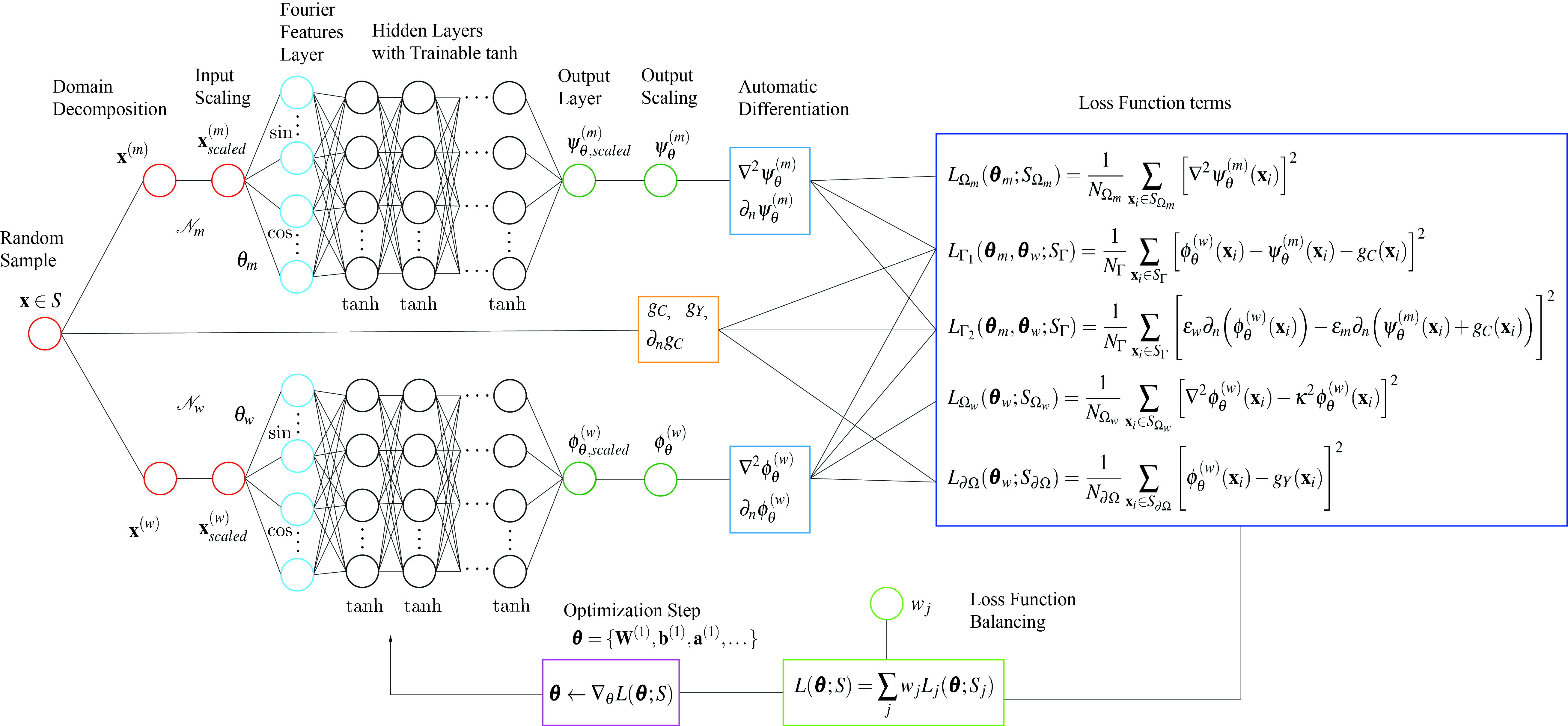}  
    \caption{Schematic of the architecture of our PINN algorithm for solving the PB equation. This method involves segregating the collocation points across the two domains, which are then fed into each branch of the neural network. Note that each branch has its own architecture, Fourier features and scaling layers. Each output contributes to the construction of the loss function $L$, which is minimized to find the optimal trainable parameters $\theta$ of the neural network.}
    \label{fig:algorithm}
\end{figure*}

We discuss improvements to the PINN architecture in this section. These enhancements lead to significant increases in the accuracy and efficiency of the computed solution, as we later demonstrate in the Results and Discussion section. The architecture of our PINN algorithm for solving the PB equation is illustrated in Fig. \ref{fig:algorithm}. The input is passed to either $\mathcal{N}_m$ or $\mathcal{N}_w$, depending on its location, and then successively passes through an input scaling layer, a Fourier feature layer, hidden layers with trainable activation function, and finally an output scaling layer. \rev{We detail each one of these enhancements next.} The output of the network, along with derivatives computed using automatic differentiation, is then used to compute the loss function. The minimization of the loss function employs a loss balancing algorithm, \rev{which we also describe in this section}. The parameters of the network are updated and the process repeated. 

\subsubsection{Scaling layers}
\label{sec:sl}
    It is well known that scaling helps convergence of the neural network training.\cite{wang2023expert} Two scaling layers each for the networks $\mathcal{N}_m$ and $\mathcal{N}_w$ are employed to normalize the inputs and outputs, improving its convergence during training. This ensures that the values entering and leaving the network are between -1 and 1, or at least close to this range. 

    The input scaling is performed before the hidden layers (or the Fourier features layer if that is used). Given an input 
    $\vb{x} = [{x}_1, \, {x}_2, {x}_3]$, 
    the input scaling is for each component 
    ${x}_i$ is, 
    \begin{equation}
         {x}_{scaled,i} = 2\frac{({x}_i-{x}_{min,i})}{({x}_{max,i}-{x}_{min,i})} -1, \quad i \in \{1,2,3\},
    \end{equation}
    where $\vb{x}_{min}=  [{x}_{min,1}, \, {x}_{min,2}, {x}_{min,3}]$ and $\vb{x}_{max}=  [{x}_{max,1}, \, {x}_{max,2}, {x}_{max,3}]$ correspond to the minimum and maximum coordinates of the collocation points in the appropriate  subdomain, and form hyperparameters (i.e., nontrainable parameters) for each network. 
    
    The output scaling is performed after the hidden layers. If the output of the last hidden layer is $y_{scaled}$ (which is a scalar for both $\mathcal{N}_m$ and $\mathcal{N}_w$), then the output scaling transform is,    
    \begin{equation}
        y = \frac{y_{scaled}+1}{2}(y_{max}-y_{min})+y_{min},
    \end{equation}
   where the values $y_{min}$ and $y_{max}$ correspond to hyperparameters (i.e., nontrainable parameters) for each network and must be estimated based on approximations of the real solution in each domain. If these values, $y_{min}$ and $y_{max}$, correspond to the actual maximum and minimum of the solution, then the output scaling tends to scale the values of $y_{scaled}$ between $-1$ and $1$. However, estimating $y_{min}$ and $y_{max}$ is not straightforward since the solution to the problem is not known a priori.
    To estimate the values of $y_{min}$ and $y_{max}$, an approximation of the potential is constructed by superimposing the known solution of the Born ion,\cite{holst1994poisson} assuming that each charge in the molecule is an independent Born ion \rev{(BI)}. More precisely, let the BI function be defined as:
    \begin{equation}
        \text{BI}(q_i,R_i) = \frac{q_i}{4\pi}\left( \frac{1}{\epsilon_w(1+\kappa_w R_i)R_i} - \frac{1}{\epsilon_m R_i}  \right)
    \end{equation}
    where $R_i$ corresponds to the radius of the $i$th charge,
    the minimum and maximum reaction potentials (which are then used to obtain $y_{min}$ and $y_{max}$) are estimated with the following equations:
    \begin{equation*}
        \begin{split}
            \psi_{max} &= \max_i\left(0,\text{BI}(q_i,R_i)\:,\:\text{BI}(q_i,R_i)+\sum_{j\neq i}\text{BI}(q_j,R_{ji}')\right)\\
            \psi_{min} &= \min_i\left(0,\text{BI}(q_i,R_i)\:,\:\text{BI}(q_i,R_i)+\sum_{j\neq i}\text{BI}(q_j,R_{ji}')\right)
        \end{split}
    \end{equation*}
    where $R_{ji}' = \norm{\vb{x}_{q,i} - \vb{x}_{q,j}}$. Note that ($y_{min},y_{max}$) correspond to reaction potential ($\psi_{min},\psi_{max}$) in $\mathcal{N}_m$, while the Coulomb potential $g_C$ needs to added in the $\mathcal{N}_w$ (see \eqref{eq:def_phi}). This approximation does not guarantee that the output scaling layer will scale the solution strictly between -1 and 1, but in practice the values are close enough.

\subsubsection{Random Fourier features layer}
\label{sec:rff}
    MLPs often suffer from a phenomenon called spectral bias, which biases the solution towards low-frequency functions, preventing the network from learning higher-frequency functions necessary to target the desired solution. This phenomenon can be mitigated using a Random Fourier Feature layer, \cite{tancik2020fourier} which maps the input signals to a high-frequency space before passing them through the neural network. Given an input $\vb{x}$, the layer is defined as follows: 
    \begin{equation}\label{eq:fourier}
        \vb{\bar{x}} = \begin{bmatrix}
            \cos(\vb{B}\vb{x})\\
            \sin(\vb{B}\vb{x})
        \end{bmatrix}
    \end{equation}
    where $\vb{B}$ is a matrix of shape $m \times d$, with $m$ being the number of Fourier features and $d$ the input dimension. The matrix $\vb{B}$ is generated from a normal distribution $\vb{B} \sim \text{Normal}(0, \sigma^2)$ and is non-trainable. This simple layer improves the PINN method’s ability to learn sharp gradients and complex solutions. \cite{wang2023expert} In this work, we used $m=128$ Fourier features, $d=3$ corresponding to points in three dimensions, and a standard deviation of $\sigma = 1$.
 
\subsubsection{Trainable activation function}
\label{sec:taf}
    Motivated by the work by Jagtap and co-authors,\cite{jagtap2020adaptive} we implemented trainable activation functions. In this approach, the activation functions in each perceptron of the neural network are associated with a trainable parameter. Specifically, in this work, we use the hyperbolic tangent function, where the trainable parameter modifies the activation function as follows:
    \begin{equation}
    \label{eq:trainable_tanh}
        \sigma(\vb{x}) = \tanh(\vb{a}\odot\vb{x}),
    \end{equation}
    where $\odot$ indicates element-wise multiplication. 
    Here, $\vb{a}$ is a vector of trainable parameters (initialized to 1), which is then  incorporated into the overall set of trainable parameters $\bm{\theta}$. 
    With this modification, the operation in the $i$th hidden layer is:
\begin{equation}\label{eq:hidden_layer_tf}
    \vb{x}^{(i)}= 
\tanh(\vb{a}^{(i)}\odot(W^{(i)}\vb{x}^{(i-1)} + \vb{b}^{(i)})),
\end{equation}
and the full set of parameters is defined as:
\begin{equation}
    \bm{\theta} = \{ \vb{W}^{(1)},\vb{b}^{(1)},\vb{a}^{(1)}, \dots, \vb{W}^{(H)},\vb{b}^{(H)},\vb{a}^{(H)}  \}
\end{equation}


\subsubsection{Loss balancing algorithm}
\label{sec:lba}
    
    To ensure that each term of the loss function contributes effectively to the modification of the trainable parameters $\bm{\theta}$, we implemented a weight-adapting algorithm.~\cite{wang2023expert} Consider the loss function as a linear combination of several loss terms:
    \begin{equation}
    \label{eq:weighted_loss}
        L(\bm{\theta};S) = \sum_{j} w_j L_j(\bm{\theta};S_j).
    \end{equation}
    Note that in our case the function $L(\bm{\theta};S)$ is given in Eq. \eqref{ec:loss_elipt_pb} and corresponds to the form given above in Eq. \eqref{eq:weighted_loss}.
    The objective is to determine the weights $w_j$ for each loss term $L_j$ such that the gradient $w_j \norm{\nabla_{\theta}L_j}$ remains constant across all terms $j$:
    \begin{equation}
        C = w_j \norm{\nabla_{\theta}L_j} \quad \forall j.
    \end{equation}
    To achieve this, we compute an estimator $\hat{w}_j$ for each term:
    \begin{equation}
        \hat{w}_j = \frac{\displaystyle \sum_i \norm{\nabla_\theta L_i}}{\norm{\nabla_\theta L_j}}. 
    \end{equation}
    
    Next, we update the old weight using a soft adjustment, where $\alpha$ is a hyperparameter controlling the update rate:
    \begin{equation}\label{eq:weight_up}
          w_{j,\text{new}} = \alpha w_{j,\text{old}} + (1-\alpha)\hat{w}_j
    \end{equation}
    
    In this work, we set $\alpha = 0.7$ to balance the influence of old and new weights, applying this adjustment every $r=1000$ iterations

\subsubsection{Complete Algorithm}

The complete algorithm used for the PINN solution of the PBE is presented in Algorithm~\ref{algoritmo}.
%
\begin{minipage}{\linewidth}
\begin{algorithm}[H]
    \caption{Algorithm for solving the PBE using PINN}\label{algoritmo}
    \begin{algorithmic}
    \Require Molecular information  and hyperparameters
    \State Generate mesh from the molecular information \Comment \rev{Fig. \ref{fig:mesh_img}}
    \State Create neural network $\mathcal{N} = \mathcal{N}_m \cup \mathcal{N}_w$ \Comment{\rev{Fig. \ref{fig:algorithm}}}
    \State Initialize trainable parameters $\bm{\theta}^{[0]}$ 
    \Comment{\rev{Eq. \eqref{eq:def_phi}}}
    \For{$k=1$ to $k=n$} \Comment{Training loop}
         \For{$i \in \{\Omega_m, \Omega_w, \partial \Omega, \Gamma\}$}
        \State Construct subdomain collocation points $S_i$ \Comment{\rev{Fig. \ref{fig:colpoints_img}}}
        \EndFor
        \State Set  $S =  S_{\Omega_m} \cup S_{\Omega_w} \cup S_{\partial \Omega} \cup S_{\Gamma}$ \Comment{Collocation points}
        \State Compute $\psi_{\theta},\phi_{\theta}$ from $\mathcal{N}(S;\bm{\theta}^{[k]})$  \Comment{Eqs. \eqref{eq:def_psi} and \eqref{eq:def_phi}}
        
        \State Compute loss function $L(\bm{\theta}^{[k]};S)$  \Comment{Eqs. \eqref{ec:loss_elipt_pb} and \eqref{eq:weighted_loss}}
       
        \State Update trainable parameters $\bm{\theta}^{[k+1]}$ \Comment{ Optimization step} 
        \If{mod($k$,r)==0}
           \For{$j \in \{\Omega_m, \Omega_w, \partial \Omega, \Gamma_1, \Gamma_2\}$}
             \State Update weight $w_j$ of loss term $L_j$  \Comment{\rev{Eq. \ref{eq:weight_up}}}
        \EndFor
        \EndIf
    \EndFor
    \Ensure Optimized parameters $\bm{\theta}^{[n]} ( = \bm{\theta} = \bm{\theta}_m\cup\bm{\theta}_w$) 
    \end{algorithmic}
\end{algorithm}
\end{minipage}
\par
\vspace*{8pt}



We implemented Algorithm~\ref{algoritmo} in a software package and named it XPPBE. 
This solver requires two inputs: a \verb|.yaml| file, which contains all the solver parameters (such as architectures, mesh parameters, and optimization methods, among others), and a \verb|.pdb| file which contains the molecular information. A more detailed explanation of these files can be found in the tutorials available in the GitHub repository.\cite{githubrepo}

\section{\label{sec:res}Results and Discussion}


\rev{We carried out a sequence of experiments to validate the PINN implementation for the RPB  outlined in Algorithm \ref{algoritmo}}. The results are organized to demonstrate the impact of each algorithm feature on the solution, therefore providing evidence of the importance of including all of them. The analysis starts with the simple Born ion, and scales up to more complicated structures, where the influence of each component is clearer. 
The physical parameters were set to a solvent permittivity of $\epsilon_w=80$, an inverse of the Debye length of $\kappa=0.125$ \AA$^{-1}$, and a solute dielectric constant of $\epsilon_m=2$, except the spherical cases, where $\epsilon_m=1$.

We explore a variety of PINN architectures to gauge the effectiveness of different features. 
The base case, termed {\it Minimal}, is the architecture presented in the section \rev{ ``A Multidomain PINN for RPB''} and 
considers a fully connected multi-layer perceptron (MLP)  with 4 hidden layers and 200 neurons per layer, with an hyperbolic tangent activation function. The enhancements considered, presented in the section \rev{``An enhanced PINN architecture for RPB''}, are adding the weight adjusting algorithm (WA) of Equation \eqref{eq:weight_up}, using trainable activation functions (TF) of Eq. \eqref{eq:trainable_tanh}, layers including Fourier features (FF) of Eq. \eqref{eq:fourier}, and  scaling of the input (SI) and output (SO) of the neural network of Fig. \ref{fig:algorithm}. A combination of these features is indicated by the  `+' sign e.g., $WA + SO$ indicates the usage of weight adjusting algorithm and output scaling. 
All trainable parameters were initialized using the Glorot normal distribution. We employed the Adam optimization algorithm with an exponentially decaying learning rate starting from 0.001.

The collocation points are created from surface and volume meshes (triangles and tetrahedrons, respectively) as described in Fig. \ref{fig:colpoints_img}. As the molecular surface definition, we used the solvent-excluded surface (SES), \cite{connolly1985molecular} which we meshed with \texttt{msms}\cite{sanner1996reduced} for the spheres and arginine, and with  \texttt{Nanoshaper}\cite{decherchi2013general} for the bigger molecules. From those surface definitions, we generated volumetric tetrahedral meshes with \texttt{TetGen}\cite{hang2015tetgen} through \texttt{PyGAMer}\cite{lee2020open}. The external spherical surface ($\partial\Omega$) was placed at a minimum distance $d_{\partial\Omega} = 3.5$ {\AA} from $\Gamma$, except the full proteins (1pgb and 1uqb in the ``Results'' section), where $d_{\partial\Omega} = 4$ {\AA}).

For comparisons, we solved the PBE using either analytical expressions, available for spherical inclusions,\cite{kirkwood1934theory,holst1994poisson} or BEM, through the software PBJ,\cite{search2022towards}\footnote{\url{https://github.com/bem4solvation/pbj}} to compute the difference in $\Delta G_{solv}$ by:
%
\begin{equation}
D_{G_{solv}} := \frac{|\Delta G_{solv,\theta} - \Delta G_{solv}^{ref}|}{|\Delta G_{solv}^{ref}|}
\end{equation}
and the difference in reaction potential ($\psi$):
%
\begin{equation}
D_\psi = \sqrt{\frac{\sum_i^{N_v}\left(\psi_\theta^{(\Gamma)}(\mathbf{x}_{v,i}) - \psi^{ref}(\mathbf{x}_{v,i})\right)^2}{\sum_i^{N_v}\left(\psi^{ref}(\mathbf{x}_{v,i})\right)^2}}
\end{equation}
evaluated at the $N_v$ vertices of the reference surface mesh ($\mathbf{x}_{v,i})$. Here, 
\begin{equation}
    \Delta G_{solv,\theta} = \frac{1}{2}\sum_i^{n_c}q_i\psi_\theta^{(m)}(\vb{x}_{q,i})
\end{equation}
is the solvation energy computed with the PINN solution,
 and the superscript $ref$ corresponds to a reference solution with either an analytical approach or BEM. 
 
 We also report training and validation losses, which correspond to the evaluation of the loss function in  Eq. \eqref{ec:loss_elipt_pb} by setting all weights to one ($w_j=1$). The training loss is computed on the collocation nodes used during training, while the validation loss is evaluated on separate points not included in the training process.

The results detailed in this section were obtained using the XPPBE software,\cite{githubrepo} an open-source TensorFlow-based Python code that can easily run PBE calculations from a PDB file. All runs were performed on a workstation with two Intel(R) Xeon(R) CPU E5-2680 v3 @ 2.50GHz with 12 cores each and 96 GB RAM memory, and a CUDA-enabled NVIDIA K40 GPU.

\subsection{Born Ion}

We consider the electrostatic potential of a 1 \AA~ radius Born ion with a centered 1$e^-$ charge, using 8 different PINN architectures. Starting from the base setup ({\it Minimal}), we systematically add other features to evaluate their effect on accuracy.
In this test, the number of collocation nodes was 1,128 in $\partial\Omega$, 916 in $\Gamma$, 8,798 in $\Omega_w$, and 2,353 in the $\Omega_m$, and the training was carried out for  20,000 iterations. 

\begin{figure}[ht]
 \centering
        \subfigure[\rev{Analytical reaction potential}]{\includegraphics[width=0.4\textwidth]{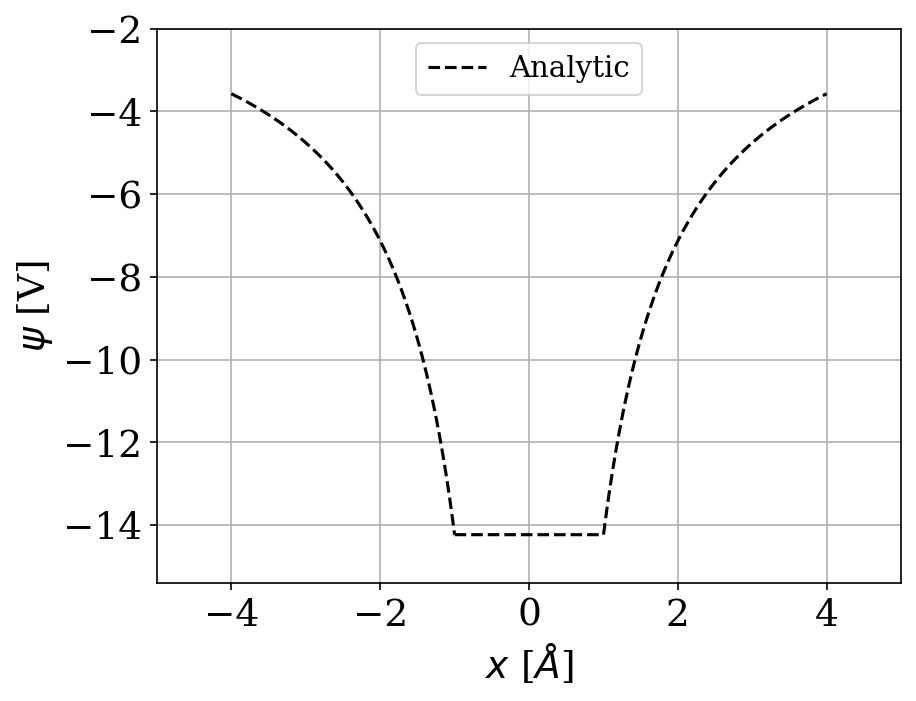}\label{fig:born_analytic}}
    \subfigure[\rev{$\abs{\psi_\theta - \psi^{ref}}$}]{\includegraphics[width=0.4\textwidth]{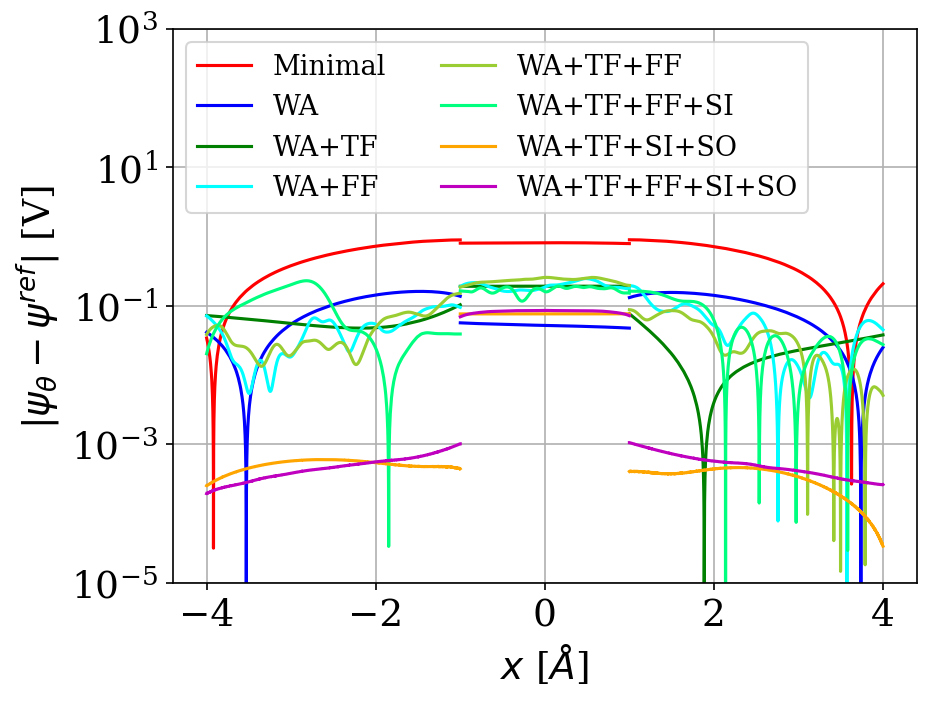}\label{fig:born_abserrors}}
  \caption{\rev{(a) Analytical solution of the reaction potential along the $x$ axis for the Born Ion, (b) Absolute difference for the reaction potential for each case along the $x$ axis with respect to the analytical solution.}} 
  \label{fig:potential_born_ion}
\end{figure}

Fig. \ref{fig:potential_born_ion} compares the reaction potential computed with PINN ($\psi_\theta$)  with an analytical solution~\cite{kirkwood1934theory}. The {\it Minimal} setup is clearly the least accurate (Fig. \ref{fig:born_abserrors}), but the results are considerably improved by including the (WA) feature\rev{, especially in $\Omega_m$ (middle region)}. Including other features induces some errors in that region, however the two last architectures demonstrate better accuracy. 

\begin{figure}[ht]
  \centering
  \subfigure[$\Delta G_{solv,\theta}$]{\includegraphics[width=0.3\textwidth]{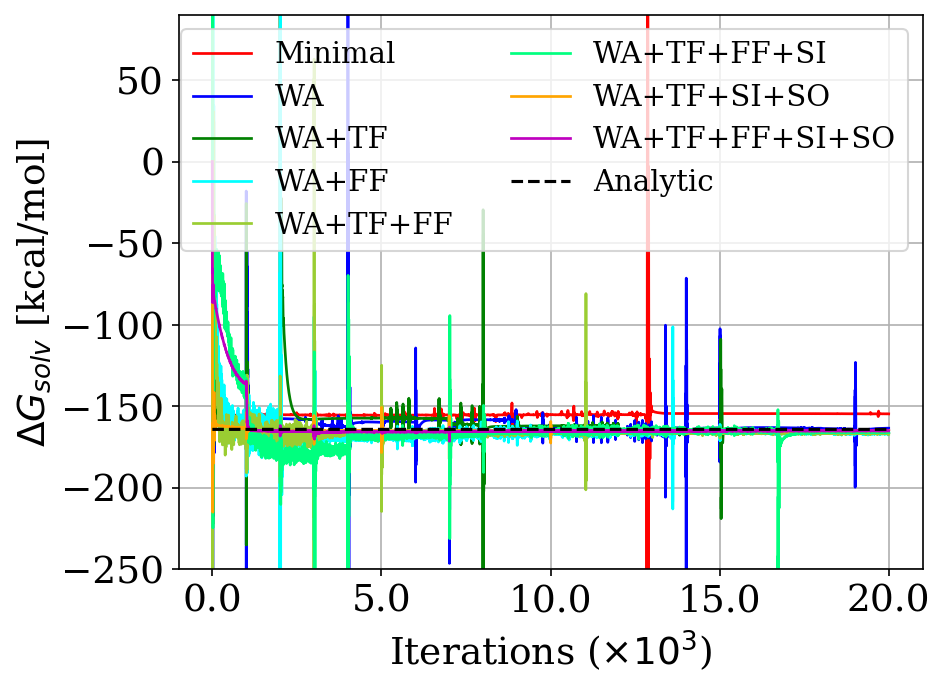}}
      \subfigure[$D_{G_{solv}}$]{\includegraphics[width=0.3\textwidth]{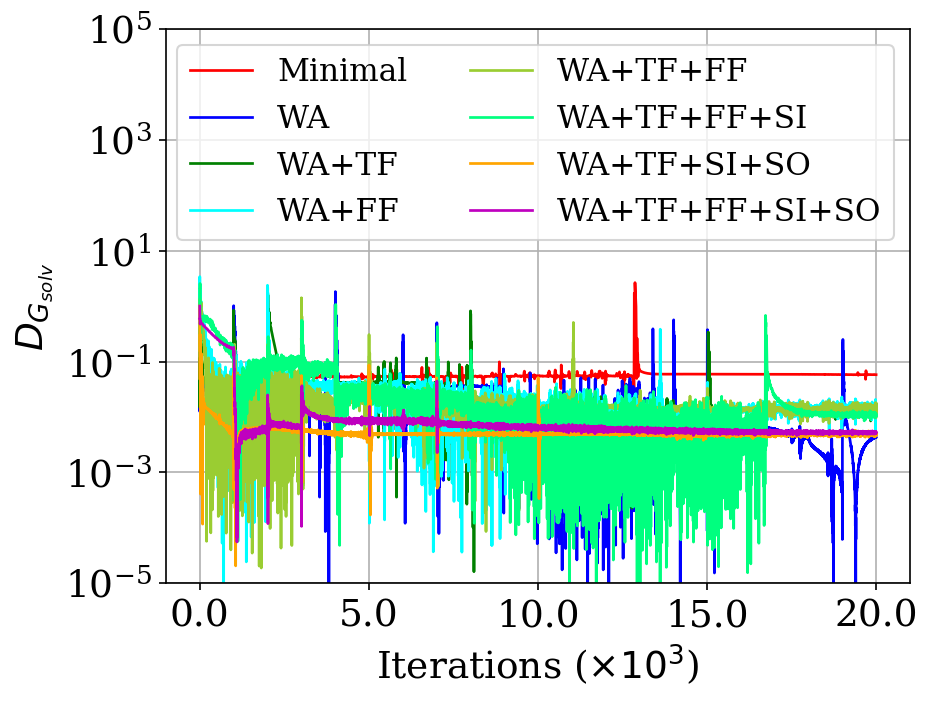}}
  \subfigure[Training Loss]{\includegraphics[width=0.3\textwidth]{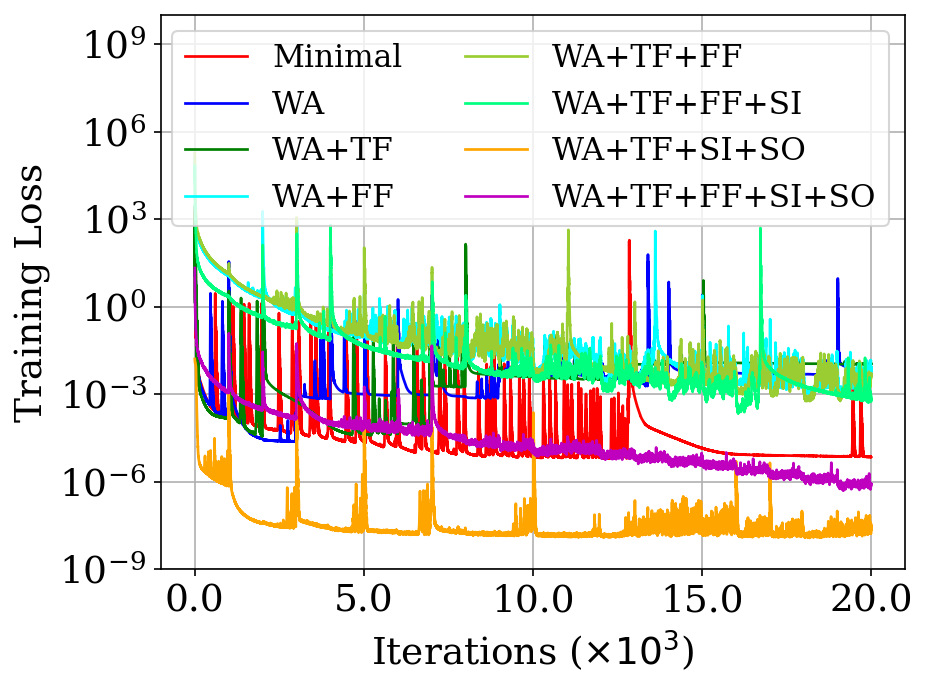}}
    \subfigure[Validation Loss]{\includegraphics[width=0.3\textwidth]{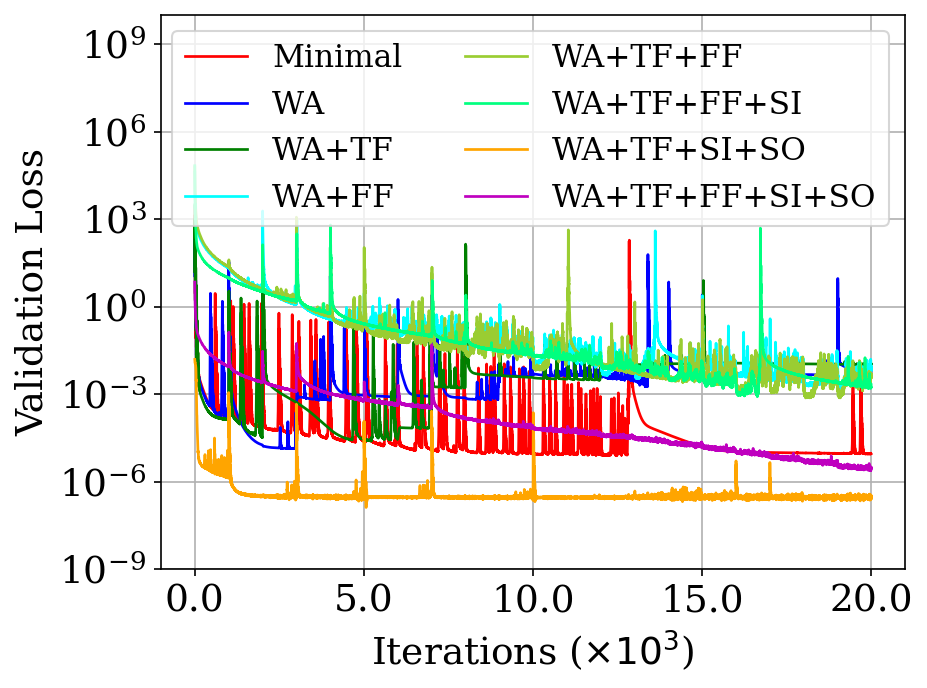}}
    \subfigure[$D_\psi$]{\includegraphics[width=0.3\textwidth]{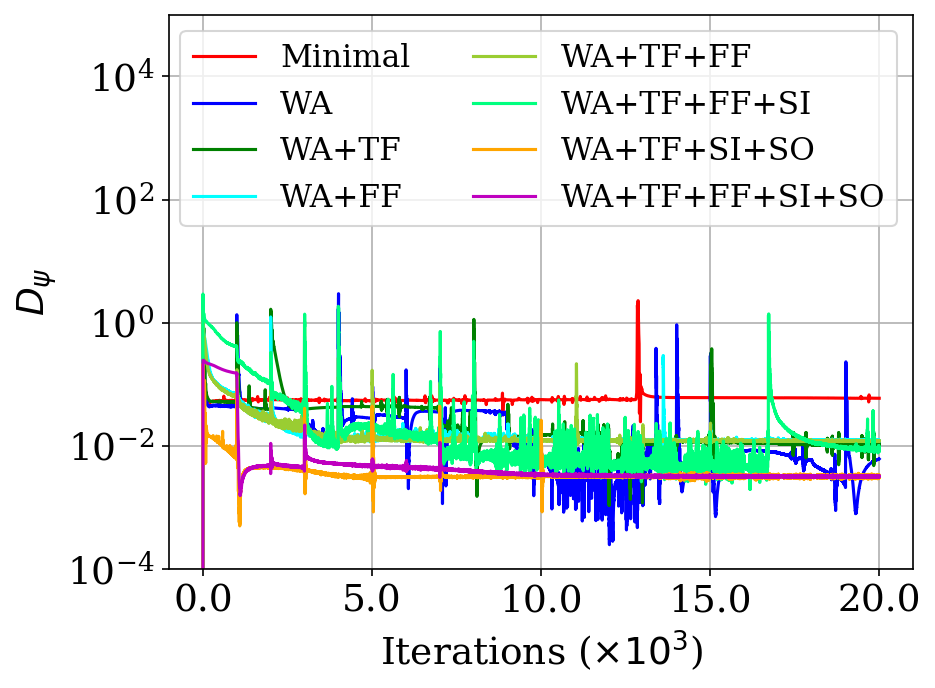}}
  \caption{Evolution of $\Delta G_{solv,\theta}$, $D_{G_{solv}}$, training and validation loss, and $D_\psi$ with iterations for the Born ion with 8 different architectures.}
  \label{fig:gsolv_tloss_born_ion}
\end{figure}

These results are further supported by the evolution of $\Delta G_{solv,\theta}$, $D_{G_{solv}}$, training and validation loss, and $D_\psi$ with iterations in Fig. \ref{fig:gsolv_tloss_born_ion}. The  figure consistently shows that the architectures (WA + TF + SI + SO) and (WA + TF + FF + SI + SO) have the lowest and smoother metrics (errors and losses) as demonstrated by the yellow and purple lines. Interestingly, the architecture with only the (WA) feature, which seemed to show excellent accuracy in Fig. \ref{fig:born_abserrors}, has a poor and noisy convergence in all metrics of Fig. \ref{fig:gsolv_tloss_born_ion}. The values for the last iteration is detailed in Table \ref{tab:results_table_born_ion}, where both $D_{G_{solv}}$ and $D_\psi$ arrive at a value of $\sim$10$^{-2}$---10$^{-3}$, even though the losses may be much smaller. These orders of magnitude for the error are consistent with previous work using PINN to solve partial differential equations.\cite{wang2023expert}

From this analysis, we conclude that the best two architectures include the following features: (WA + TF + SI + SO) and (WA + TF + FF + SI + SO), and we further analyze them in more complicated settings.

  \begin{table}[htbp]
    \centering
    \caption{Results for the Born ion after 20,000 iterations with 8 alternative architectures. Analytical solvation energy was -164.19 kcal/mol.}
    \begin{tabular}{|c|c|c|c|c|}
      \hline
      Case  & $D_{G_{solv}}$  & $D_\psi$ & Training & Validation   \\
      &  &  & loss & loss  \\
      \hline
      Minimal    & 5.72E-02 & 5.98E-02 & 7.10E-06 & 9.10E-06  \\
      WA    & 3.67E-03 & 6.18E-03 & 6.04E-03 & 5.86E-03  \\
      WA+TF    & 5.91E-02 & 6.15E-02 & 7.93E-06 & 9.78E-06  \\
      WA+FF    & 1.32E-02 & 1.22E-02 & 6.86E-03 & 7.45E-03  \\
      WA+TF+FF    & 1.81E-02 & 1.22E-02 & 4.21E-03 & 3.76E-03  \\
      WA+TF+FF+SI    & 1.13E-02 & 8.40E-03 & 8.27E-04 & 2.12E-03  \\
      WA+TF+SI+SO    & 5.38E-03  & 3.10E-03 & 2.44E-08  &  2.71E-07 \\
      WA+TF+FF+SI+SO    & 6.04E-03 & 3.27E-03 & 8.19E-07 & 2.97E-06  \\
      \hline
      \end{tabular}%
    \label{tab:results_table_born_ion}%
  \end{table}%

  \subsection{Spherical molecule with off-centered charge}

  The only difference between the best-performing architectures for the Born ion is the incorporation of the Fourier features (FF). Here, we further analyze the impact of FF  using a slightly more challenging test: a 1 \AA~spherical inclusion with an off-centered $+1e^-$ charge placed 0.45 \AA~away from the center. This case also has an analytical solution for comparison.\cite{kirkwood1934theory} 

\begin{figure}[ht]
  \centering
          \subfigure[\rev{Analytical reaction potential}]{\includegraphics[width=0.4\textwidth]{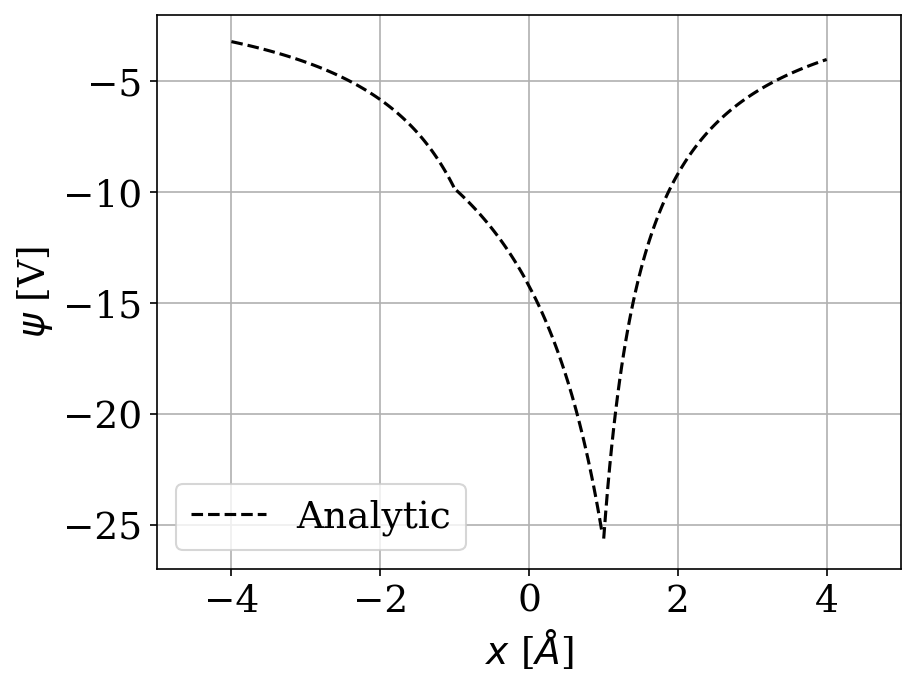}\label{fig:noncentered_analytic}}
    \subfigure[\rev{$\abs{\psi_\theta - \psi^{ref}}$}]{\includegraphics[width=0.4\textwidth]{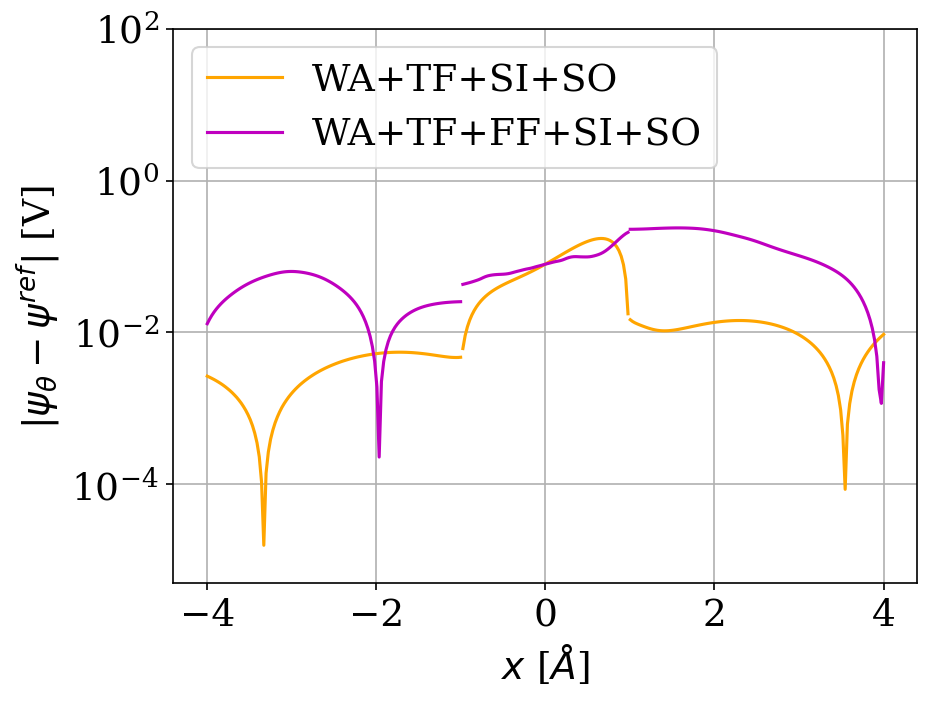}\label{fig:noncentered_abserrors}}
  \caption{\rev{(a) Analytical solution of the reaction potential along the $x$ axis for a sphere with an off-centered charge, (b) Absolute difference for the reaction potential for each case along the $x$ axis with respect to the analytical solution.}}
  \label{fig:potential_noncentered}
\end{figure}

\begin{figure}[ht]
    \centering
    \subfigure[$\Delta G_{solv,\theta}$]{\includegraphics[width=0.3\textwidth]{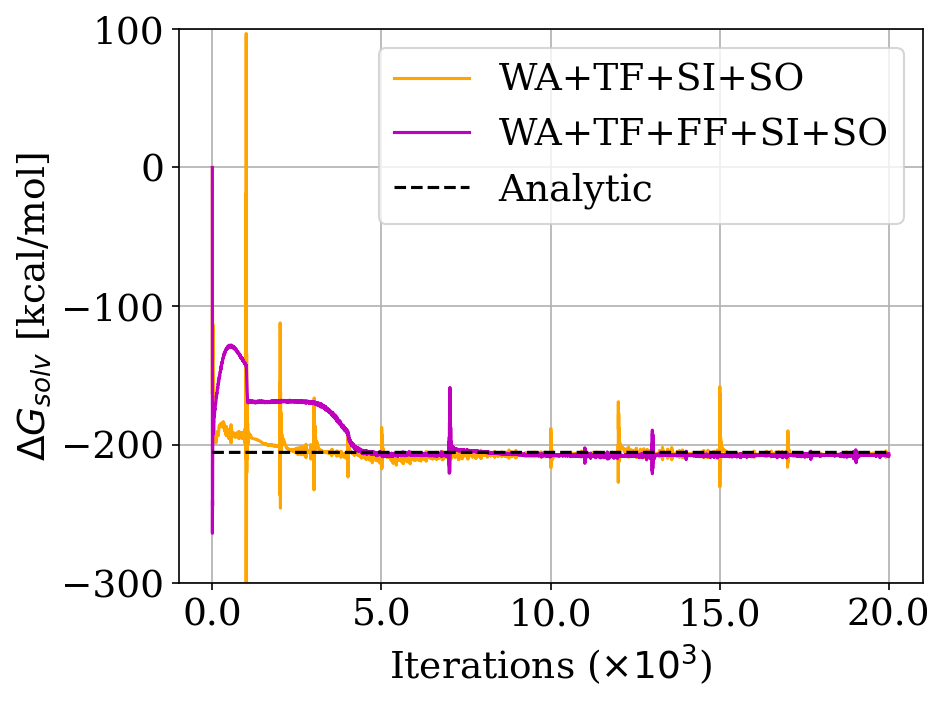}}
            \subfigure[$D_{G_{solv}}$]{\includegraphics[width=0.3\textwidth]{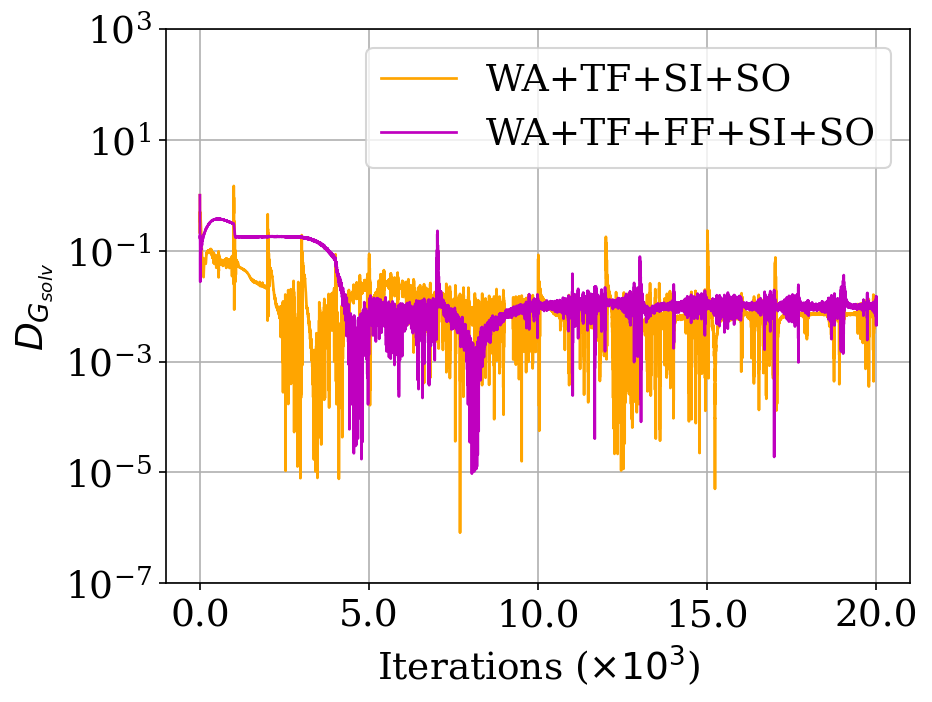}\label{fig:noncentered_Gsolv_error}}
    \subfigure[Training Loss]{\includegraphics[width=0.3\textwidth]{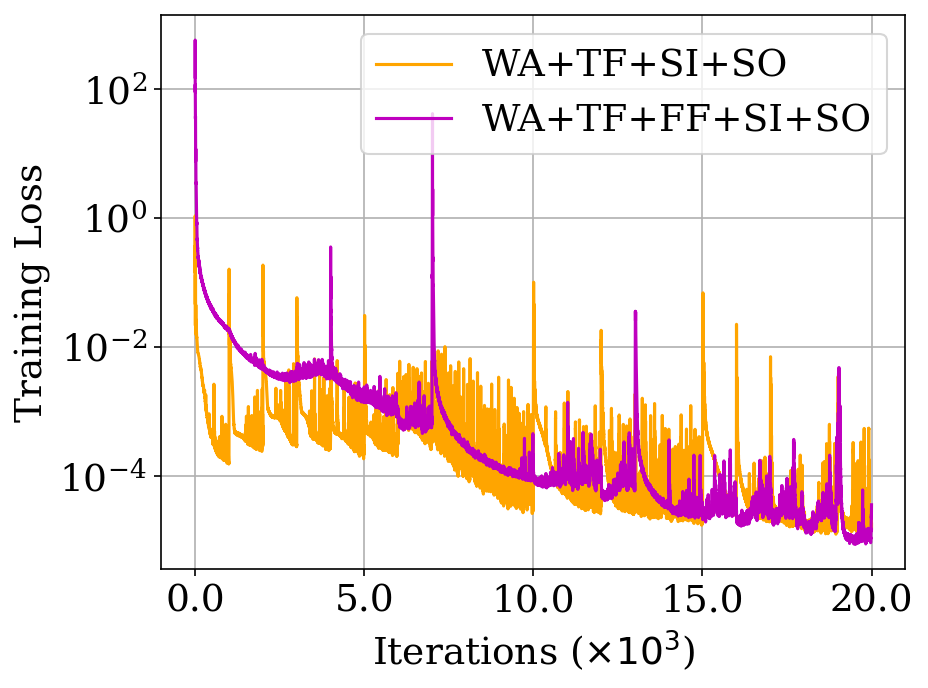}\label{fig:noncentered_losses}}
        \subfigure[Validation Loss]{\includegraphics[width=0.3\textwidth]{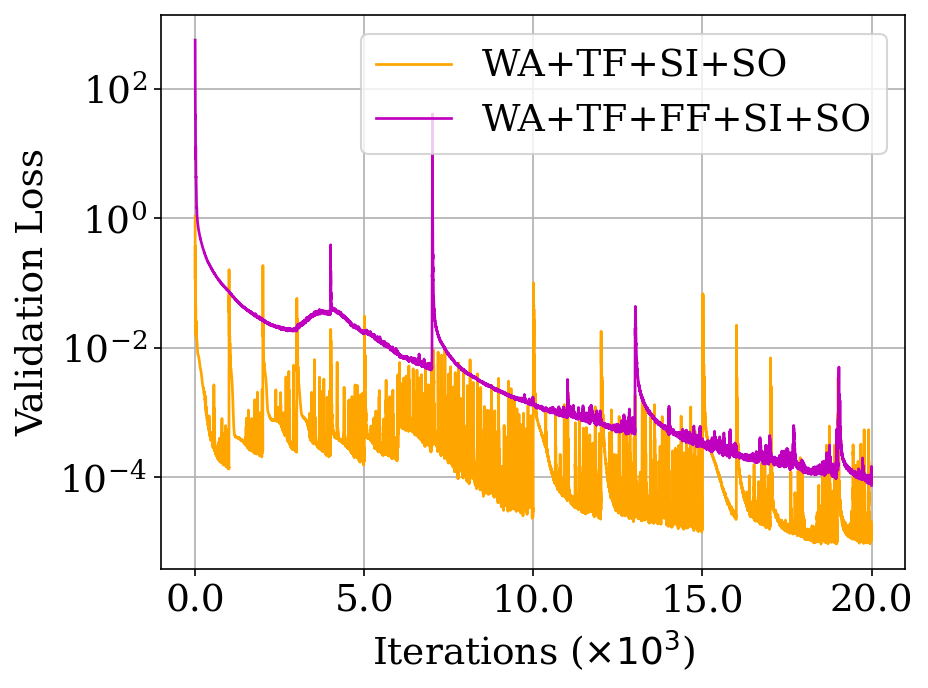}\label{fig:noncentered_losses_val}}
            \subfigure[$D_\psi$]{\includegraphics[width=0.3\textwidth]{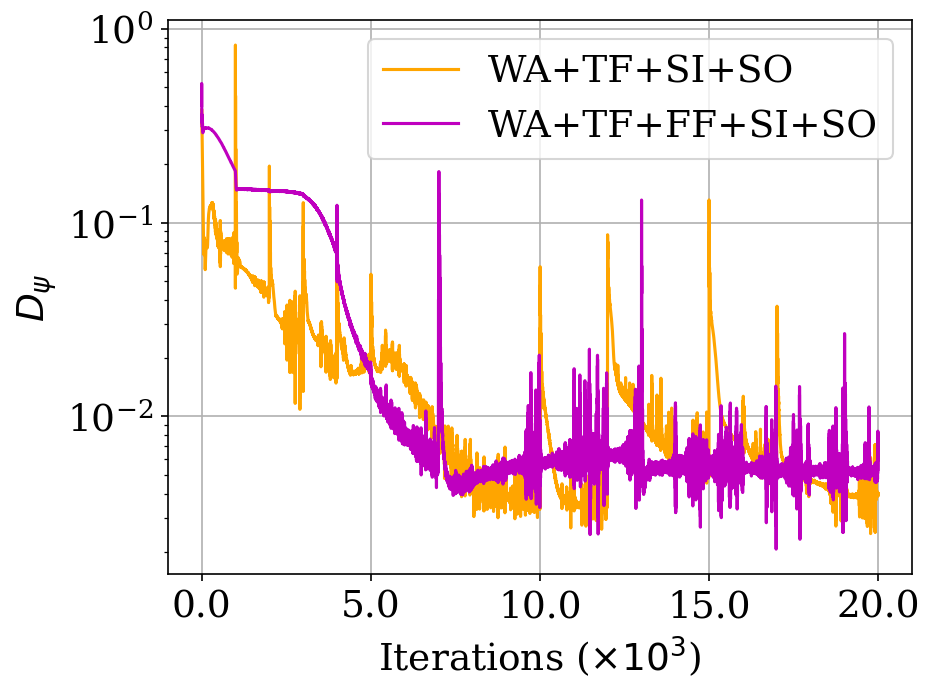}}
    \caption{Evolution of $\Delta G_{solv,\theta}$, $D_{G_{solv}}$, training and validation loss, and $D_\psi$ with iterations for the sphere with an off-centered charge with 2 different architectures.}
    \label{fig:gsolv_tloss_noncentered}
  \end{figure}

The results are shown in Figs. \ref{fig:potential_noncentered} and \ref{fig:gsolv_tloss_noncentered}, and Table \ref{tab:results_table_noncentered}. 
Even though the results in Fig. \ref{fig:potential_noncentered} and Table \ref{tab:results_table_noncentered} are not conclusive regarding the impact of including the Fourier features, Fig. \ref{fig:gsolv_tloss_noncentered} gives more information. The evolution of $\Delta G_{solv,\theta}$, $D_{G_{solv}}$, and the losses is less noisy when considering the Fourier features (purple line), and hence, we decided to continue our study using the architecture with all features (WA + FF + TF + SI + SO) included.

  \begin{table}[htbp]
    \centering
    \caption{Results for the sphere with an off-centered charge after 20,000 iterations. The analytical solution was -205.55 kcal/mol.}
    \begin{tabular}{|c|c|c|c|c|c|}
      \hline
      Case  & $D_{G_{solv}}$  & $D_\psi$ & Training & Validation   \\
      &  &  & loss & loss  \\
      \hline
      WA+TF+SI+SO    & 8.44E-03 & 3.92E-03 & 1.51E-05 & 1.38E-05 \\
      WA+TF+FF+SI+SO    & 5.54E-03 & 6.57E-03 & 3.53E-05 & 7.33E-05 \\
      \hline
      \end{tabular}%
    \label{tab:results_table_noncentered}%
  \end{table}%

%
%
%
%
%

  \subsection{Arginine}

\subsubsection{Density of collocation nodes} 

Moving to a more realistic setting, we now use  PINN  on a single arginine\footnote{\url{https://www.rcsb.org/ligand/ARG}} (27 atoms), and study how the density of collocation nodes at the surface affects the quality of the solution.
Table \ref{tab:arg_coll_node_study} describes the number of collocation nodes in 4 cases ({\it Coarse}, {\it Medium}, {\it Fine}, and {\it Finest}). 
The collocation node distributions were obtained from surface triangulations with 0.5, 1.0, 2.0, and 4.0 vertices per \AA$^2$ on $\Gamma$, respectively, 0.59 vertices per \AA$^2$ on $\partial\Omega$, and  tetrahedrons that conform to said triangulations, which have a maximum volume limit of 0.05 \AA$^3$ for $\Omega_m$ and 0.6 \AA$^3$ for $\Omega_s$.
  \begin{table}[ht]
    \centering
    \caption{Number of collocation nodes in density study of arginine}
      \begin{tabular}{|c|c|c|c|c|}
        \hline
         Density & $\Omega_m$ & $\Omega_w$  & $\Gamma$ & $\partial\Omega$  \\
        \hline
         Coarse & 3383 & 10413 & 282 & 1238\\
         Medium & 4418 & 10440 & 372 & 1238\\
         Fine   & 6017 & 11596 & 624 & 1238 \\
         Finest & 7568 & 15089 & 1318 & 1238\\
      \hline
      \end{tabular}%
    \label{tab:arg_coll_node_study}%
  \end{table}%

\begin{figure}[ht]
    \centering
    \subfigure[Coarse]{\includegraphics[width=0.3\textwidth]{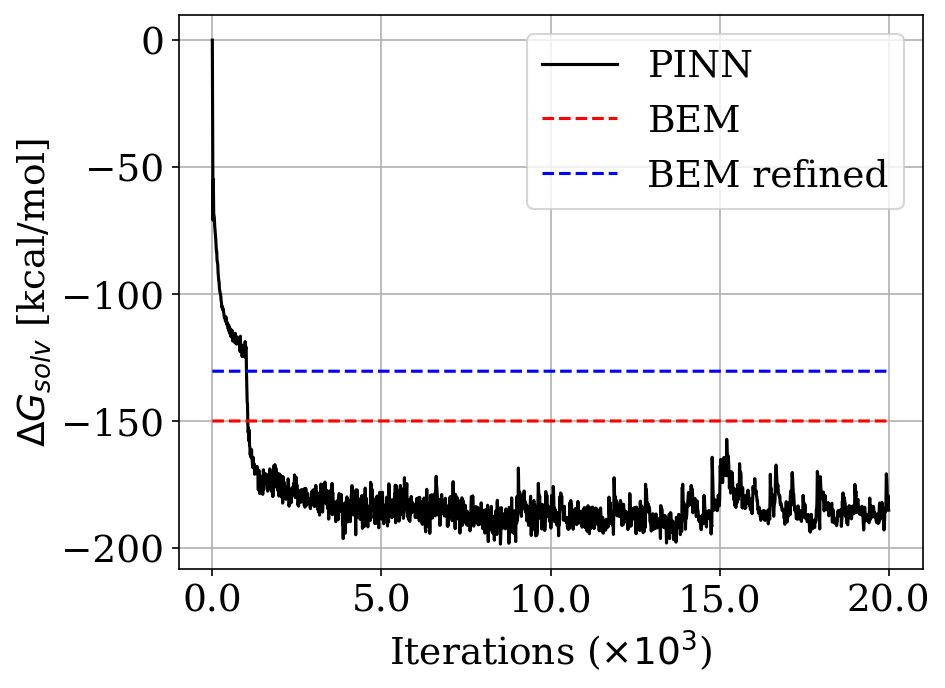}\label{fig:arg_coarse}}
    \subfigure[Medium]{\includegraphics[width=0.3\textwidth]{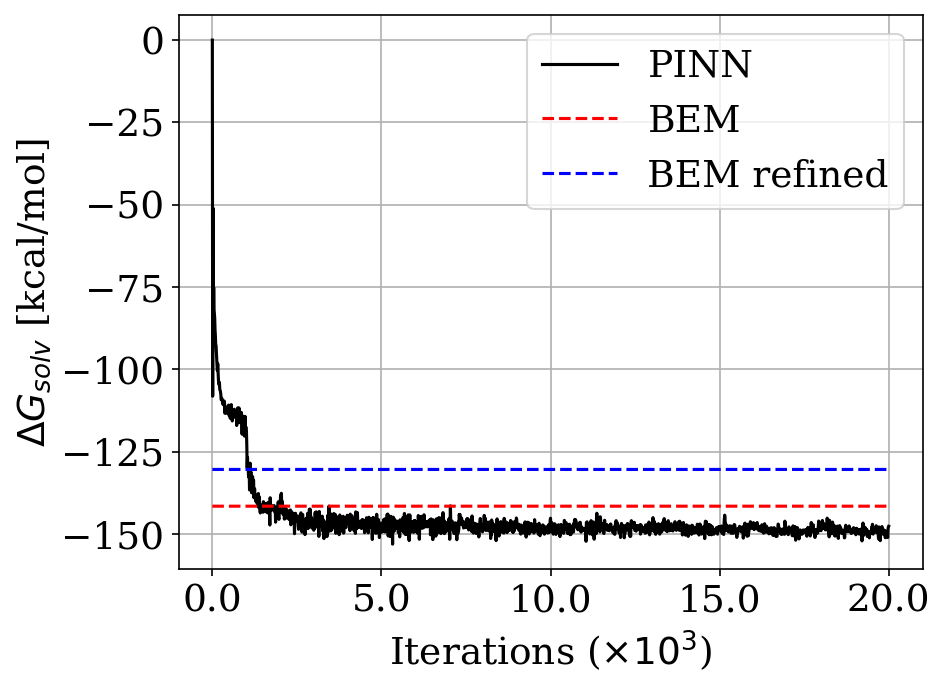}\label{fig:arg_medium}}
    \subfigure[Fine]{\includegraphics[width=0.3\textwidth]{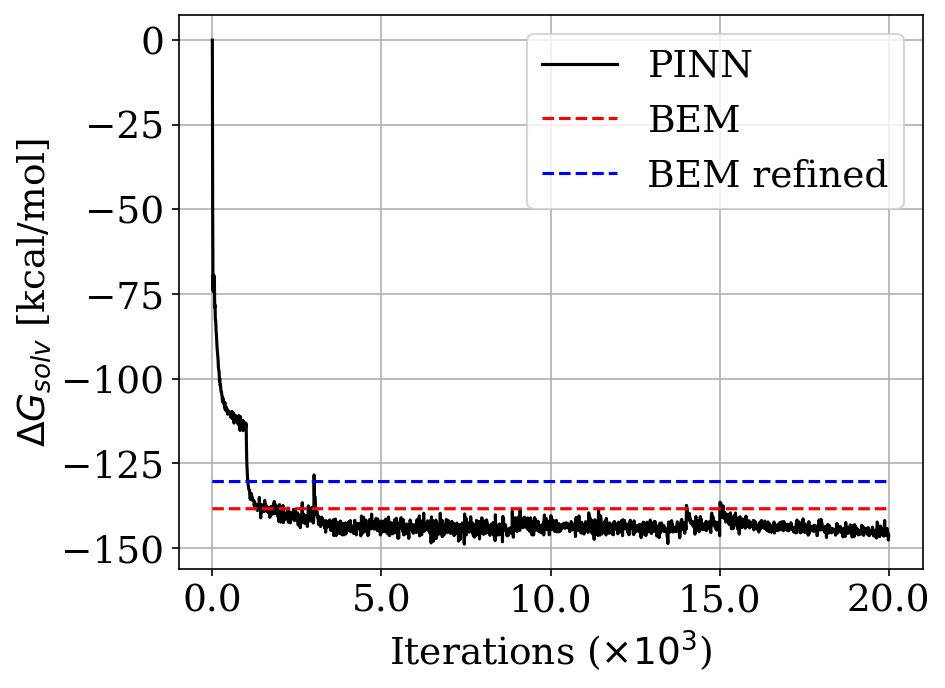}\label{fig:arg_fine}}
    \subfigure[Finest]{\includegraphics[width=0.3\textwidth]{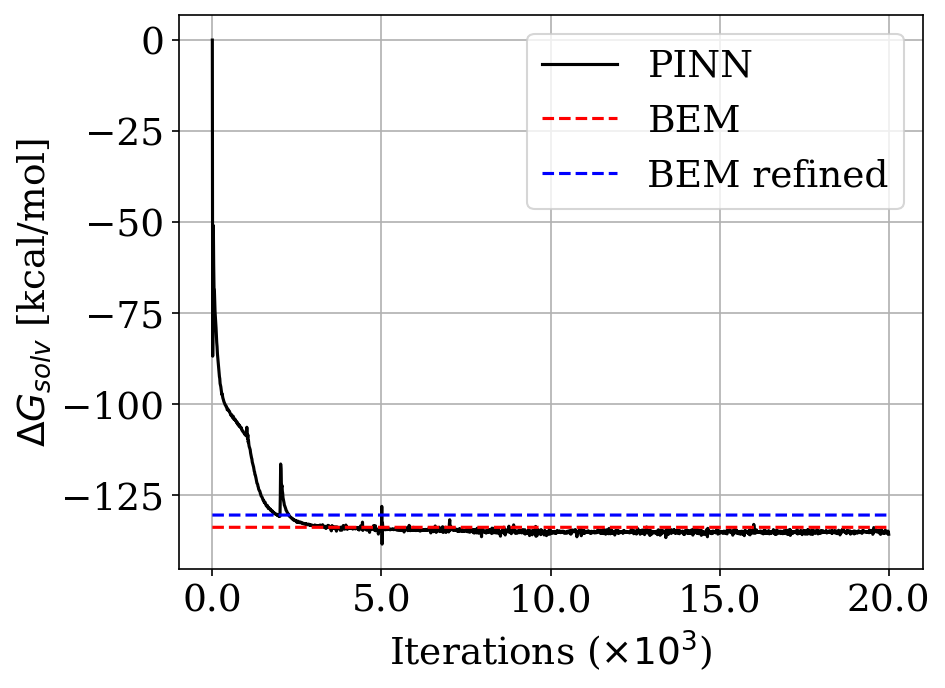}\label{fig:arg_finest}}
    \caption{Solvation energy history with iterations (black line). The red line is a BEM solution computed on the surface mesh that was used to generate the collocation nodes. The blue line corresponds to a fine-mesh BEM solution. }
    \label{fig:arg_gsolv_mesh}
  \end{figure}

Table \ref{tab:res_arg_coll} shows the results for the different collocation point densities. 
Similar to standard numerical methods, all indicators improve as the number of collocation nodes increases.
Note that $D_{G_{solv}}$ is computed against a BEM solution that uses the equivalent surface mesh for each case ({\it ie.} the reference solution and definition of $\Gamma$ is different in each case). 
Regardless, $D_{G_{solv}}$ and $D_\psi$ decrease with the number of collocation nodes, indicating that it is converging to the numerical solution computed with BEM. 
The latter statement is more evident in Fig. \ref{fig:arg_gsolv_mesh}, where the black line converges to the red line, which corresponds to the BEM solution computed on the mesh that was used to generate the collocation points. As a reference, the blue line in Fig. \ref{fig:arg_gsolv_mesh} is a BEM solution with a surface mesh that is 4 times finer than the {\it Finest} case. 

\begin{figure}[ht]
    \centering
    \subfigure[Coarse]{\includegraphics[width=0.3\textwidth]{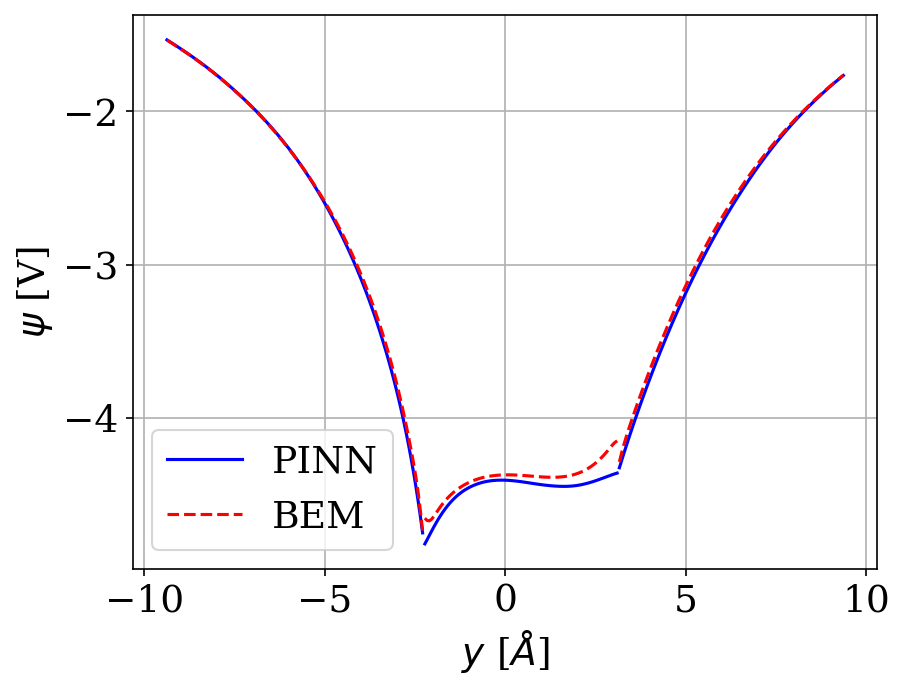}}
    \subfigure[Medium]{\includegraphics[width=0.3\textwidth]{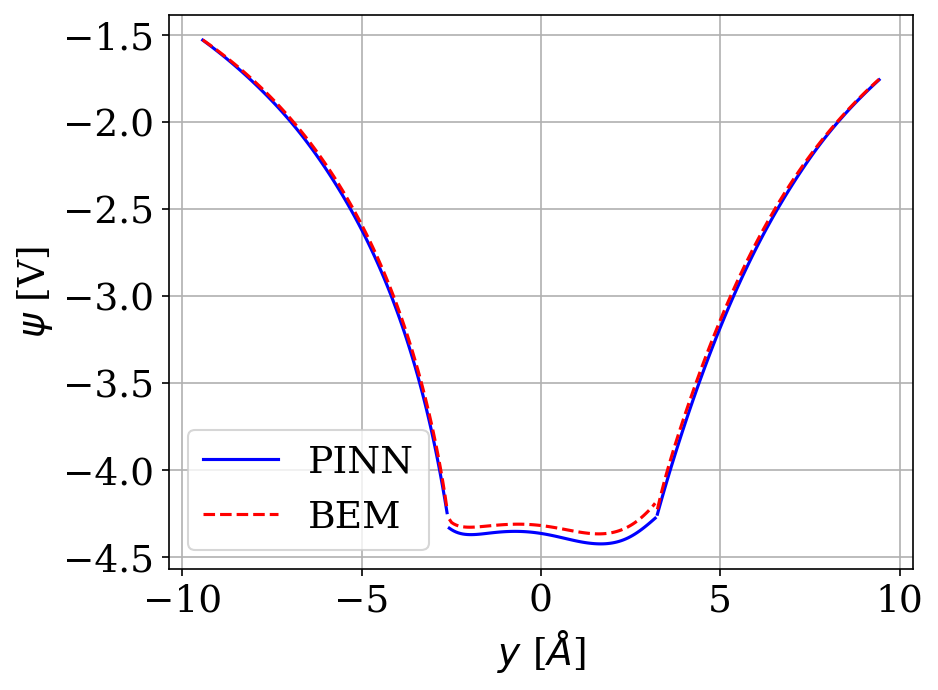}}
    \subfigure[Fine]{\includegraphics[width=0.3\textwidth]{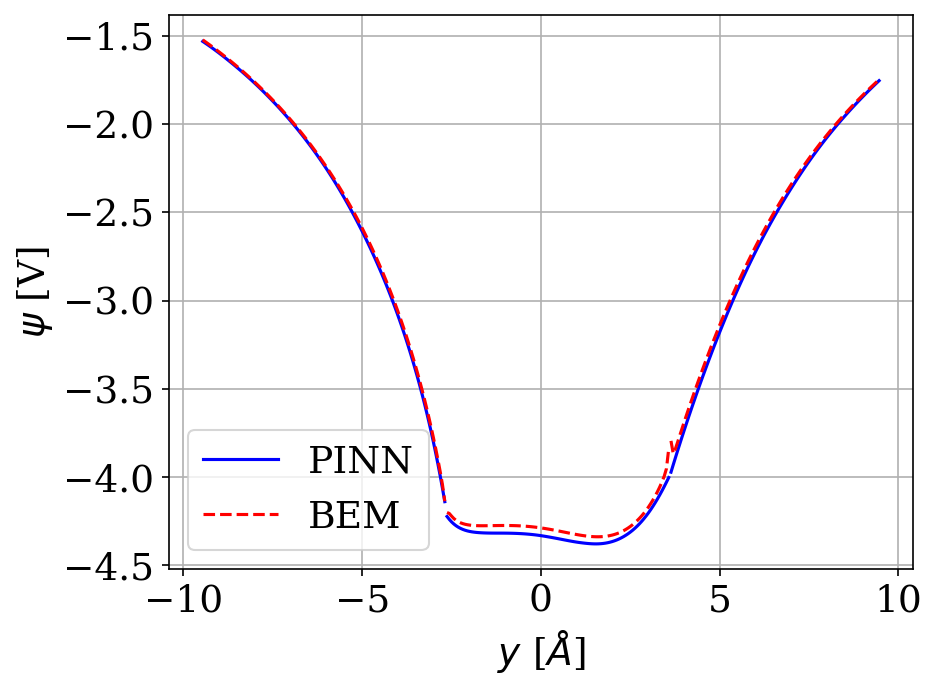}}
    \subfigure[Finest]{\includegraphics[width=0.3\textwidth]{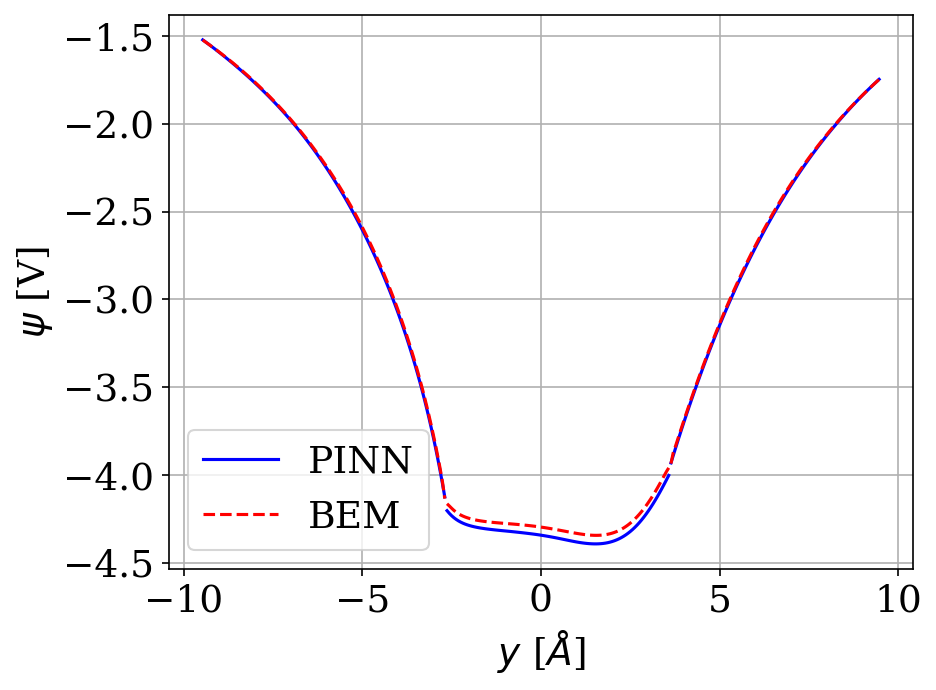}\label{fig:arg_ang2_mesh_finest}}
    \caption{Reaction potential ($\psi$) for arginine with different node refinements, along the $y$ axis. Red line: BEM solution. Blue line: PINN solution.}
    \label{fig:arg_ang2_mesh}
  \end{figure}

Fig. \ref{fig:arg_ang2_mesh} shows the reaction potential in the $y$ axis. From these results, it is evident that the coarsest PINN calculation (Fig \ref{fig:arg_coarse}) struggles to adapt to the BEM solution, specially near the interface, however, this improves substantially for the {\it Medium} density in Fig. \ref{fig:arg_medium}. For the two finest cases in Figs. \ref{fig:arg_fine} and \ref{fig:arg_finest}, the reaction potential seems to have already converged to the BEM solution

\begin{figure}[ht]
    \centering
    \subfigure[Coarse]{\includegraphics[width=0.3\textwidth]{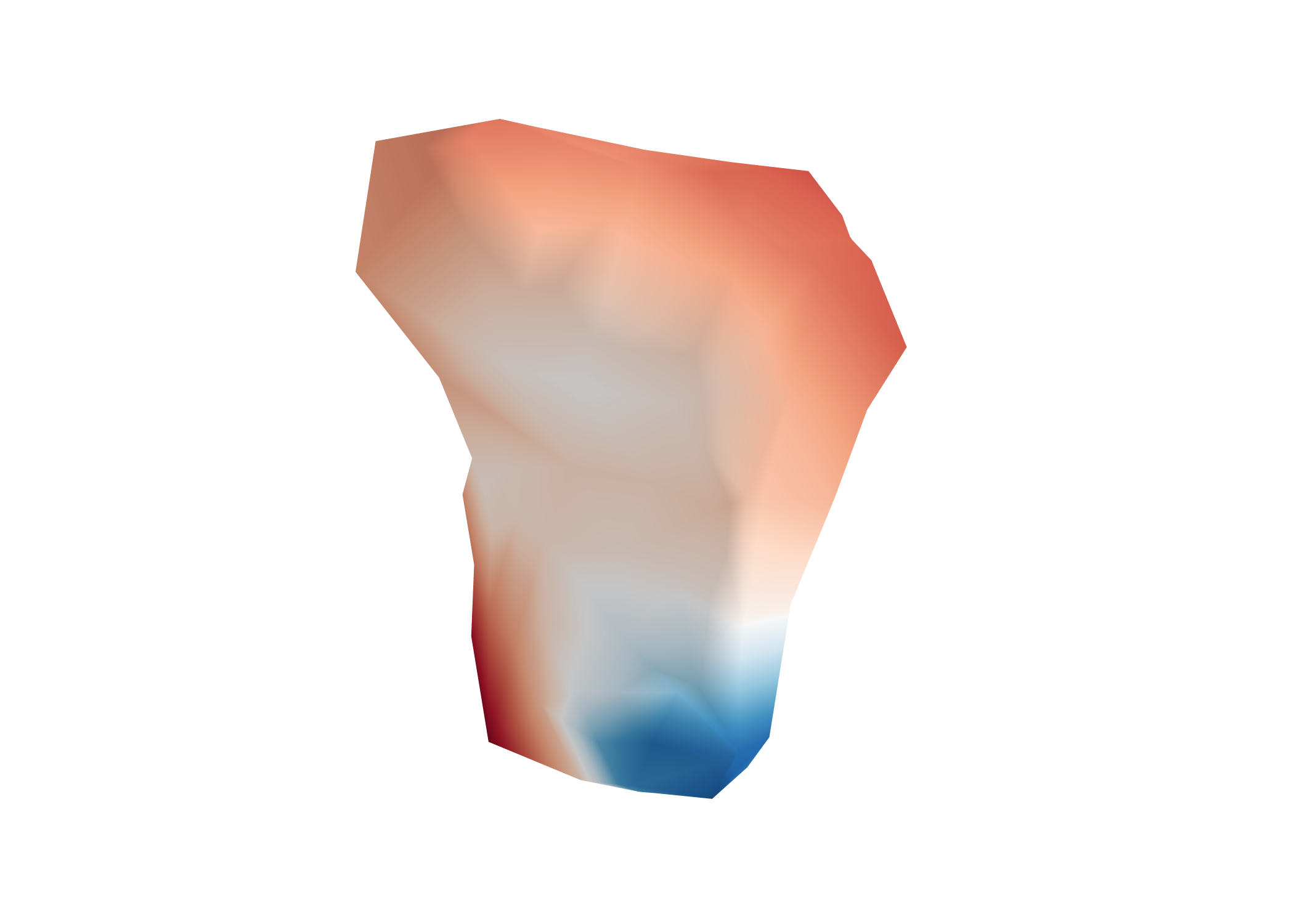}}
    \subfigure[Medium]{\includegraphics[width=0.3\textwidth]{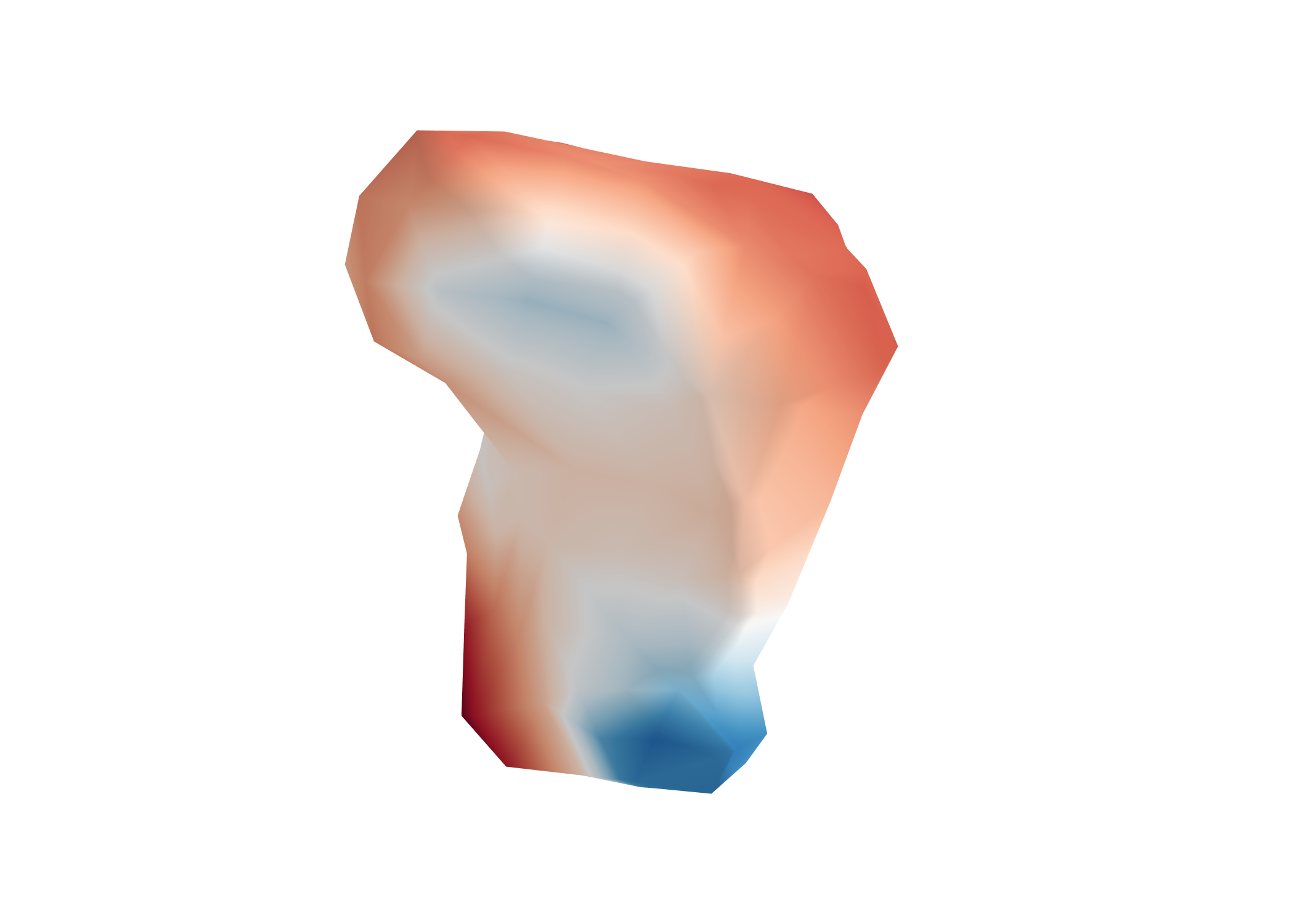}}
    \subfigure[Fine]{\includegraphics[width=0.3\textwidth]{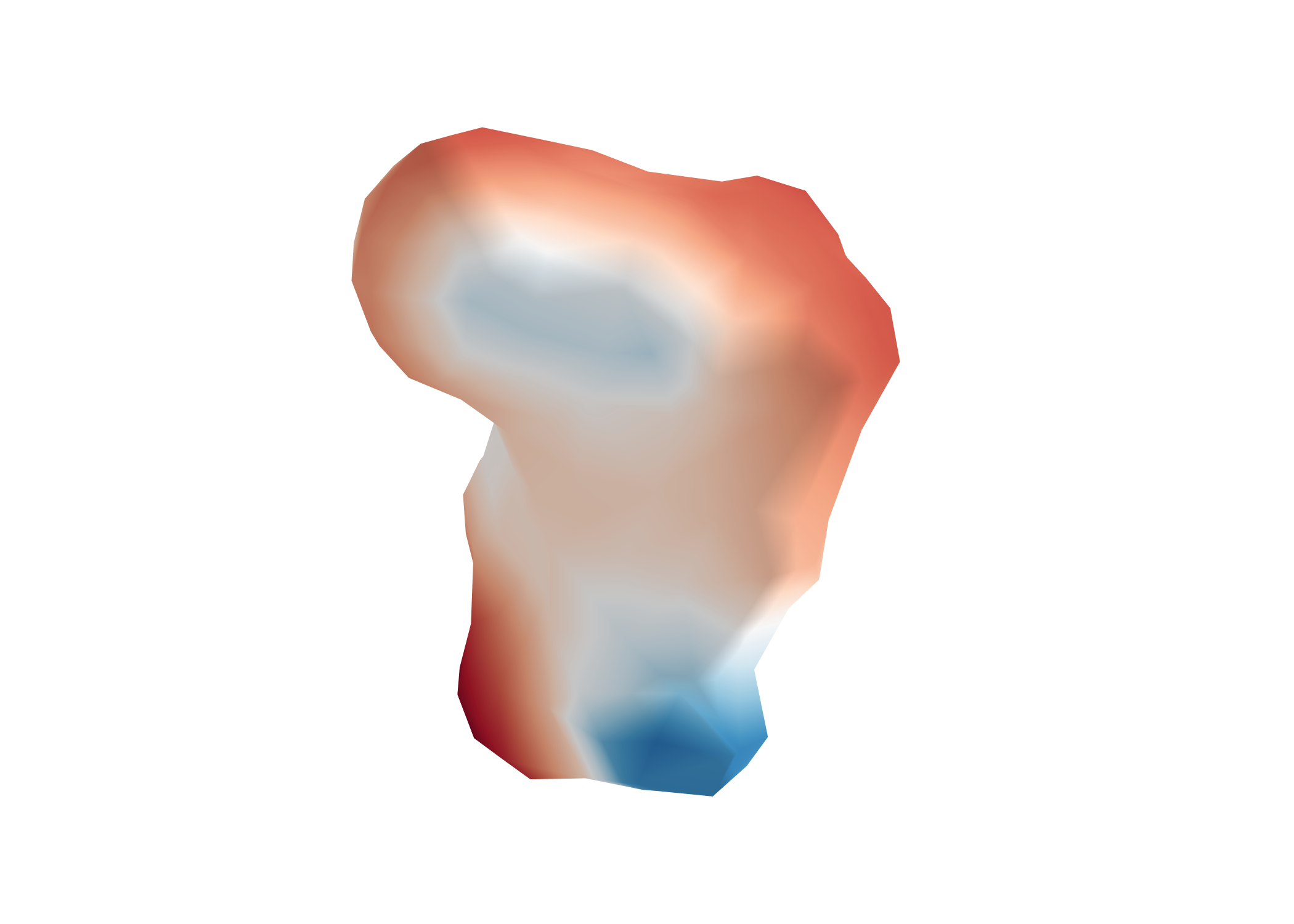}}
    \subfigure[Finest]{\includegraphics[width=0.3\textwidth]{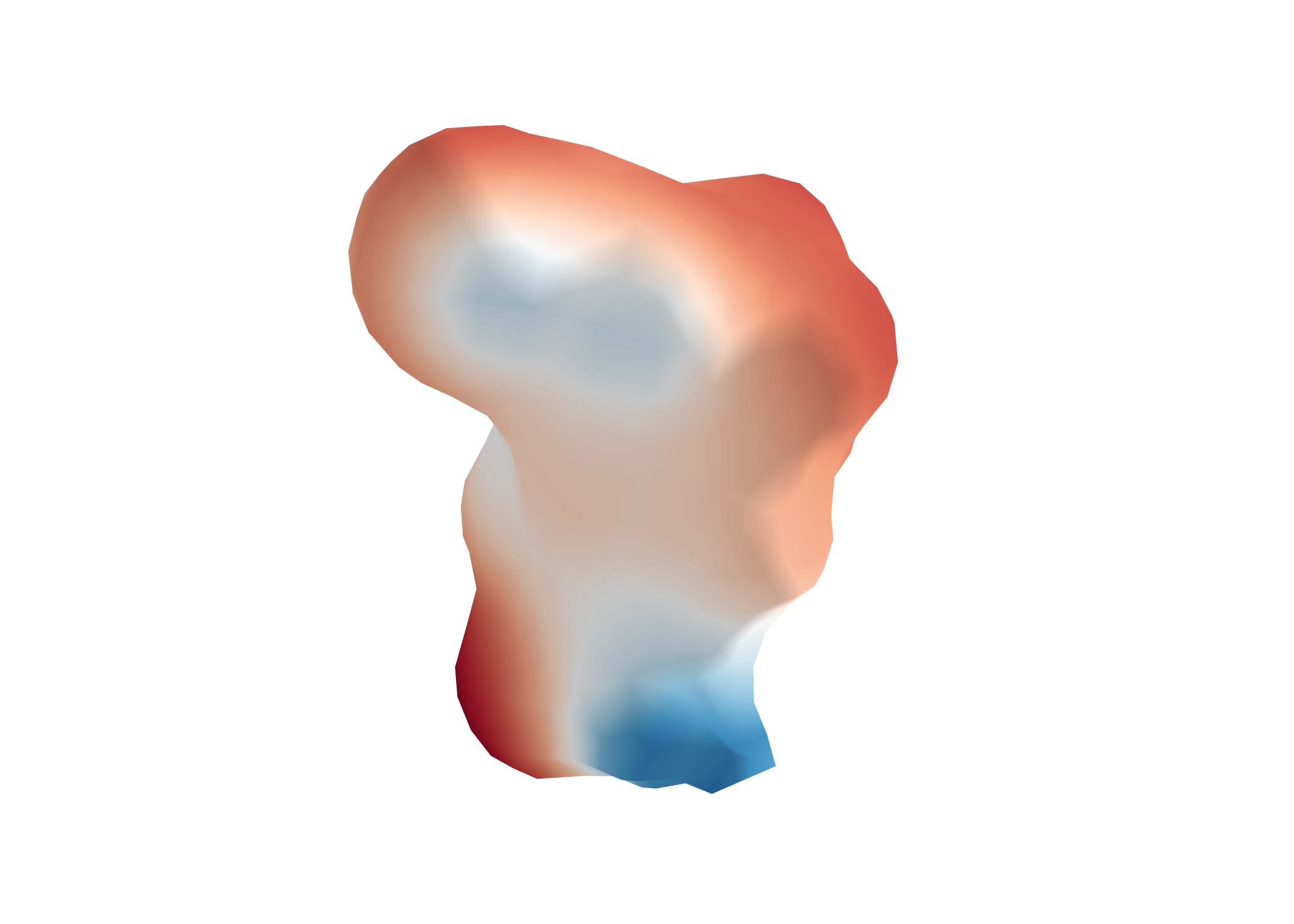}}
            \subfigure{\includegraphics[width=0.3\textwidth]{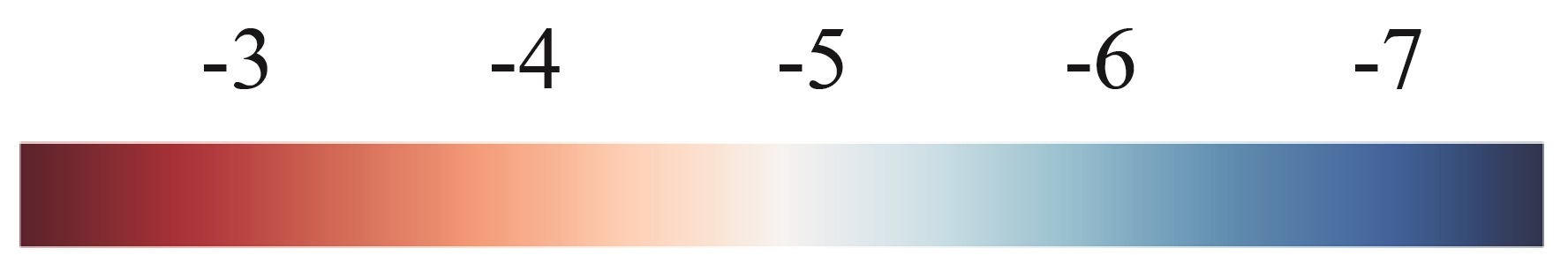}}
    \caption{Reaction potential ($\psi^{(\Gamma)}_\theta$, in Volts) on the molecular surface of arginine, for different collocation node density.}
    \label{fig:arg_3d_mesh}
  \end{figure}

\begin{figure}[ht]
    \centering
    \subfigure[Coarse]{\includegraphics[width=0.3\textwidth]{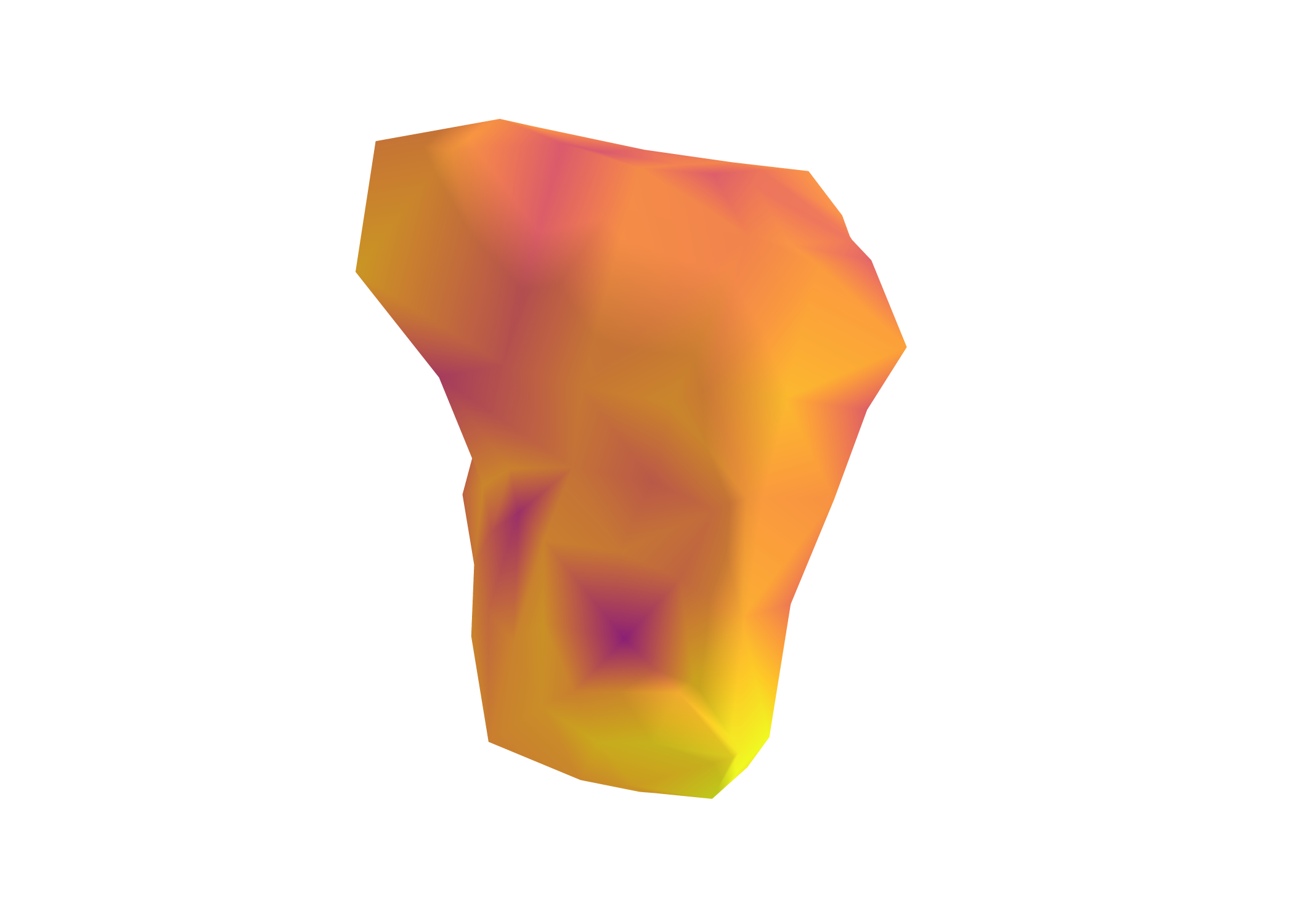}}
    \subfigure[Medium]{\includegraphics[width=0.3\textwidth]{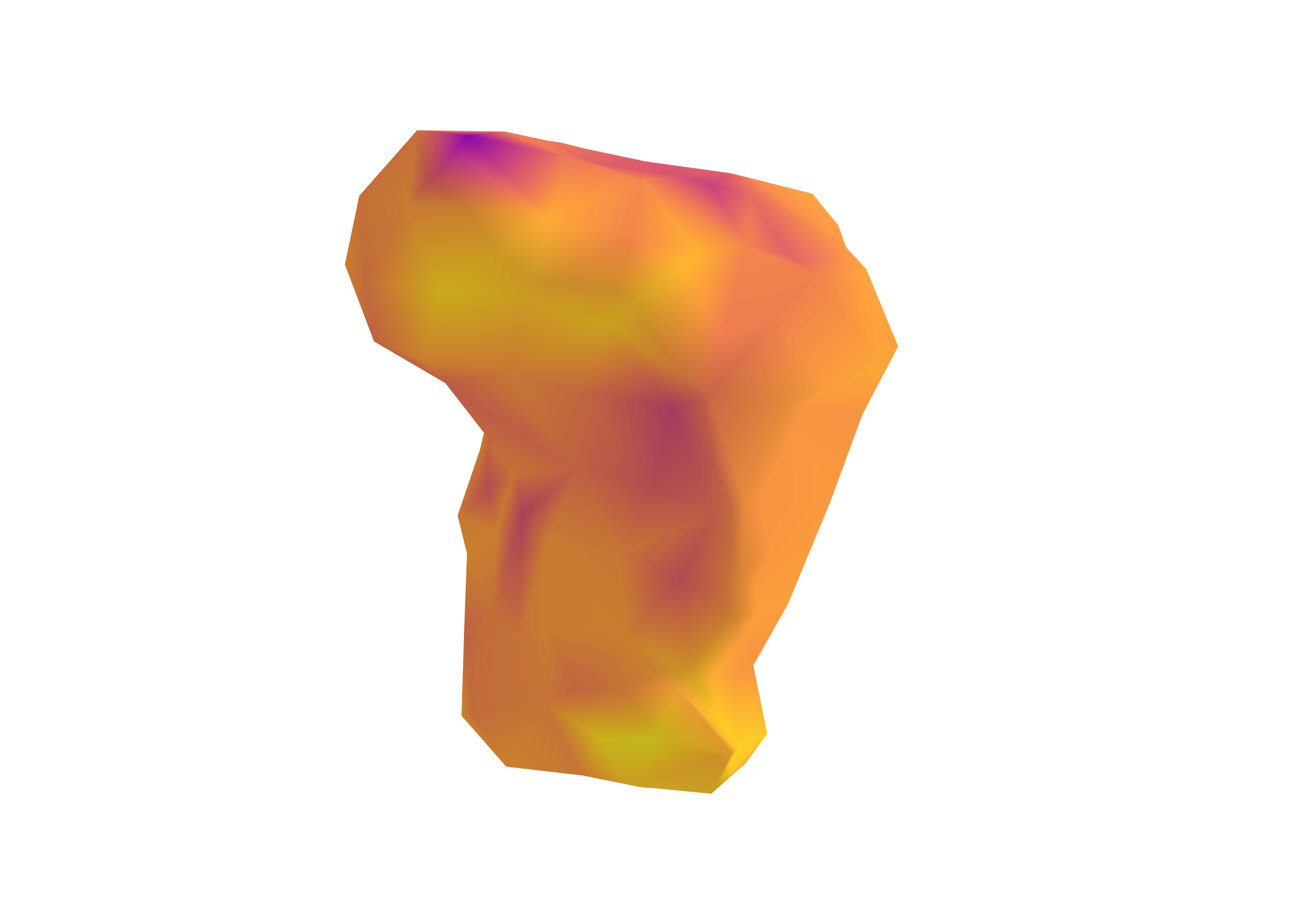}}
    \subfigure[Fine]{\includegraphics[width=0.3\textwidth]{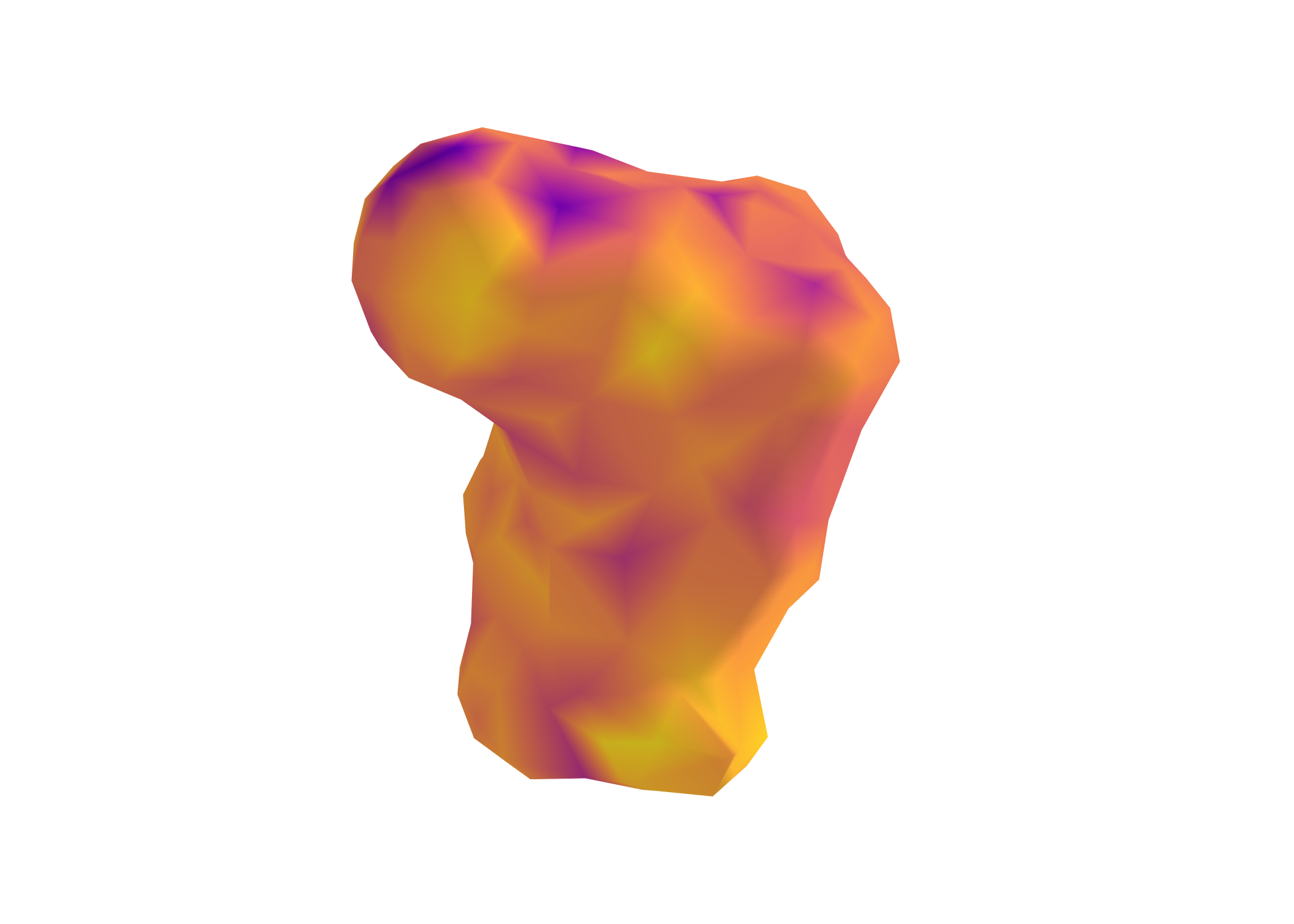}}
    \subfigure[Finest]{\includegraphics[width=0.3\textwidth]{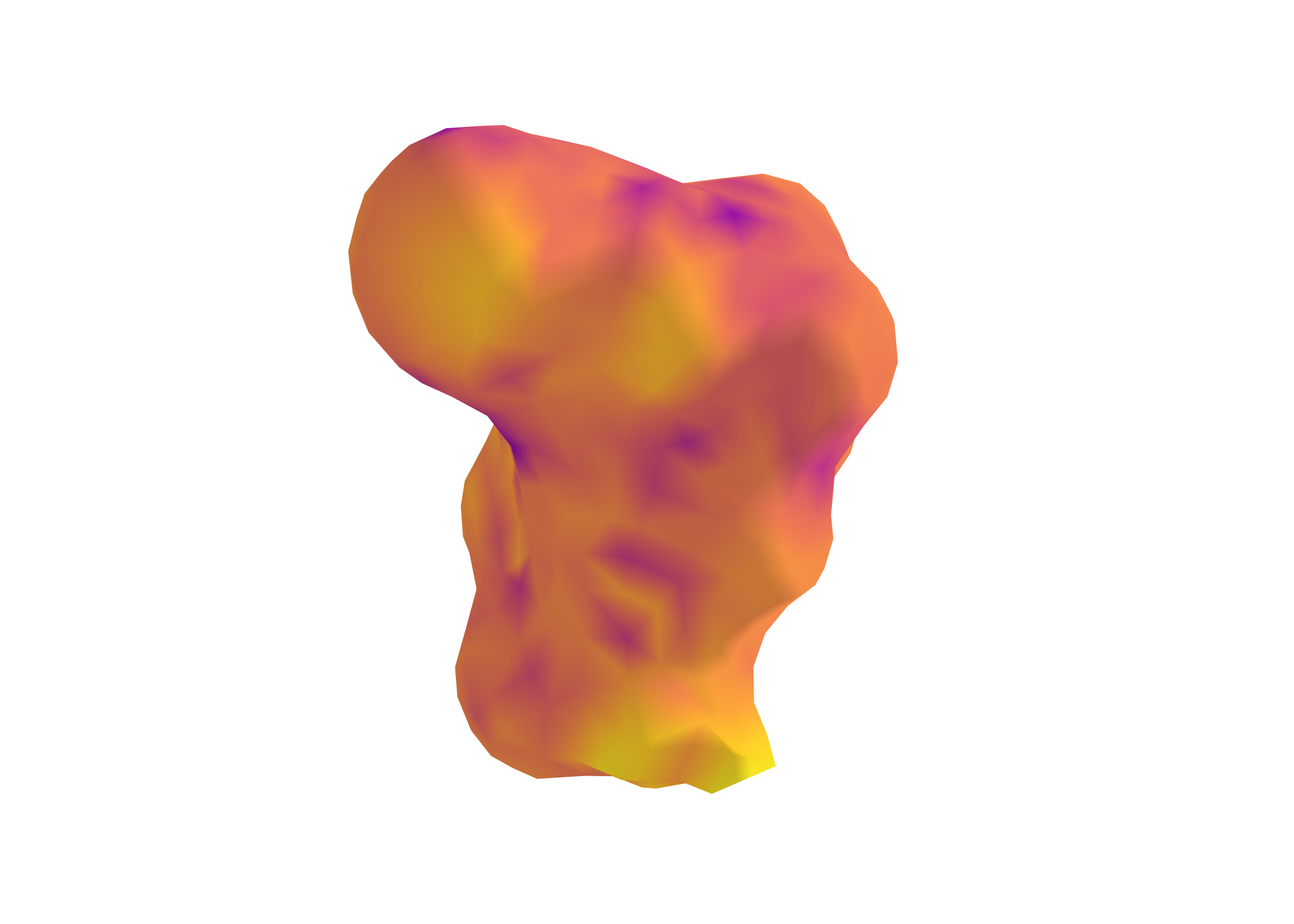}}
        \subfigure{\includegraphics[width=0.3\textwidth]{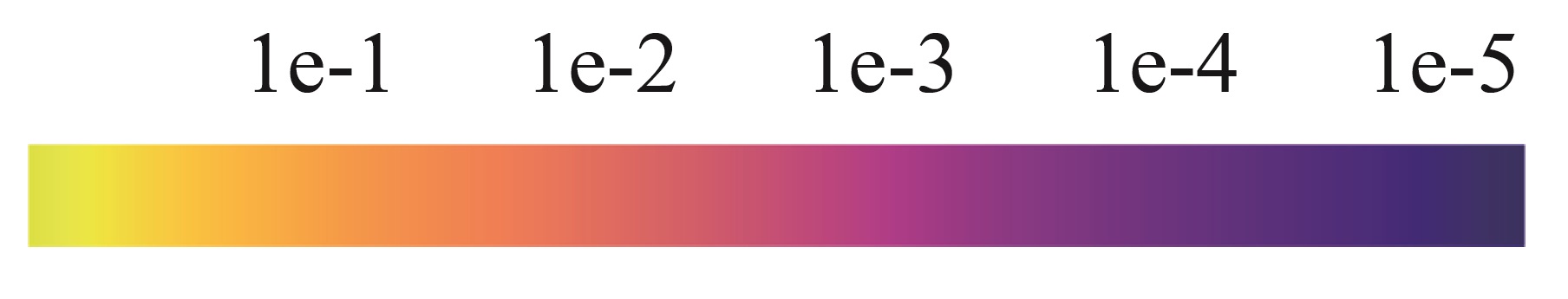}}
    \caption{Absolute difference (in Volts) in reaction potential between PINN and BEM on the molecular surface of arginine, for different collocation node density.}
    \label{fig:arg_3d_mesh_error}
  \end{figure}

Figs. \ref{fig:arg_3d_mesh} and \ref{fig:arg_3d_mesh_error} show the reaction potential on the molecular surface ($\Gamma$). Similar to Fig. \ref{fig:arg_ang2_mesh}, there are no notable differences in the reaction potential of Fig. \ref{fig:arg_3d_mesh} beyond the {\it Medium} density. However, the absolute difference plots of Fig. \ref{fig:arg_3d_mesh_error} do show differences for the finer cases, which clearly present more purple regions than the {\it Medium} density. This indicates closer agreement between PINN and BEM as the node density increases, something that is also evidenced by Table \ref{tab:res_arg_coll}. 

  \begin{table}[htbp]
    \centering
    \caption{Results for collocation node density study of arginine after 20,000 iterations.}
      \begin{tabular}{|c|c|c|c|c|c|}
        \hline
        Case  & $\Delta G_{solv,\theta}$ &  $D_{G_{solv}}$  & $D_\psi$ & Training & Validation   \\
         & kcal/mol & &  & loss & loss  \\
        \hline
      Coarse  & -185.1& 2.34E-01& 2.20E-02 & 1.58E-04 & 1.22E-04  \\
      Medium  & -147.7& 4.31E-02 & 1.41E-02 & 8.61E-05 & 9.90E-05  \\
      Fine  & -146.3 & 5.68E-02 & 1.01E-02 & 2.73E-05 & 1.32E-04  \\
      Finest & -135.7& 1.42E-02 & 8.21E-03 & 3.72E-06 & 9.84E-06  \\
      \hline
      \end{tabular}%
    \label{tab:res_arg_coll}%
  \end{table}%

The surface and volume meshes used to generate the collocation points are related because the tetrahedrons in $\Omega_m$ and $\Omega_w$ conform to the triangles in $\Gamma$ and $\partial\Omega$. However, there is no clear reason why the surface and volume collocation points should be coupled.
To decouple them, we performed two extra calculations where, starting from the collocation points from the {\it Finest} calculation, we sample only a subset of the volume collocation points, while using all surface nodes. 
This way, the geometrical details of $\Gamma$ remain constant, while decreasing the sampling size in the volume, and hence, the computational cost of each iteration.

Table \ref{tab:arg_vol_sample} details the sampling size, which correspond to 30\% and 60\% of the volume nodes. 
Results are presented in Table \ref{tab:res_arg_sample} at 20,000 iterations (like Table \ref{tab:res_arg_coll}) and 35,000 iterations.
The logic behind exploring results with more iterations is that, as only a subset of the tetrahedral volumes are considered to generate the collocation nodes, more iterations may be necessary to correctly sample the whole space. 
The latter intuition is somewhat true, as
the $D_{G_{solv}}$ and $D_\psi$ improve from 20,000 to 35,000 iterations for both {\it Finest 30\%} and {\it Finest 60\%}. However, there is no significant improvement in the indicators from Table \ref{tab:res_arg_sample} compared to the {\it Finest} case in Table \ref{tab:res_arg_coll}, with $D_{G_{solv}}$ and $D_\psi$ being in the order of 10$^{-2}$---10$^{-3}$, consistent with  other results in this work. This demonstrates that sampling a subset of volume nodes while maintaining the surface nodes is an effective strategy towards lowering the computational cost without sacrificing accuracy. 

\begin{table}[ht]
  \centering
  \caption{Number of collocation nodes in each domain and surface, when sampling a subset of the volume nodes.}
    \begin{tabular}{|c|c|c|c|c|}
      \hline
       Simulation & $\Omega_m$  & $\Omega_w$  & SES & $\partial\Omega$  \\
      \hline
       Finest 30\% & 2282 & 4589 & 1318 & 1238\\
       Finest 60\% & 4494 & 9090 & 1318 & 1238\\
    \hline
    \end{tabular}%
  \label{tab:arg_vol_sample}%
\end{table}%

\begin{table}[htbp]
  \centering
  \caption{Results for arginine with subset of volume collocation nodes.}
    \begin{tabular}{|c|c|c|c|c|c|}
      \hline
      Case & Iter.   & $D_{G_{solv}}$ & $D_\psi$ & Train. & Val.   \\
      & &  &  &  loss & loss  \\
      \hline
      Finest 30\% & 20000 & 1.10E-02 & 8.05E-03 & 5.25E-06 & 8.10E-06  \\
      Finest 30\% & 35000  & 8.72E-03 & 7.78E-03 & 2.84E-06 & 1.28E-05  \\
      Finest 60\% & 20000 & 9.85E-03 & 9.23E-03 & 4.31E-06 & 5.27E-06  \\
      Finest 60\% & 35000  & 9.63E-03 & 6.71E-03 & 3.82E-06 & 1.45E-05  \\
      \hline
    \end{tabular}%
  \label{tab:res_arg_sample}%
\end{table}%


\subsubsection{Incorporating experimental measurements}
One attractive feature of PINN that sets it apart from standard numerical techniques is the freedom to add loss functions with information from other sources, like different models or experimental measurements (see Eq. \eqref{eq:loss_alternative}). This is specially exciting in molecular electrostatics, as recent advances in NMR spectroscopy are capable of measuring effective electrostatic potentials ($\phi_{ENS}$) around hydrogen atoms of a molecule.\cite{yu2021novo} This quantity is computed from a simulation as
\begin{equation}\label{eq:phi_ens}
    \phi_{ENS} = -\frac{k_BT}{e}\ln{\frac{\Gamma_{2,+}}{\Gamma_{2,-}}}
\end{equation}
where the $\Gamma_{2,+}$ and $\Gamma_{2,-}$ are the rate of transverse magnetization, which is:
\begin{equation}\label{eq:phi_ens_gamma2}
    \Gamma_{2,\pm} = C_0 \int_{\Omega_w} r^{-6} \exp{-\frac{\pm e \phi}{k_BT}} \odif{V},
\end{equation}
 $C_0$ being a constant that depends on NMR parameters, but is irrelevant to our case, as it is canceled out in the ratio $\Gamma_{2,+}/\Gamma_{2,-}$ of Eq. \eqref{eq:phi_ens}.

 We performed calculations on arginine using the same setup as {\it Finest} in Table \ref{tab:arg_coll_node_study}, this time including the following loss term:
\begin{equation}\label{eq:loss_exp}
    L_{\phi_{ENS}}(\bm{\theta}_w;S_{\Omega_w}) = \frac{1}{N_{H_{at}}}\sum_{H_{at} \in \text{atoms}} \Bigg[\phi_{ENS}^\theta(H_{at,i}) - \phi_{ENS}^{exp} \Bigg]^2
\end{equation}
where $\phi_{ENS}^{exp}$ is the experimental measurement, $\phi_{ENS}^{\theta}$ the PINN calculation, and the sum is over the $N_{H_{at}}$ hydrogen atoms ($H_{at}$) of the molecule where the measurement is performed. In this case, we used $\phi_{ENS}$ computed with the PBE (using BEM through PBJ\cite{search2022towards}) as $\phi_{ENS}^{exp}$, as it was shown to be a good approximation.\cite{yu2021novo}
To correctly compute the integral in Eq. \eqref{eq:phi_ens_gamma2}, we had to extend the domain $\Omega_w$ to $d_{\partial\Omega}=7$ \AA.

During the training process, the significant oscillations in the electrostatic potential led to large values in the exponential term of Eq. \eqref{eq:phi_ens_gamma2}, causing the solution to diverge. To mitigate this issue, the exponential function was approximated using its series expansion: $\exp(x) = 1.0 + x + x^2/2! + x^3/3! + x^4/4!$.

\begin{figure}[ht]
  \centering
  \includegraphics[width=0.4\textwidth]{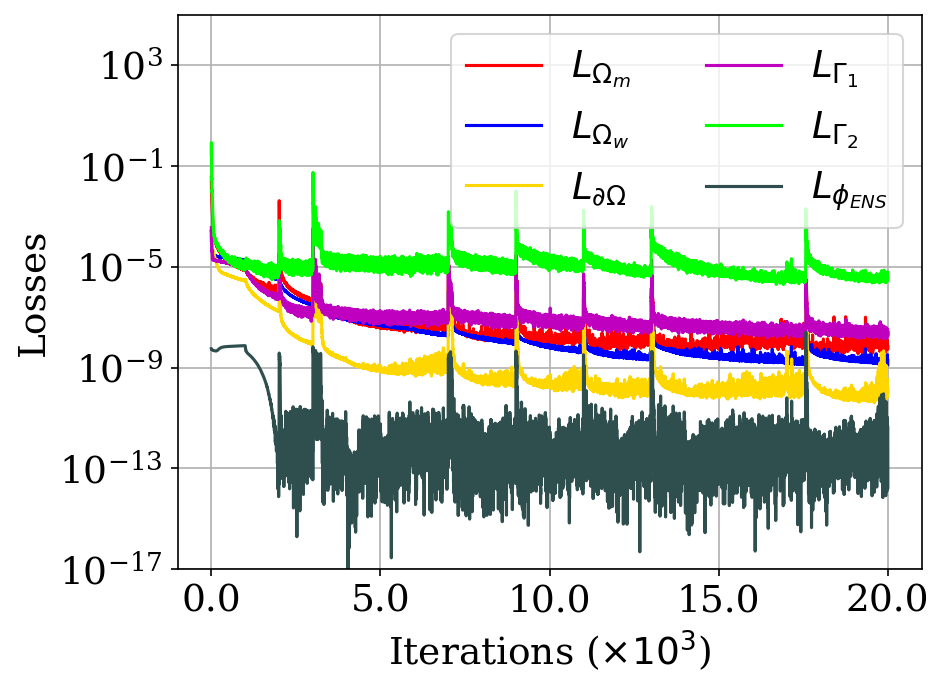}
  \caption{Evolution of losses for arginine including Eq. \eqref{eq:loss_exp}}
  \label{fig:experimental}
\end{figure}

\begin{figure}[ht]
  \centering
  \includegraphics[width=0.4\textwidth]{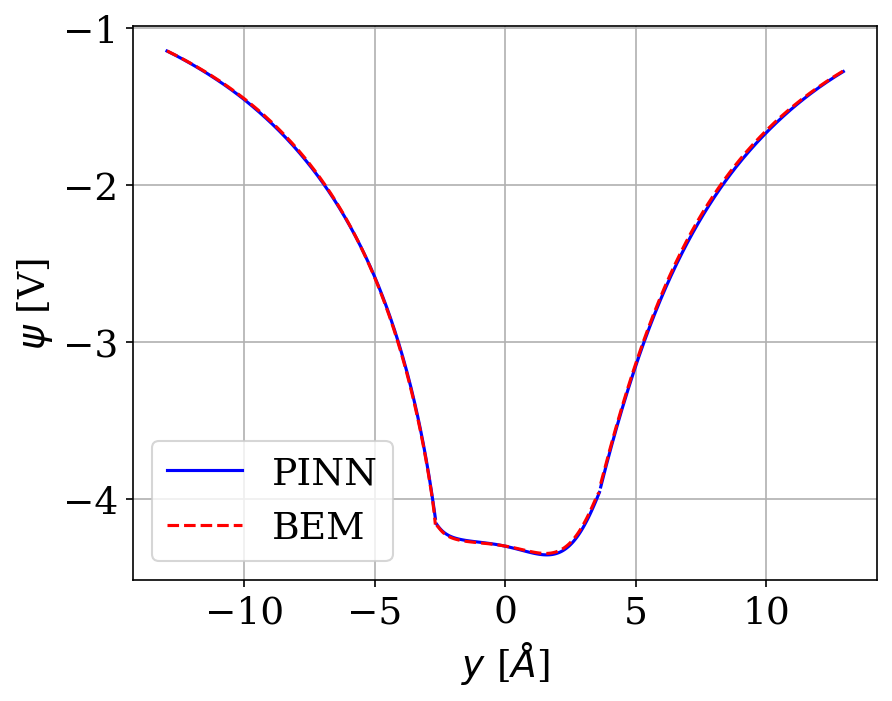}
  \caption{Reaction potential ($\psi$) for arginine along the $y$ axis using experimental measurements.}
  \label{fig:psi_experimental}
\end{figure}

The results for this setup are presented in Figs. \ref{fig:experimental} and \ref{fig:psi_experimental}, and Table \ref{tab:arg_exp}. 
Fig. \ref{fig:experimental} shows that the loss function decreases with increasing number of iterations for arginine, with
results in Table \ref{tab:arg_exp} that are comparable with the {\it Finest} case in Table \ref{tab:res_arg_coll}. Fig. \ref{fig:psi_experimental} also shows a similar behavior to the case without the experimental loss function in Fig. \ref{fig:arg_ang2_mesh_finest}. Moreover, the relative differences in $\phi_{ENS}$ between PINN and BEM for two hydrogens in the arginine structure in Table \ref{tab:phi_ens_error} are also in the $\sim$10$^{-2}$---10$^{-3}$ range. Even though the experimental loss in Fig. \ref{fig:experimental} is noisy, it is consistently lower than the others losses, indicating that PINN incorporates the experimental $\phi_{ENS}$ appropriately.

\begin{table}[htbp]
\centering
\caption{Results for arginine considering a loss function that includes experimental measurements (Eq. \eqref{eq:loss_exp}) after 20,000 iterations.}
  \begin{tabular}{|c|c|c|c|c|}
    \hline
    $\Delta G_{solv,\theta}$ & $D_{G_{solv}}$  & $D_\psi$ & Training & Validation \\
    kcal/mol &  &  & loss & loss \\
    \hline
  -134.86 & 8.24E-03 & 1.09E-02 & 7.88E-06 & 1.42E-05  \\
  \hline
  \end{tabular}%
\label{tab:arg_exp}%
\end{table}%

\begin{table}[ht]
    \centering
    \caption{Relative difference in the prediction of $\phi_{ENS}$ for two hydrogens in arginine.}
    \begin{tabular}{|c|c|}
        \hline
        Hydrogen &\multirow{2}{*}{$\frac{|\phi_{ENS}^\theta - \phi_{ENS}^{exp}|}{\phi_{ENS}^{exp}}$} \\
        number & \\
        \hline
        1 & 9.80E-03 \\
        2 & 1.11E-02 \\
        \hline
    \end{tabular}
    \label{tab:phi_ens_error}
\end{table}


\subsection{\label{sec:proteins}Full proteins}

To show the applicability of PINN in real problems, we computed the electrostatic potential and $\Delta G_{solv}$ of the immunoglobulin-binding domain of protein G (PDB code 1pgb,\cite{gallagher1994two} 855 atoms) and ubiquitin (PDB code 1ubq,\cite{vijay1987structure} 1231 atoms). Following the conclusions from the results for the sphere and arginine, we considered all the features detailed in section \rev{``An enhanced PINN architecture for RPB''} ({\it ie.} WA+TF+FF+SI+SO), and sampling a subset of the tetrahedral volumes to generate the collocation nodes, resulting in the parameters detailed by Table \ref{tab:large_coll_points}. The mesh settings that led to Table \ref{tab:large_coll_points} are shown in Table \ref{tab:large_mesh}.

\begin{table}[ht]
\centering
\caption{Number of collocation nodes for full proteins.}
  \begin{tabular}{|c|c|c|c|c|}
    \hline
     Case & $\Omega_m$ & $\Omega_w$  & $\Gamma$ & $\partial\Omega$  \\
    \hline
     1pgb & 33,252 & 110,941 & 10,092 & 23,976\\
     1ubq & 44,751 & 159,199 & 13,064 & 33,122\\
  \hline
  \end{tabular}%
\label{tab:large_coll_points}%
\end{table}%

\begin{table}[ht]
\centering
\caption{Mesh settings to generate collocation nodes for 1pgb and 1ubq.}
  \begin{tabular}{|c|c|c|c|c|}
    \hline
     \multicolumn{2}{|c|}{Surf. dens.}  & \multicolumn{2}{c|}{Max. vol.} &     \\
      \multicolumn{2}{|c|}{vert/\AA$^2$}  & \multicolumn{2}{c|}{size \AA$^3$} &     \\
     \cline{1-4}
      $\Gamma$ & $\partial\Omega$  & $\Omega_m$ & $\Omega_w$ &  sample  \\
    \hline
      1.8 & 1.6 & 0.6 & 2.0 & 70\% \\ 
  \hline
  \end{tabular}%
\label{tab:large_mesh}%
\end{table}%



\begin{figure}[ht]
  \centering
  \subfigure[Solvation energy]{\includegraphics[width=0.3\textwidth]{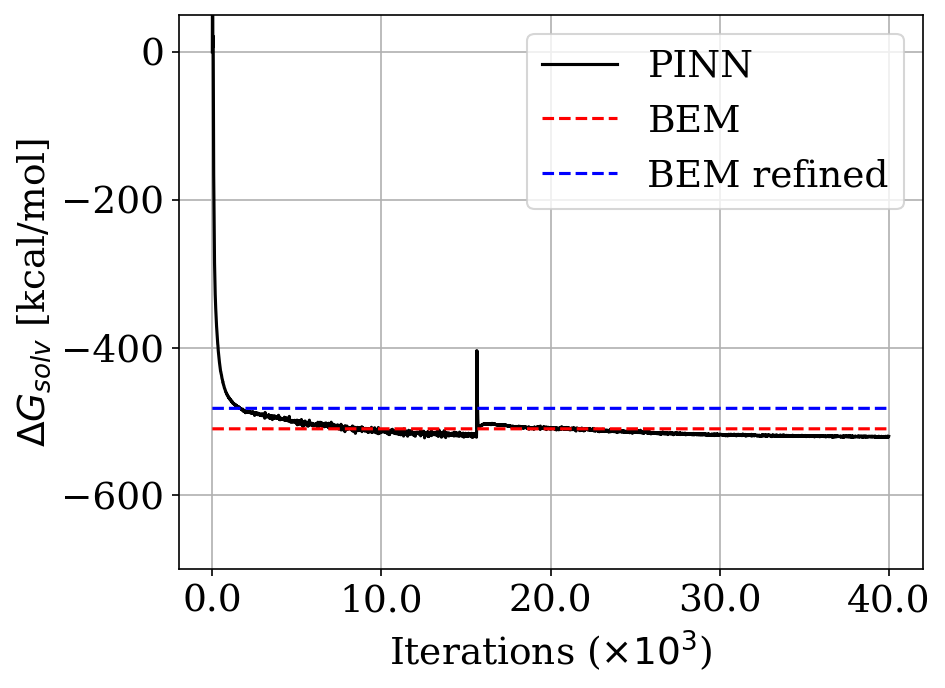}}
  \subfigure[Losses]{\includegraphics[width=0.3\textwidth]{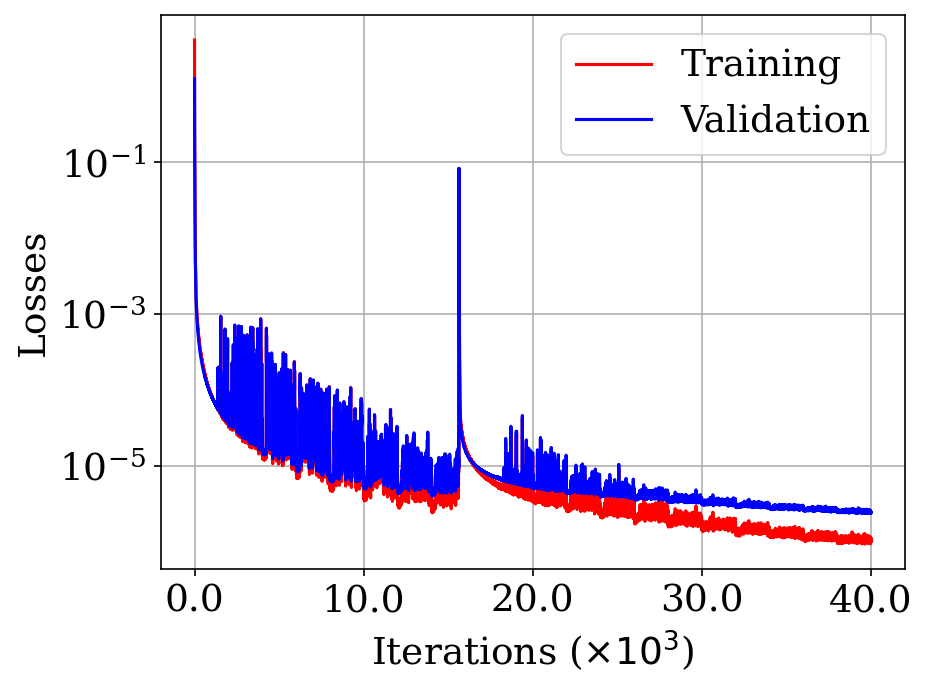}}
    \subfigure[Loss terms]{\includegraphics[width=0.3\textwidth]{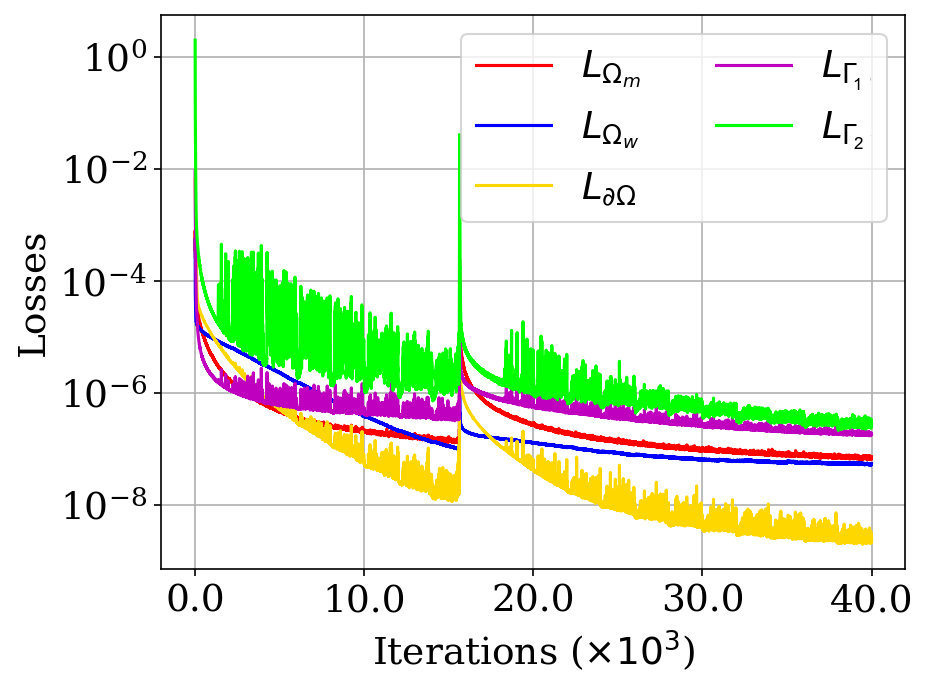}}
  \caption{Solvation energy and losses evolution for 1pgb}
  \label{fig:1pgb_loss}
\end{figure}

\begin{figure}[ht]
  \centering
  \subfigure[Solvation energy]{\includegraphics[width=0.3\textwidth]{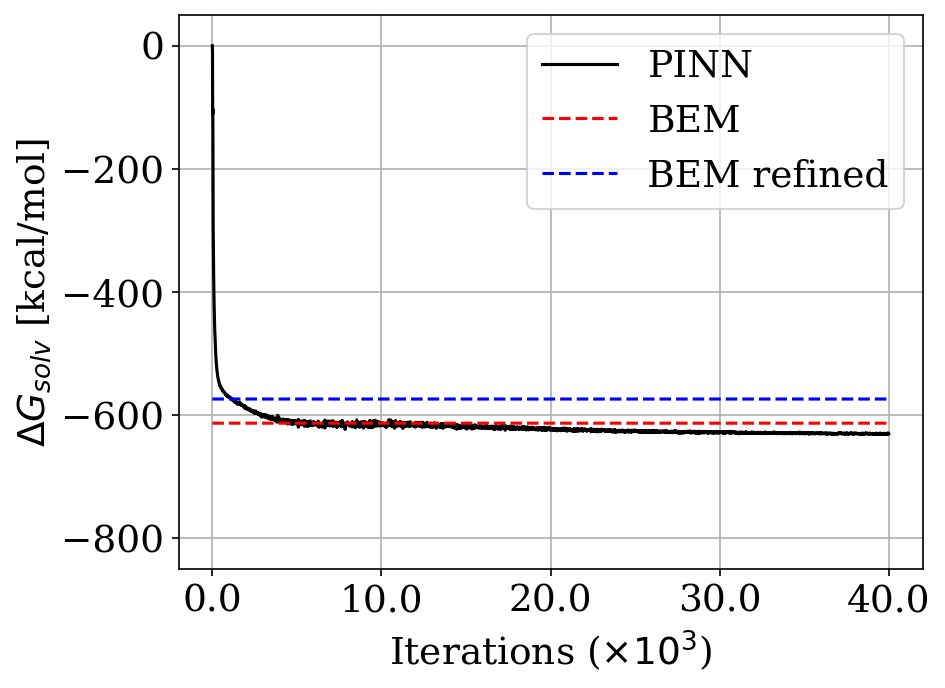}}
  \subfigure[Losses]{\includegraphics[width=0.3\textwidth]{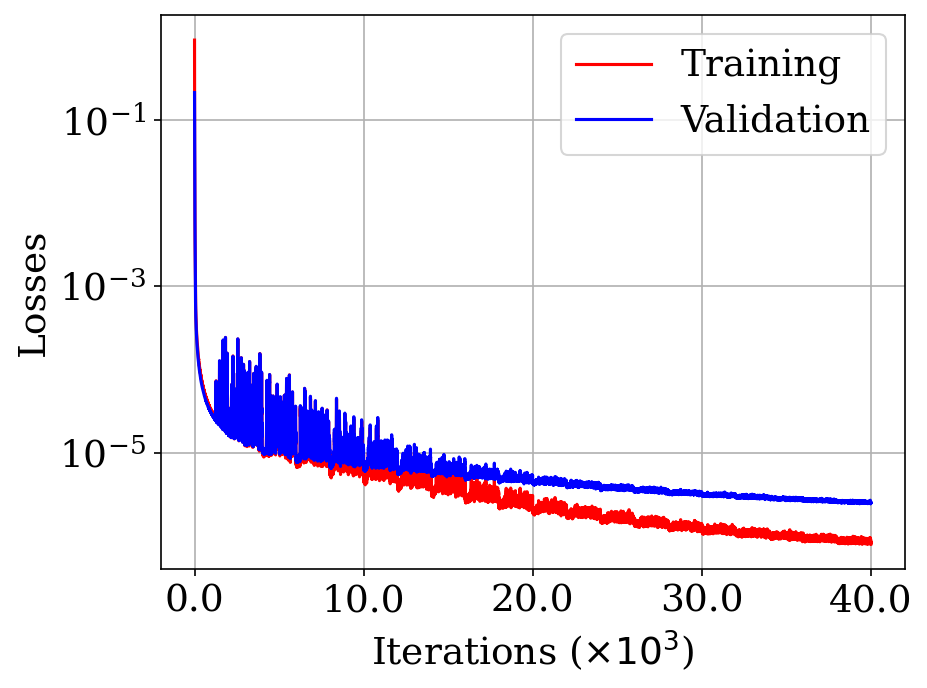}}
      \subfigure[Loss terms]{\includegraphics[width=0.3\textwidth]{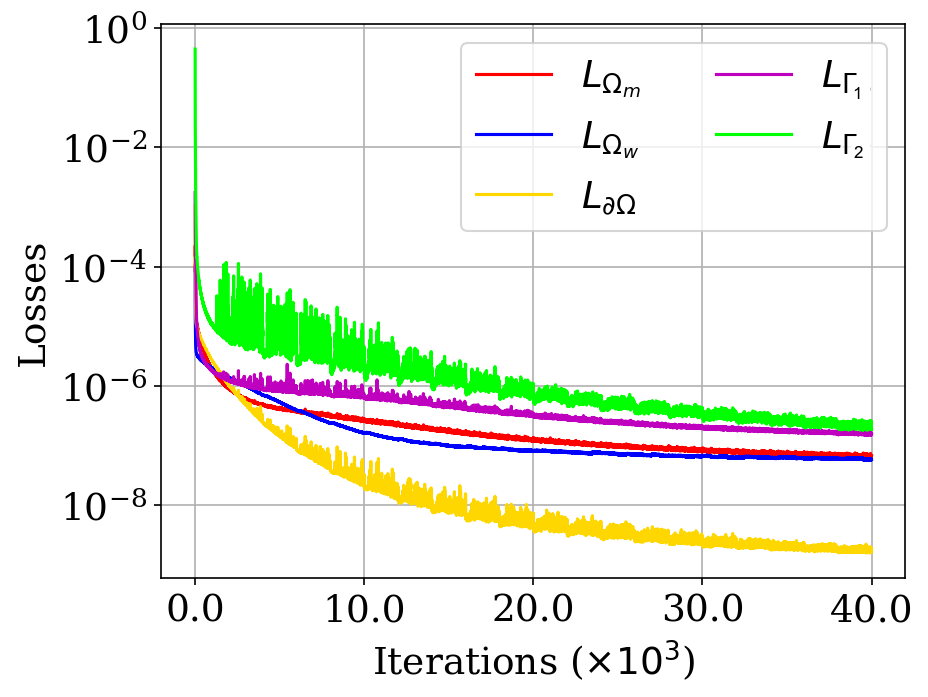}}
  \caption{Solvation energy and losses evolution for 1ubq}
  \label{fig:1ubq_loss}
\end{figure}

Table \ref{tab:res_proteins} contains the results for 1pgb and 1ubq after 40,000 iterations. The difference with a reference BEM solution  (using the same surface mesh as in the creation of the collocation nodes) is of the order of 10$^{-2}$ in both $\Delta G_{solv}$ and surface potential, which is similar to arginine in Table \ref{tab:res_arg_coll}. Considering these calculations run for 40,000 iterations, it is not surprising that the training and validation losses go lower than arginine, to 10$^{-6}$, however, by looking at the evolution of $\Delta G_{solv,\theta}$ in Figs. \ref{fig:1pgb_loss} and \ref{fig:1ubq_loss}, we see it has stagnated and has reasonable results by 7,000 iterations or less.  

\begin{figure}[ht]
  \centering
  \subfigure[$x$ axis]{\includegraphics[width=0.4\textwidth]{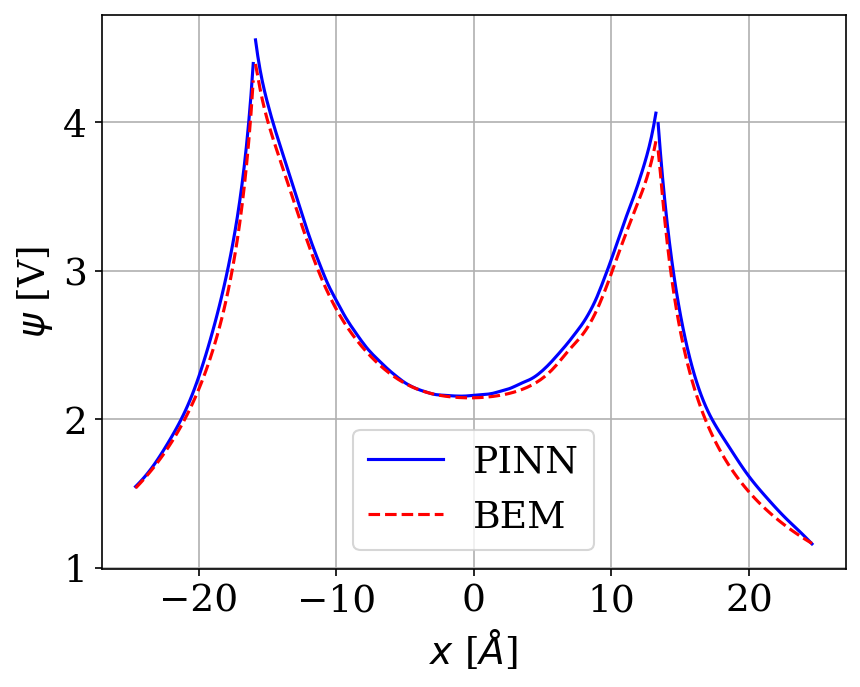}}
  \subfigure[$y$ axis]{\includegraphics[width=0.4\textwidth]{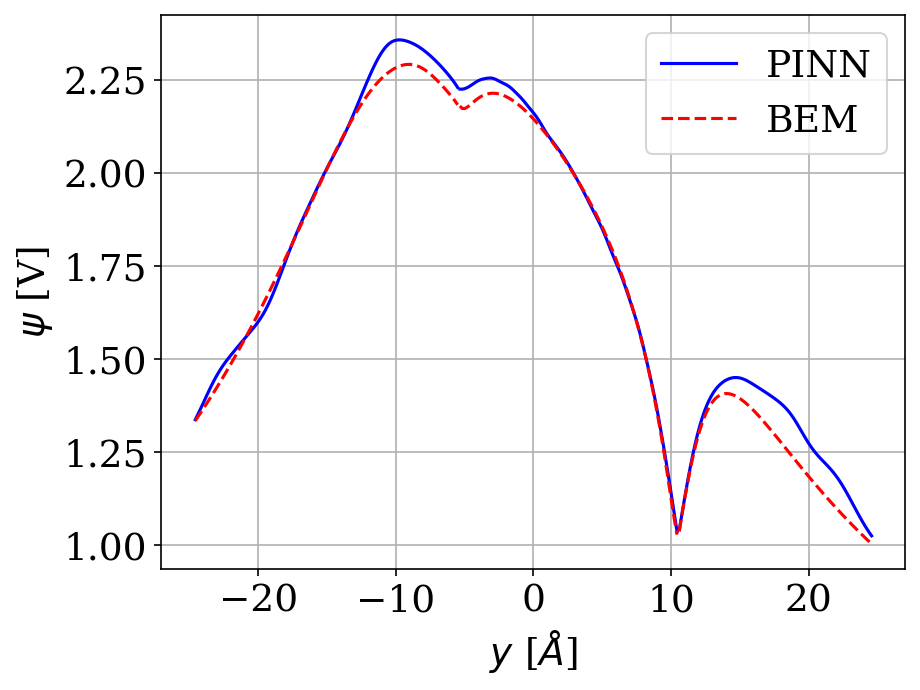}}
  \caption{Reaction potential ($\psi$) along the $x$ and $y$ axis for 1pgb comparing PINN (blue) and BEM (red).}
  \label{fig:1pgb_phi}
\end{figure}

\begin{figure}[ht]
  \centering
  \subfigure[$x$ axis]{\includegraphics[width=0.4\textwidth]{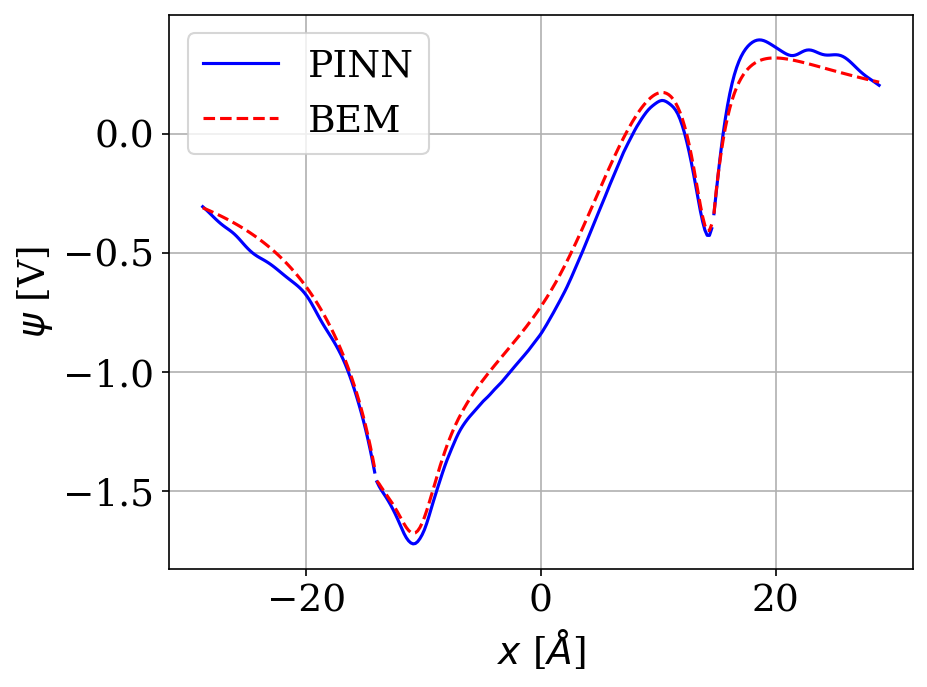}}
  \subfigure[$y$ axis]{\includegraphics[width=0.4\textwidth]{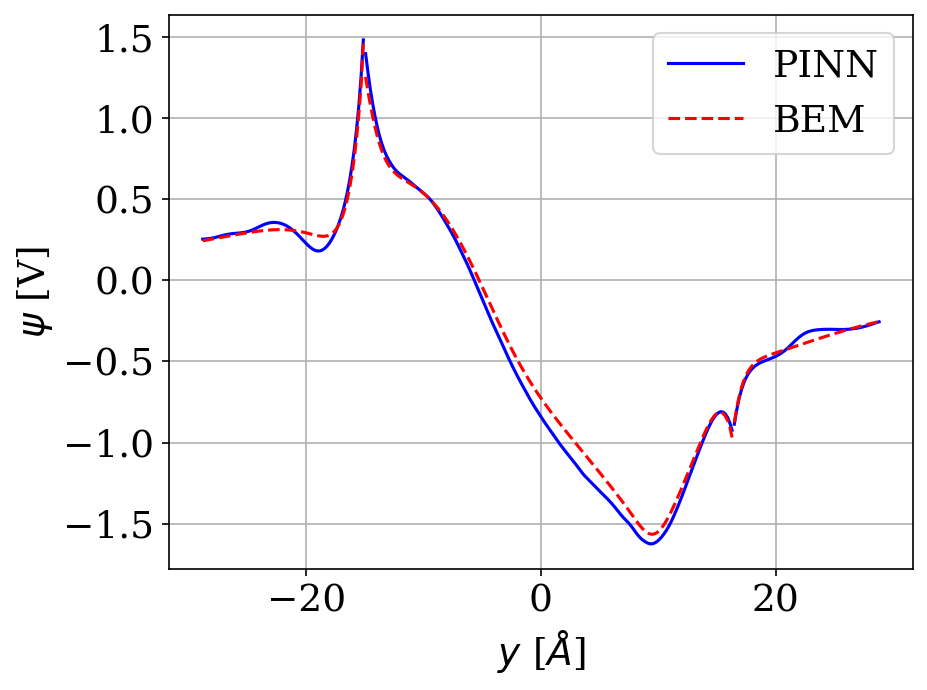}}
  \caption{Reaction potential ($\psi$) along the $x$ and $y$ axis for 1ubq comparing PINN (blue) and BEM (red).}
  \label{fig:1ubq_phi}
\end{figure}

Figs. \ref{fig:1pgb_phi} and \ref{fig:1ubq_phi} show the electrostatic potential along the $x$ and $y$ axis for 1pgb and 1ubq, respectively. Regardless of the high number of iterations and low difference in energy, differences in reaction potential are more evident. However, PINN is capable of reproducing the main features of the solution appropriately, like large gradient changes across the interface.

\begin{figure}[ht]
  \centering
  \subfigure[Reaction potential]{\includegraphics[width=0.4\textwidth]{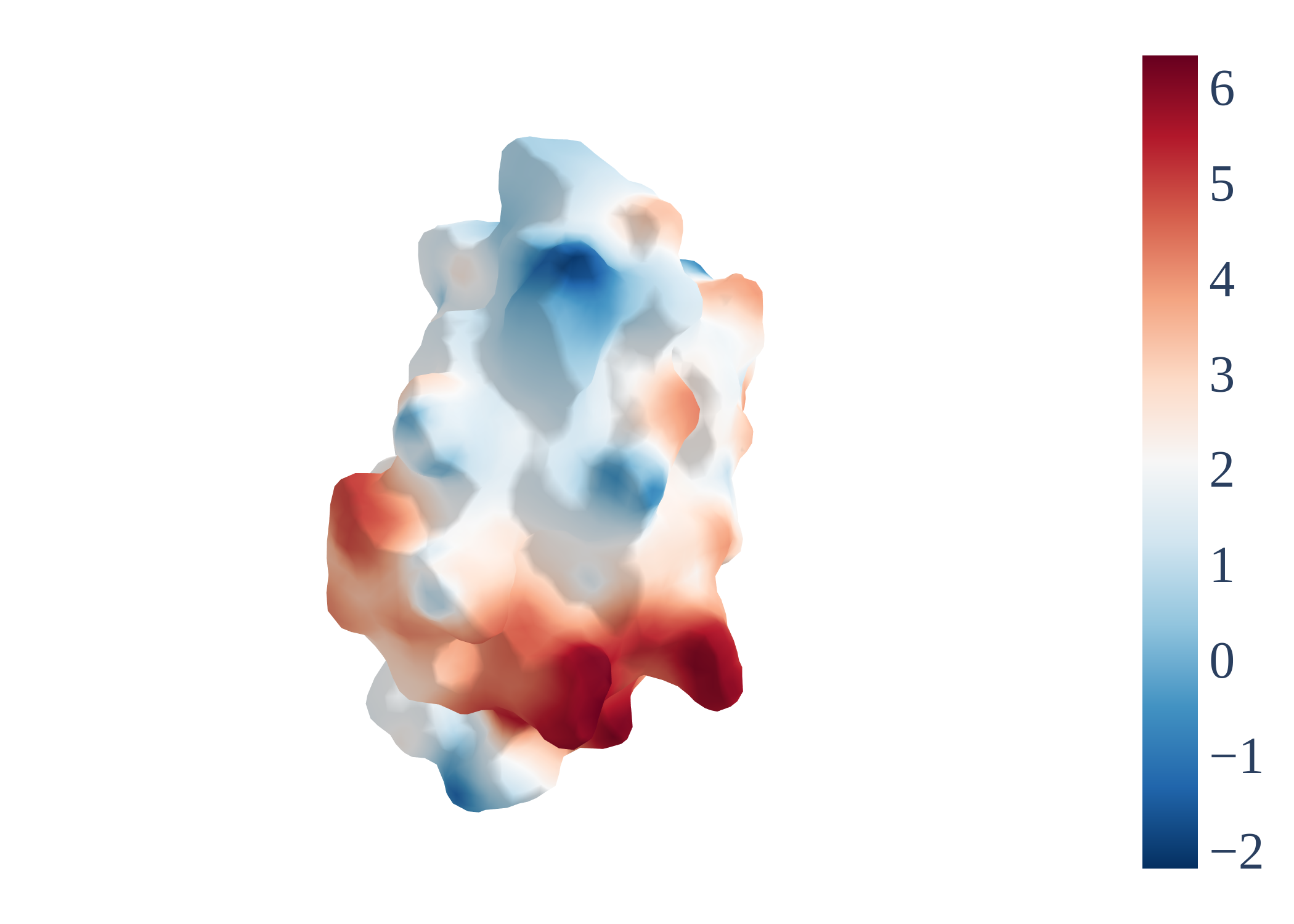}}
  \subfigure[Absolute error]{\includegraphics[width=0.4\textwidth]{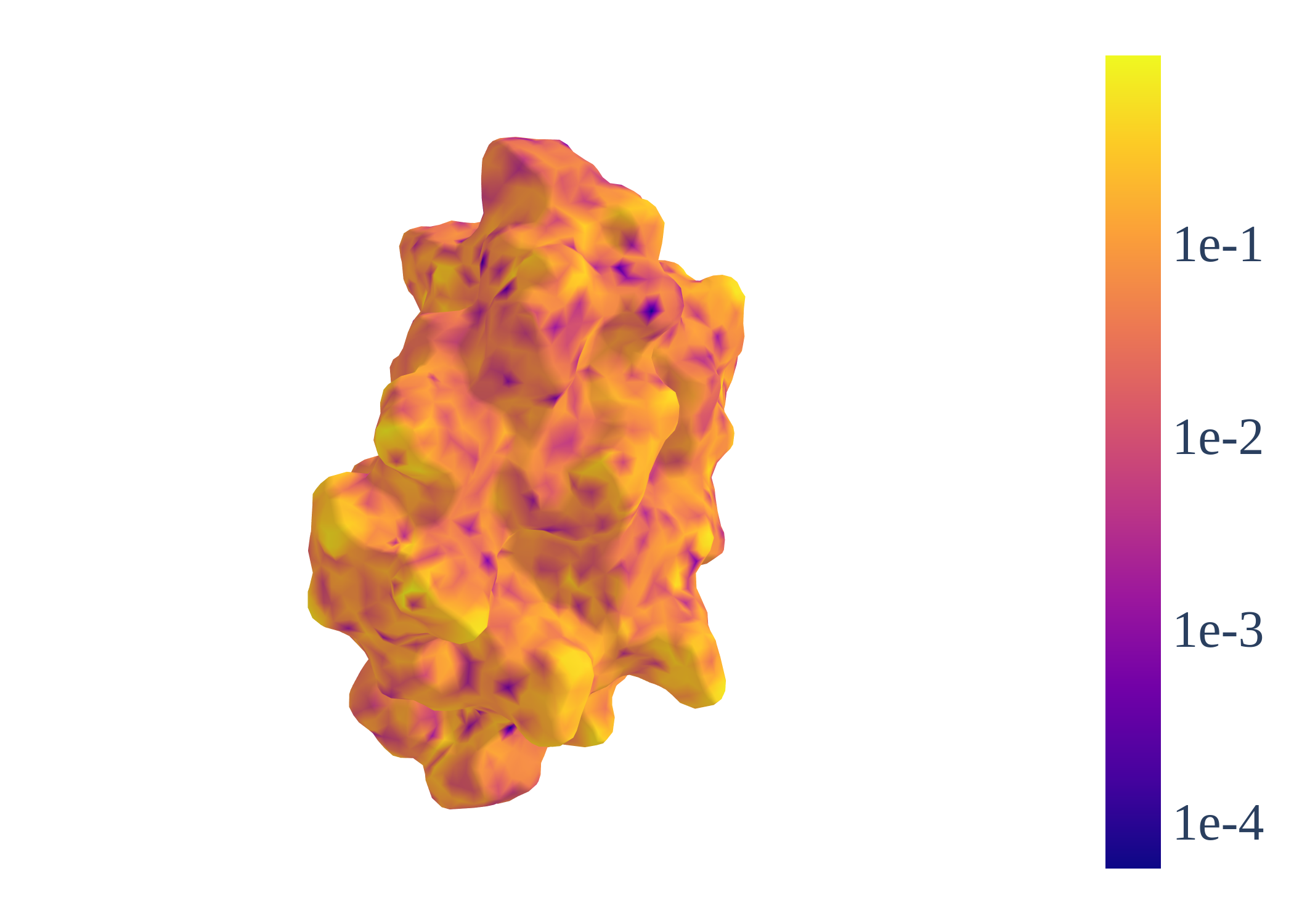}}
  \caption{Reaction potential ($\psi$) and absolute error (in Volts) at $\Gamma$ for 1pgb.}
  \label{fig:1pgb_phi_surf}
\end{figure}

\begin{figure}[ht]
  \centering
  \subfigure[Reaction potential]{\includegraphics[width=0.4\textwidth]{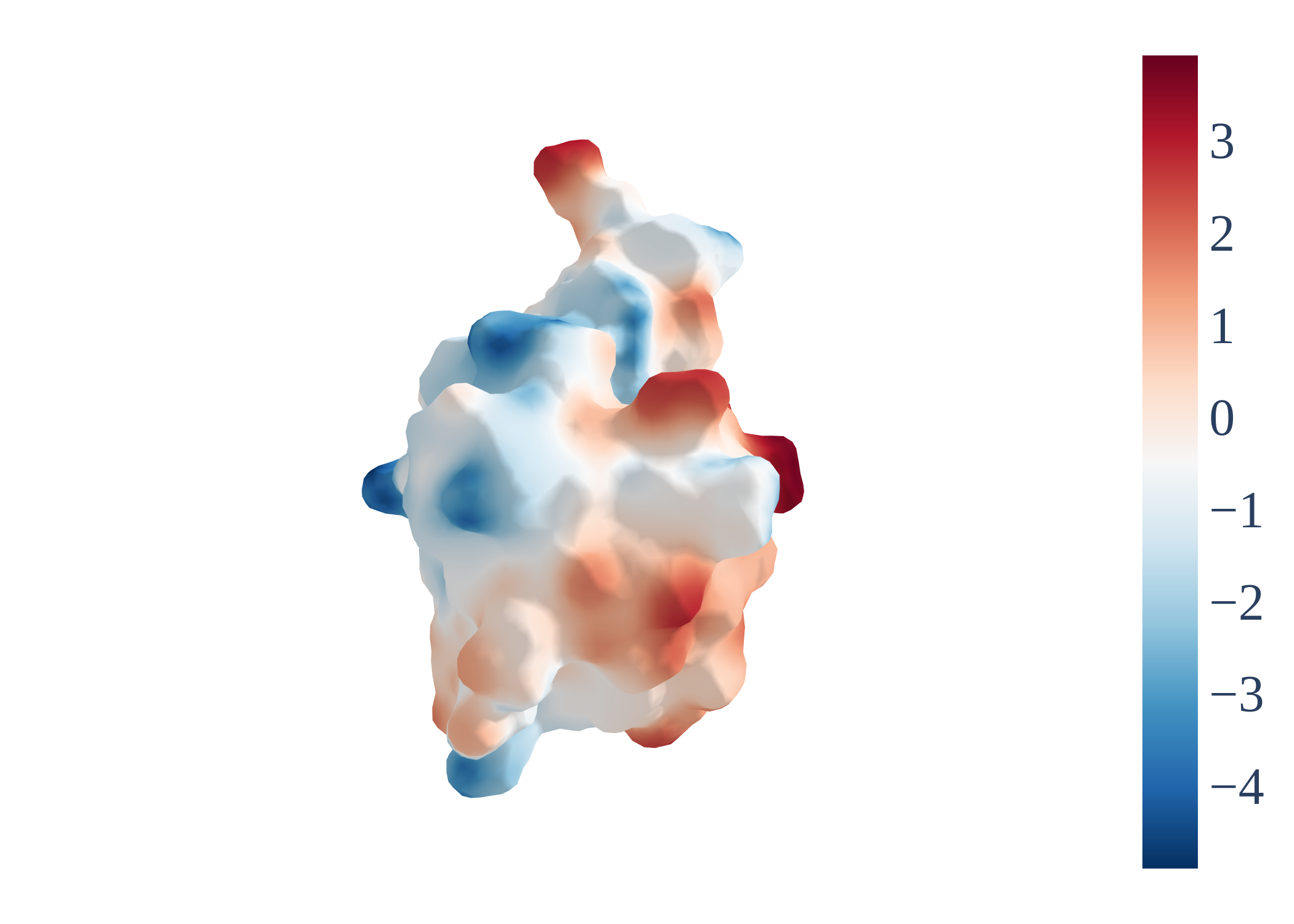}}
  \subfigure[Absolute error]{\includegraphics[width=0.4\textwidth]{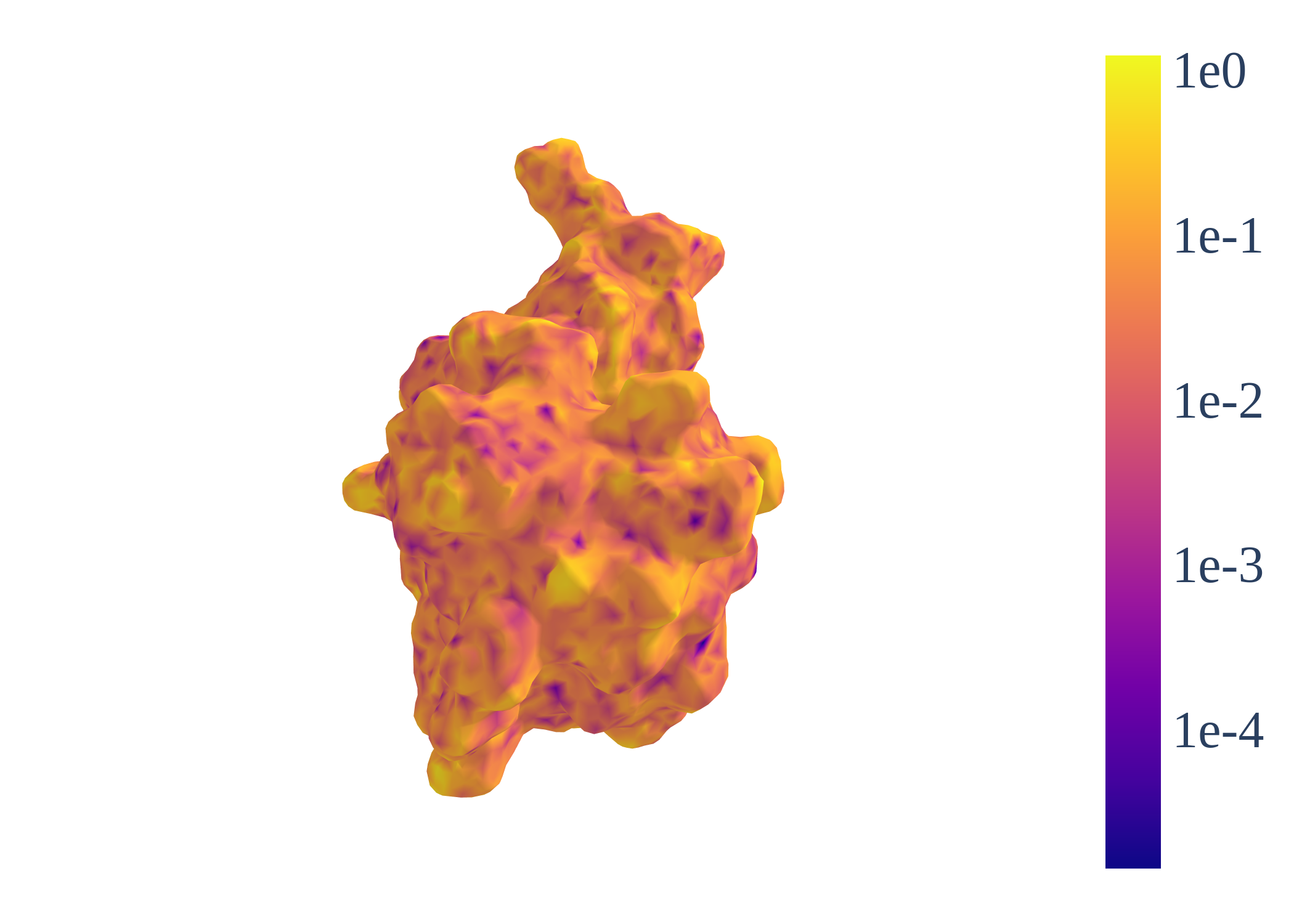}}
  \caption{Reaction potential ($\psi$) and absolute error (in Volts) at $\Gamma$ for 1ubq}
  \label{fig:1ubq_phi_surf}
\end{figure}

The low difference between BEM and PINN in the surface reaction potential ($\psi^{(\Gamma)}$) of Table \ref{tab:res_proteins} is further illustrated by the excellent agreement between the BEM and PINN solutions  in  Figs. \ref{fig:1pgb_phi_surf} and \ref{fig:1ubq_phi_surf}. This remarkable result opens possibilities to use PINN in applications where the electrostatic potential on the surface is key, like the detection of binding pockets in drug design.

\begin{table}[htbp]
  \centering
  \caption{Results for full proteins after 40,000 iterations.}
    \begin{tabular}{|c|c|c|c|c|c|}
      \hline
        Case  & $\Delta G_{solv,\theta}$ & $D_{G_{solv}}$  & $D_\psi$ & Training & Validation   \\
         &[kcal/mol]&  &  & loss & loss  \\
      \hline
    1pgb & -520.14 & 1.99E-02 & 3.41E-02 & 1.06E-06 & 2.50E-06  \\
    1ubq & -630.10 & 2.82E-02 & 4.69E-02 & 7.99E-07 & 2.52E-06  \\
    \hline
    \end{tabular}%
  \label{tab:res_proteins}%
\end{table}%

\section{\label{sec:chall}Challenges and future work}

The results the \rev{``Results and Discussion''} section are evidence that PINN is a viable alternative to solve the linear PBE in real proteins. Although this is a promising statement, in our exploration we identified a few challenges moving towards making PINN a mainstream tool in electrostatic calculations. We hope this section will inspire researchers to address them, in order to take full advantage of neural networks in computing the electrostatic potential in molecular systems. 

All results in this work correspond to calculations of the linearized version of the PBE. This is a good approximation for a large family of molecules, including many proteins, however, it becomes problematic for highly charged systems, like RNA and DNA, and the nonlinear PBE in Equation \eqref{eq:pbe} is required. 
\rev{Exploring and designing a PINN architecture for computation of electrostatic potential and solvation energy for highly charged molecules will form a future direction of research.}

Another area of potential improvement is the optimization algorithm. In this work we used the common Adam optimizer, however, there are well-known alternatives that may converge faster, like the limited memory Broyden–Fletcher–Goldfarb–Shanno (L-BFGS) algorithm.\cite{liu1989limited} 
Unfortunately the L-BFGS algorithm is not implemented in TensorFlow. We consider that exploring other optimization algorithms and memory efficiency strategies is a line of future work that may make PINN more competitive in front of traditional numerical methods.  

\rev{Incorporating other experimental information is also an interesting future challenge. Our results indicate that $\phi_{ENS}$ is successfully incorporated into the PINN framework, however, the PB equation already provides a good approximation of $\phi_{ENS}$.\cite{yu2021novo} Experimental data that does not agree well with the PB equation would generate a competition between the different loss functions, making it harder to solve numerically.}



\section{\label{sec:conc}Conclusions}
This work presents a thorough investigation of PINN to solve the PBE in molecular electrostatics. Starting from a basic implementation of PINN for an interface linear PBE problem using MLP, we explored the impact of different enhancement techniques using spherical test cases with available analytical results. We conclude that the best performing neural network architecture includes layers that scale the input and output, and add random Fourier features; and also that considers a trainable activation function, and a weight balancing algorithm. We further test this architecture on realistic molecular geometries, like a single arginine, protein G (1pgb), and ubiquitin (1ubq), where we also propose a collocation node sampling strategy that decreases the computational cost. In all cases, PINN was capable of reproducing the details of the electrostatic potential field correctly, and the solvation energy converged to a reference value up to $\sim$10$^{-2}-10^{-3}$\rev{, which is in the order of previous work using PINN for PDEs.} We find that including all the features described in this work is crucial for PINN to appropriately solve the linear PBE.

We also explore the possibility to consider experimental information to our model. Using arginine as a test case, we were able to incorporate a loss function that includes NMR-measured effective electrostatic potentials near hydrogen atoms. This is an important result, as it makes PINN stand out with respect to standard numerical methods that do not have this capability.


\rev{We note that there are numerous traditional methods for solving the PB equation (FDM, FEM, BEM, etc.), whereas the aim of this work is to explore the recently developed ML techniques. As such, we leave a direct comparison with the traditional solvers for future work. 
Currently, PINN is usually slower than standard numerical schemes in most applications,\cite{chuang2022experience,grossmann2024can} however, they have great potential to improve their performance, as it has happened in complex problems like weather modeling~\cite{KuEtAl23,BXZCGT23} where  a 10,000$\times$ speedup is observed.}

\section*{Data Availability Statement}
All data and software required to reproduce the results of this manuscript can be found in the GitHub repository \url{https://github.com/MartinAchondo/XPPBE}.

\begin{acknowledgement}

MAM thanks the support from ANID (Chile) through Beca de Mag\'ister Nacional 22230566.
CDC acknowledges the support from CCTVal through ANID PIA/APOYO AFB 220004 and Universidad T\'ecnica Federico Santa Mar\'ia through from Proyectos Internos PI-LIR-23-03.
shown in this document.

\end{acknowledgement}





\bibliography{main}

\end{document}